\documentclass[12pt,draftcls,onecolumn]{IEEEtran}     
\usepackage{color}
\ifCLASSINFOpdf
   \usepackage[pdftex]{graphicx}
   \DeclareGraphicsExtensions{.pdf,.png,.eps}
\else

\fi

\usepackage[tight,footnotesize]{subfigure}

\usepackage[font=footnotesize]{subfig}

\usepackage{subfig}
\usepackage{subfigure}
\usepackage{mathrsfs}
\usepackage{epsfig}
\usepackage{color}
\usepackage{bm}
\usepackage{multirow}
\usepackage{mathtools}

\begin{document}
\title{Bayesian Compressive Sensing for Ultrawideband Inverse Scattering in Random Media}
\author{A.~E. Fouda and F.~L. Teixeira 
\thanks{AEF was with ElectroScience Laboratory, Department
of Electrical and Computer Engineering, The Ohio State University, Columbus, OH
USA. He is presently with Sensor Physics and Technology Group, Halliburton Energy Services, Houston, TX, USA.}
\thanks{FLT is with ElectroScience Laboratory, Department
of Electrical and Computer Engineering, The Ohio State University, Columbus, OH
USA}
\thanks{e-mail:fouda.1@osu.edu, teixeira@ece.osu.edu}
}


\maketitle

\begin{abstract}
We develop an ultrawideband (UWB) inverse scattering technique for reconstructing continuous random media based on Bayesian compressive sensing. In addition to providing maximum a posteriori estimates of the unknown weights, Bayesian inversion provides estimate of the confidence level of the solution, as well as a systematic approach for optimizing subsequent measurement(s) to maximize information gain. We impose sparsity priors directly on spatial harmonics to exploit the spatial correlation exhibited by continuous media, and solve for their posterior probability density functions efficiently using a fast relevance vector machine. We linearize the problem using the first-order Born approximation which enables us to combine, in a single inversion, measurements from multiple transmitters and ultrawideband frequencies. We extend the method to high contrast media using the distorted-Born iterative method. We apply time-reversal strategies to adaptively focus the inversion effort onto subdomains of interest, and hence reduce the overall inversion cost. The proposed techniques are illustrated in a number of canonical scenarios including crosshole and borehole sensing. 

\end{abstract}


\IEEEpeerreviewmaketitle

\section{Introduction}

\IEEEPARstart{T}{he} goal of inverse scattering is to estimate unknown parameters of target(s) of interest from noisy (cluttered) measurements. Target parameters may include location, size, orientation, and material properties \cite{OceanAcustInv, EMMarine, MelesGPR1, MelesGPR2, InvDORT, DevaneyInvScatt}. Electromagnetic inverse scattering finds many applications in medical imaging \cite{BreastImagBook, BreastReview, BayesInfBrain, HabashyBiomedInv, HagnessBreast1, HagnessBreast2}, through-wall imaging \cite{thruwallSciencelong, UWBthruwall, THRUWALLLI}, non-destructive testing \cite{EMNDT, MwaveNDT1, MwaveNDT2}, ground penetrating radar \cite{TRGPRPlumb, TRGPRAsif, RappaportStatisticalGPR, PotinGPR} and geophysical exploration in general \cite{HabashyCrossWell, ThreDEMBook, MelesGPR1, MelesGPR2}.

In Bayesian-based inversion, both target's parameters and clutter are modeled as random variables with certain probability density functions (PDFs). The inversion algorithm combines (any) a priori information on the target's parameters, with physics-based forward-problem PDFs, and array acquisitions to produce a posteriori PDFs of the unknowns \cite{TarantolaBook, BayesianInvProbBook, LemmBook, MassaContSourc, MassaBorn, BayesInfBrain, OceanAcustInv, EMMarine}. This approach provides means for measuring the confidence interval of the inversion and adaptively optimizing subsequent measurement(s) \cite{BayesExpDes, BCSCarin}. Bayesian inference applied to compressive sensing was presented in \cite{BCSCarin, MTCarin}, where sparsity priors were imposed on a compressible (sparse) set of unknowns. That problem was solved efficiently using the relevance vector machine (RVM) technique \cite{TippingSparse, TippingFastRVM}. Recently, Bayesian compressive sensing has been applied in microwave imaging of sparse discrete scatterers using single frequency data, and the contrast source formulation \cite{MassaContSourc}, or the first order Born approximation \cite{MassaBorn}. Bayesian compressive sensing, combined with signal subspace methods, for imaging discrete targets has been presented in \cite{MarengoCS1, MarengoCS2, MarengoCS3}. 

In non-Bayesian inverse scattering techniques, a cost function, to be iteratively minimized, is defined, and optimization techniques, such as the conjugate gradient method \cite{HaradaConjGrad}, are used to guide the iterations. This approach is computationally costly, since it requires forward problem solution to compute the cost function and check convergence at each iteration step \cite{MelesGPR1, MelesGPR2,  Moghadam1, Weedon1, MoraInv, WangDBIM}. Bayesian inversion alleviate the need for that since it has a `built-in' measure for accuracy through the confidence level it provides. In addition, Bayesian compressive sensing (BCS) solved using the RVM, as presented in \cite{TippingSparse, BCSCarin}, provides an elegant, closed-form solution for the posterior PDF; and therefore, there is no need for (costly) numerical computations of higher-order integrations that are done otherwise using Markov Chain Monte Carlo and Gibbs sampling \cite{PDFSampling, Gibbssampling}, as in \cite{OceanAcustInv, EMMarine, BayesInfBrain, ThreDEMBook} for example. Finally, BCS with RVM does not require inversion of the projection matrix relating measurements to model parameters. Note that, this matrix may not be square, where for example the number of measurements is less than the number of unknowns, and it can be ill-conditioned, making it highly sensitive to noise. 

In this paper, we develop BCS-assisted ultrawideband (UWB) inverse scattering techniques for reconstructing continuous random media properties. We exploit frequency decorrelation of the UWB interrogating signal to produce a statistically stable inversion, which does not depend on the particular realization of the (random) clutter but only on its statistical properties \cite{Fouda2, Fouda4}. We start by presenting a summary of BCS as applied to a linear regression model following \cite{TippingSparse, BCSCarin, MassaContSourc}. Then, we apply that model to the electromagnetic inverse scattering problem under a first-order Born approximation, as in \cite{MassaBorn}, but extended here to incorporate, in a single inversion, ultrawideband multistatic measurements, and apply it to continuous random media rather than to (sparse) discrete scatterers. We present several examples based on numerical simulations to assess the performance of the proposed technique under different scenarios. Next, we present an adaptive sensing approach for determining successive measurement locations so as to maximize the differential information gain. After that, a new technique denoted as time-reversal-assisted localized-inversion (TRALI) is introduced to reduce the computational cost of the inversion algorithm by adaptively focusing the inversion effort onto sub-domains of interest. The method is then extended to high contrast media (i.e., those that do not conform with the first-order Born approximation) by introducing the Bayesian Distorted-Born Iterative Method (BDBIM). 

It should be pointed out that the term `compressive sensing' is used here in a broad sense
to refer to problems where the unknown function (target locations and properties in our case) can be expressed as a sparse set of weights w.r.t. some expansion bases. This is done to conform to the prior usage in
 \cite{MassaContSourc, MassaBorn} and does not match the more formal usage of the term compressive sensing that refers to recovering certain signals from sparse data acquisitions (i.e., using less samples or measurements than dictated by the Nyquist criterion) \cite{CandesCS, PotterCS}. In particular, in this work as well as in \cite{MassaContSourc, MassaBorn}, the number of measurements can be larger than the number of unknowns. 

\section{Bayesian Compressive Sensing using the Relevance Vector Machine}

Consider a linear regression model, where a vector $\textbf{y}$ of $N$ noisy measurements is related to a vector $\textbf{w}$ of $M$ (unknown) weights through the linear relationship  
\begin{equation}
\label{equnCH6_1} 
\textbf{y}=\textbf{B}\textbf{w}+\textbf{n}
\end{equation}
where $\textbf{B}$ is the projection matrix, and $\textbf{n}$ is a vector of additive noise. We are seeking maximum a posteriori (MAP) estimates for the weights as follows   
\begin{equation}
\label{equnCH6_2} 
\mathop{\hat{\textbf{w}}=\arg \max }\limits_{\textbf{w}} \left(p(\textbf{w}\left|\textbf{y}\right. )\right)
\end{equation}
The main challenge is trying to avoid `over-fitting' the \textit{noisy} measurements \cite{TippingSparse}. 
From Bayes' rule, the posterior PDF is given by
\begin{equation}
\label{equnCH6_3} 
p(\textbf{w}\left|\textbf{y}\right. )=\frac{p(\textbf{y}\left|\textbf{w}\right. )p(\textbf{w})}{p(\textbf{y})} 
\end{equation}
Let's consider each term of the above PDF; we start with the prior $p(\textbf{w})$. If we have a priori knowledge that the weights vector is sparse, meaning that only few number of weights are non-zero, then a reasonable choice for the prior will be a sparsity prior such as the Laplace PDF.
However, using such prior, a closed form solution for the posterior PDF cannot be obtained \cite{BCSCarin}. An alternative approach, introduced in \cite{TippingSparse}, is to use hierarchical priors by defining $p(\textbf{w})$ through a vector of hyperparameters $\bm{\alpha}$ as follows 
\begin{equation}
\label{equnCH6_5} 
p(\textbf{w})=\int p(\textbf{w}\left|\bm{\alpha} \right. ) p(\bm{\alpha})d\bm{\alpha} 
\end{equation}
where the conditional PDF is defined as 
\begin{equation}
\label{equnCH6_6} 
p(\textbf{w}\left|\bm {\alpha}\right.)=\prod _{i=1}^{M}{\mathscr{N}} (w_{i} \left|0,\alpha _{i}^{-1} \right.)
\end{equation}
in which the hyperparameters are the reciprocals of the variances of the zero-mean normal distributions. The hyperparameters themselves are assumed to be distributed according to the following Gamma distribution
\begin{equation}
\label{equnCH6_7} 
p(\bm{\alpha} )=\prod _{i=1}^{M}\Gamma (\alpha _{i} \left|a,b\right. ) 
\end{equation}
with $a$ and $b$ being the scale parameters of the Gamma distribution. The resulting prior in (\ref{equnCH6_5}) is a student-\textit{t} distribution that, with appropriate choice of scale parameters, is highly peaked at zero, thus favoring sparsity \cite{TippingSparse} \footnote{A reasonable choice, adopted in \cite{TippingSparse}, is to set scale parameters $a$ and $b$ to zero. In this case, $p(\ln(\bm{\alpha}))$ is uniform, i.e. the hyperparameters become scale invariant.}.
Now, consider the likelihood $p(\textbf{y}\left|\textbf{w}\right.)$. Assuming independent, zero-mean, Gaussian noise with variance $\sigma _{n}^{2}$, the likelihood can be written as  
\begin{equation}
\label{equnCH6_8} 
p(\textbf{y}\left|\textbf{w}\right. )=\int p(\textbf{y}\left|\textbf{w}\right.,\sigma _{n}^{2} ) p(\sigma _{n}^{2} )d\sigma _{n}^{2} 
\end{equation}
where
\begin{equation}
\label{equnCH6_9} 
p(\textbf{y}\left|\textbf{w}\right.,\sigma _{n}^{2} )=(2\pi \sigma _{n}^{2} )^{-N/2} \exp (\frac{-1}{2\sigma _{n}^{2} } \left\| \textbf{y}-\textbf{B}\textbf{w}\right\| ^{2} )
\end{equation}
and the reciprocal of the noise variance is distributed according to the following Gamma distribution with parameters $c$ and $d$
\begin{equation}
\label{equnCH6_10} 
p(\sigma _{n}^{2} )=\Gamma (\sigma _{n}^{-2} \left|c,d\right. )
\end{equation}
Combining (\ref{equnCH6_5}), (\ref{equnCH6_8}) and (\ref{equnCH6_9}), the posterior becomes
\begin{equation}
\label{equnCH6_11} 
p(\textbf{w}\left|\textbf{y}\right. )=\int\!\!\!\int \frac{p(\textbf{y}\left|\textbf{w}\right. ,\sigma _{n}^{2} )p(\sigma _{n}^{2} )p(\textbf{w}\left|\bm{\alpha} \right. )p(\bm{\alpha} )}{p(\textbf{y})} d\bm{\alpha} d \sigma _{n}^{2} 
\end{equation}
which can be simplified to
\begin{equation}
\label{equnCH6_12} 
p(\textbf{w}\left|\textbf{y}\right. )=\int\!\!\!\int p(\textbf{w},\bm{\alpha} ,\sigma _{n}^{2} \left|\textbf{y}\right. )d\bm{\alpha} d \sigma _{n}^{2} 
\end{equation}
$p(\textbf{w},\bm{\alpha} ,\sigma _{n}^{2} \left|\textbf{y}\right. )$ is the joint posterior PDF of all unknowns, and can be factorized as follows 
\begin{equation}
\label{equnCH6_13} 
p(\textbf{w},\bm{\alpha} ,\sigma _{n}^{2} \left|\textbf{y}\right. )=p(\textbf{w}\left|\textbf{y}\right. ,\bm{\alpha} ,\sigma _{n}^{2} )p(\bm{\alpha} ,\sigma _{n}^{2} \left|\textbf{y}\right.)
\end{equation}
The first term in the r.h.s. of (\ref{equnCH6_13}) can be expanded as 
\begin{equation}
\label{equnCH6_14} 
p(\textbf{w}\left|\textbf{y}\right. ,\bm{\alpha} ,\sigma _{n}^{2} )=\frac{p(\textbf{y}\left|\textbf{w}\right. ,\sigma _{n}^{2} )p(\textbf{w}\left|\bm{\alpha} \right. )}{p(\textbf{y}\left|\bm{\alpha} ,\sigma _{n}^{2} \right.)}
\end{equation}
where
\begin{equation}
\label{equnCH6_15} 
p(\textbf{y}\left|\bm{\alpha} ,\sigma _{n}^{2} \right. )=\int p(\textbf{y}\left|\textbf{w}\right. ,\sigma _{n}^{2} )p(\textbf{w}\left|\bm{\alpha} \right. ) d\textbf{w}\\
=\frac{\exp \left[-\frac{1}{2} (\textbf{y}^{T} \textbf{C}^{{-1}} \textbf{y})\right]}{(2\pi )^{N/2} \sqrt{\left|\textbf{C}\right|} }
\end{equation}
with $\textbf{C}\coloneqq \sigma _{n}^{2} \textbf{I}+\textbf{B}\textbf{A}^{-1} \textbf{B}^{T}$, and $\textbf{A}\coloneqq \rm{diag}(\boldsymbol{\alpha})$.
Using (\ref{equnCH6_9}), (\ref{equnCH6_6}) and (\ref{equnCH6_15}) in (\ref{equnCH6_14}), 
\begin{equation}
\label{equnCH6_16} 
p(\textbf{w}\left|\textbf{y}\right.,\bm{\alpha},\sigma _{n}^{2})=(2\pi )^{-(N+1)/2} \left|\bm{\Sigma} \right|^{-1/2}\\
\times \exp \left(-\frac{1}{2} (\textbf{w}-\bm{\mu} )^{T} \bm{\Sigma} ^{-1} (\textbf{w}-\bm{\mu} )\right)
\end{equation}
where $\boldsymbol{\Sigma} =(\sigma _{n}^{-2} \textbf{B}^{T} \textbf{B}+\textbf{A})^{-1}$ and $\boldsymbol{\mu} =\sigma _{n}^{-2} \boldsymbol{\Sigma} \textbf{B}^{T}\textbf{y}$.
This is the sought posterior PDF of the weights once the hyperparameters $\bm{\alpha}$, that are embedded in $\textbf{A}$, and the noise variance $\sigma _{n}^{2}$ are determined. 
Towards this end, the second term in the r.h.s. of (\ref{equnCH6_13}) can be approximated by an impulse centered around the most probable (MP) values of $\bm{\alpha}$ and $\sigma _{n}^{2}$, as follows
\begin{equation}
\label{equnCH6_17} 
p(\bm{\alpha} ,\sigma _{n}^{2} \left|\textbf{y}\right. )\approx \delta (\bm{\alpha} -\left\{\bm{\alpha} \right\}_{MP} ,\sigma _{n}^{2} -\left\{\sigma _{n}^{2} \right\}_{MP} )
\end{equation}
Note that $p(\bm{\alpha} ,\sigma _{n}^{2} \left|\textbf{y}\right. )\propto p(\textbf{y}\left|\bm{\alpha} ,\sigma _{n}^{2} \right. )p(\bm{\alpha} )p(\sigma _{n}^{2} )$, and by properly adjusting the scale parameters $a$, $b$, $c$ and $d$, $\left\{\bm{\alpha} \right\}_{MP}$ and $\left\{\sigma _{n}^{2} \right\}_{MP}$ can be obtained by only maximizing the marginal likelihood $p(\textbf{y}\left|\bm{\alpha} ,\sigma _{n}^{2} \right. )$, i.e. 
\begin{equation}
\label{equnCH6_19} 
\left(\left\{\bm{\alpha} \right\}_{MP} ,\left\{\sigma _{n}^{2} \right\}_{MP} \right)=\mathop{\arg \max }\limits_{\bm{\alpha} ,\sigma _{n}^{2} } \left\{\frac{\exp \left[-\frac{1}{2} (\textbf{y}^{T} \textbf{C}^{-1} \textbf{y})\right]}{(2\pi )^{N/2} \sqrt{\left|\textbf{C}\right|} } \right\}
\end{equation}
This is known as  type-II maximum-likelihood process, and can be efficiently solved using the fast relevance vector machine (RVM) presented in \cite{TippingFastRVM}. Finally, from (\ref{equnCH6_16}), the MAP estimates and the corresponding covariance matrix are given by
\begin{equation}
\label{equnCH6_20} 
\hat{\textbf{w}}=\left. \bm{\mu} \right|_{\left(\left\{\bm{\alpha} \right\}_{MP} ,\left\{\sigma _{n}^{2} \right\}_{MP} \right)}
\end{equation}
and
\begin{equation}
\label{equnCH6_21} 
\textrm{cov}(\textbf{w})=\left. \bm{\Sigma} \right|_{\left(\left\{\bm{\alpha} \right\}_{MP} ,\left\{\sigma _{n}^{2} \right\}_{MP} \right)}
\end{equation}

\section{UWB Inverse Scattering based on the Born Approximation}
So far, we have discussed a Bayesian compressive sensing (BCS) solution for a generic \textit{linear} regression model relating noisy measurements to \textit{sparse} weights. To apply this model to the EM inverse scattering problem, we first have to be able to write the scattered field as a linear combination of the model weights. One way to do this, is by using the contrast source formulation presented in \cite{MassaContSourc}. Another way, is by using the first order Born approximation as presented in \cite{MassaBorn}. In this work, we choose to use the latter approach for two reasons that will be clarified shortly. 

Under the first order Born approximation, the scattered field at spatial location $\textbf{r}$ and frequency $\omega_k$ resulting from incident field $E^{inc}_{t}$ generated by transmitter $t$, is given by
\begin{equation}
\label{equnCH6_22} 
E^{s}_{t} (\textbf{r},\omega_k)=\int _{D}\tau (\textbf{r}',\omega_k)E^{inc}_{t} (\textbf{r}',\omega_k)G(\textbf{r},\textbf{r}',\omega_k)d\textbf{r}'
\end{equation}
where $G$ is the 2-D scalar Green's function, and $D$ is the support of the scattering object. $\tau$ is the complex contrast function given by
\begin{equation}
\label{equnCH6_23} 
\tau (\textbf{r},\omega_k)=\left[\epsilon _{r} (\textbf{r})-\left\langle \epsilon_{r}\right\rangle \right]-j\left[\frac{\sigma (\textbf{r})-\left\langle \sigma\right\rangle}{\omega_k \epsilon _{0} } \right]\\
=\Delta \epsilon _{r}(\textbf{r})-j\frac{1}{\omega_k \epsilon _{0} }\Delta \sigma(\textbf{r})
\end{equation}
with $\epsilon_{r}$ and $\sigma$ being the relative permittivity and conductivity, respectively, and $\left\langle \epsilon_{r}\right\rangle$ and $\left\langle \sigma\right\rangle$ the corresponding mean values of the background medium. Real and imaginary parts of the scattered field, recorded at $N_s$ sensors, can be stacked in a column vector as follows
\begin{equation}
\label{equnCH6_24} 
\textbf{e}_{t,k}^{s} =\left[\begin{array}{c} {\textrm{Re}\{ E_{t}^{s}(\textbf{r}_1,\omega_k) \} } \\ {\vdots } \\ {\textrm{Re}\{  E_{t}^{s}(\textbf{r}_{N_s},\omega_k) \} \} } \\ {\textrm{Im}\{ E_{t}^{s}(\textbf{r}_1,\omega_k) \} } \\ {\vdots } \\ {\textrm{Im}\{ E_{t}^{s}(\textbf{r}_{N_s},\omega_k) \} } \end{array}\right]
\end{equation}
Discretizing the domain of investigation into $N_p$ pixels, with pixel size $D_p$, and assuming pulse-basis function expansion for the contrast \cite{MassaBorn}, the projection matrix can be constructed as follows
\begin{equation}
\label{equnCH6_25} 
\textbf{G}_{t,k} =\left[\begin{array}{cccccc} {\textrm{Re}\{ g_{1,t}(\textbf{r}_1,\omega_k) \} } & {\cdots } & {\textrm{Re}\{ g_{N_p,t}(\textbf{r}_1,\omega_k) \} } & {\textrm{Im}\{ \frac{g_{1,t}(\textbf{r}_1,\omega_k)}{\omega _{k} \epsilon _{0}} \}} & {\cdots } & \textrm{Im}\{ \frac{g_{N_p,t}(\textbf{r}_1,\omega_k)}{\omega _{k} \epsilon _{0}} \} \\ {\vdots } & {\ddots } & {} & {} & {} & {} \\ {\textrm{Re}\{ g_{1,t}(\textbf{r}_{N_{s}},\omega_k) \} } & {} & {} & {} & {} & {} \\ {\textrm{Im}\{ g_{1,t}(\textbf{r}_{1},\omega_k) \} } & {\cdots } & {\textrm{Im}\{ g_{N_p,t}(\textbf{r}_1,\omega_k) \} } & -\textrm{Re}\{ \frac{g_{1,t}(\textbf{r}_{1},\omega_k)}{\omega _{k} \epsilon _{0}}  \}  & {\cdots } & -\textrm{Re}\{ \frac{g_{N_p,t}(\textbf{r}_1,\omega_k)}{\omega _{k} \epsilon _{0}} \} \\ {\vdots } & {\ddots } & {} & {} & {} & {} \\ {\textrm{Im}\{ g_{1,t}(\textbf{r}_{N_{s}},\omega_k) \} } & {} & {} & {} & {} & {} \end{array}\right]
\end{equation}
where
\begin{equation}
\label{equnCH6_26} 
g_{p,t}(\textbf{r}_n,\omega_k) =\int _{D _{p} }E^{inc}_{t} (\textbf{r}',\omega_k)G(\textbf{r}_{n} ,\textbf{r}',\omega _{k} )d\textbf{r}'
\end{equation}
Equation (\ref{equnCH6_22}) can now be written in matrix form as follows
\begin{equation}
\label{equnCH6_27} 
\textbf{e}_{t,k}^{s} =\textbf{G}_{t,k}\left(\textbf{F}^{-1} \textbf{t} \right)+\textbf{n}_{t,k}
\end{equation}
where $\textbf{F}^{-1} \textbf{t}$ is the \textit{real} contrast vector, given by
\begin{equation}
\label{equnCH6_28} 
\textbf{F}^{-1} \textbf{t} =\left[\begin{array}{c} {\Delta \epsilon _{r}(\textbf{r}_{1}) } \\ {\vdots } \\ {\Delta \epsilon _{r}(\textbf{r}_{N_{p}}) } \\ {\Delta \sigma(\textbf{r}_{1}) } \\ {\vdots } \\ {\Delta \sigma(\textbf{r}_{N_{p}}) } \end{array}\right]
\end{equation}
$\textbf{F}^{-1}$ is the inverse Fourier transform matrix, and $\textbf{t}$ is the vector of the spatial harmonics of the real contrast function. 
$\textbf{G}_{t,k}\textbf{F}^{-1}$ can now be perceived as the projection matrix $\textbf{B}$ in (\ref{equnCH6_1}) and the unknown weights are the spatial harmonics $\textbf{t}$. Once the covariance matrix of $\textbf{t}$ is solved for by the RVM, the covariance matrix of the real contrast vector can be computed as $\textrm{cov}(\textbf{F}^{-1} \textbf{t})= \textbf{F}^{-1} \textrm{cov}(\textbf{t})(\textbf{F}^{-1})^T$. 

Formulating the problem in terms of the spatial harmonics, rather than the contrast function itself, has the following advantages: 1. Spatial harmonics conform better with the sparsity requirement of the model, since the solution is likely to possess an amount of spatial correlation that would make the contrast function more sparse in the spatial harmonics domain than in the spatial domain. 2. Spatial harmonics provide better regularization for the solution; for example, one can choose to solve for a subset of the spatial harmonics according to the problem specifics and/or the available resources.
 
Measurements corresponding to illuminations from different transmitters can be stacked in one inversion and, assuming negligible dispersion over the utilized frequency band, multifrequency measurements
can be stacked in the same inversion as well (since $\textbf{t}$ is frequency independent under this assumption). This yields what we call multistatic UWB Bayesian inversion.  In UWB inversion, low frequencies are more sensitive to lower spatial harmonics, whereas high frequencies are more sensitive to higher spatial harmonics. The highest spatial harmonic that can be resolved depends on the maximum frequency that can be used without violating the Born approximation. Increasing the number of uncorrelated measurements, whether from sufficiently spaced sensors and/or frequency samples, makes the inversion statistically more stable against random noise and/or clutter \cite{Fouda2, Fouda4}.  

Being able to define transmitter- and frequency-independent unknowns is a consequence of the adopted Born approximation. In the contrast source formulation in \cite{MassaContSourc}, the unknowns are the equivalent currents of the contrast function, which vary with frequency and incident field. For high contrast media, iterative inversion approaches can be used, as discussed in Section VI ahead.

\section{Results}

\begin{figure}[!t]
\centering
\includegraphics[width=6.4in]{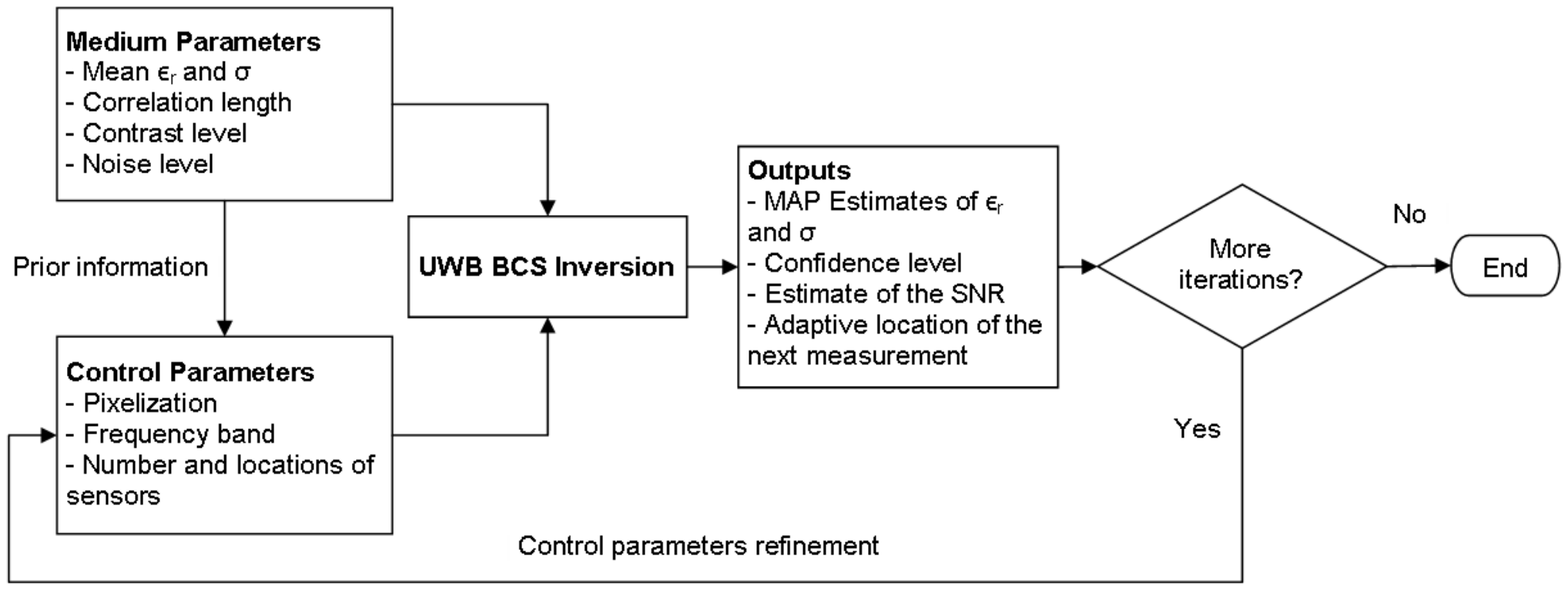}
\caption{Flowchart of the UWB BCS inversion process.}
\label{FigCH6_00}
\end{figure}

The UWB BCS inversion process is summarized in Fig. \ref{FigCH6_00}. The pointwise distribution of the medium can be characterized by some parameters such as the average electrical properties, the correlation length and the contrast level of the fluctuations. Another parameter dictated by the problem is the signal-to-noise ratio (SNR) of the measurements. A priori knowledge of any of the medium parameters can be used to select the user-controlled parameters. Those include the domain of interest (DOI) pixelization, the utilized frequency band, and the number and location of sensors. The UWB BCS inversion is then invoked, and the output contrast level and confidence level are used to decide whether more iterations are needed. In that case, posterior information can be fedback to refine the user-controlled parameters of the next iteration.     

\begin{figure}
\centering
\subfigure[]{\includegraphics[width=3.0in,trim= 100 120 100 120,clip=true]{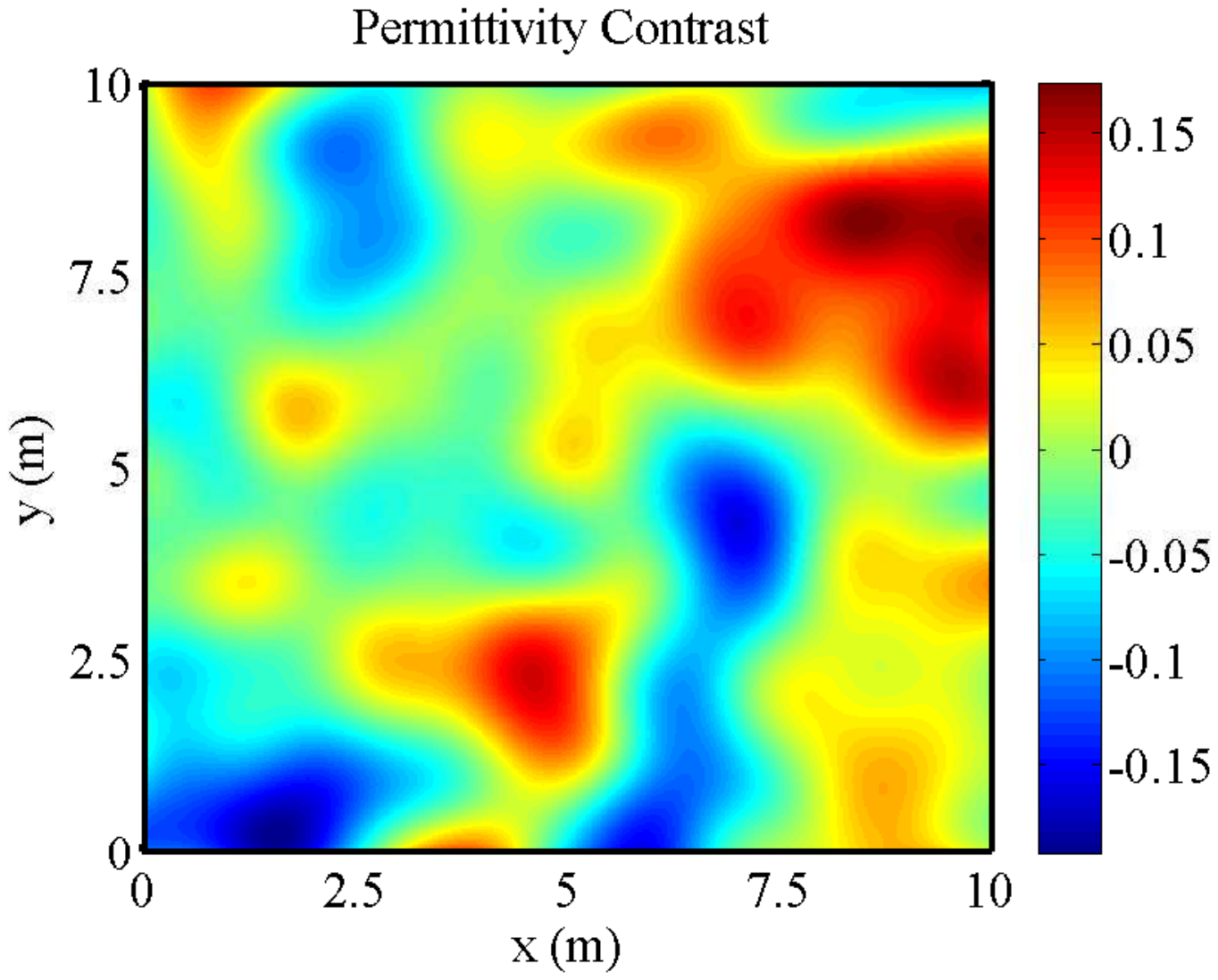}}
\subfigure[]{\includegraphics[width=3.0in,trim= 100 120 100 120,clip=true]{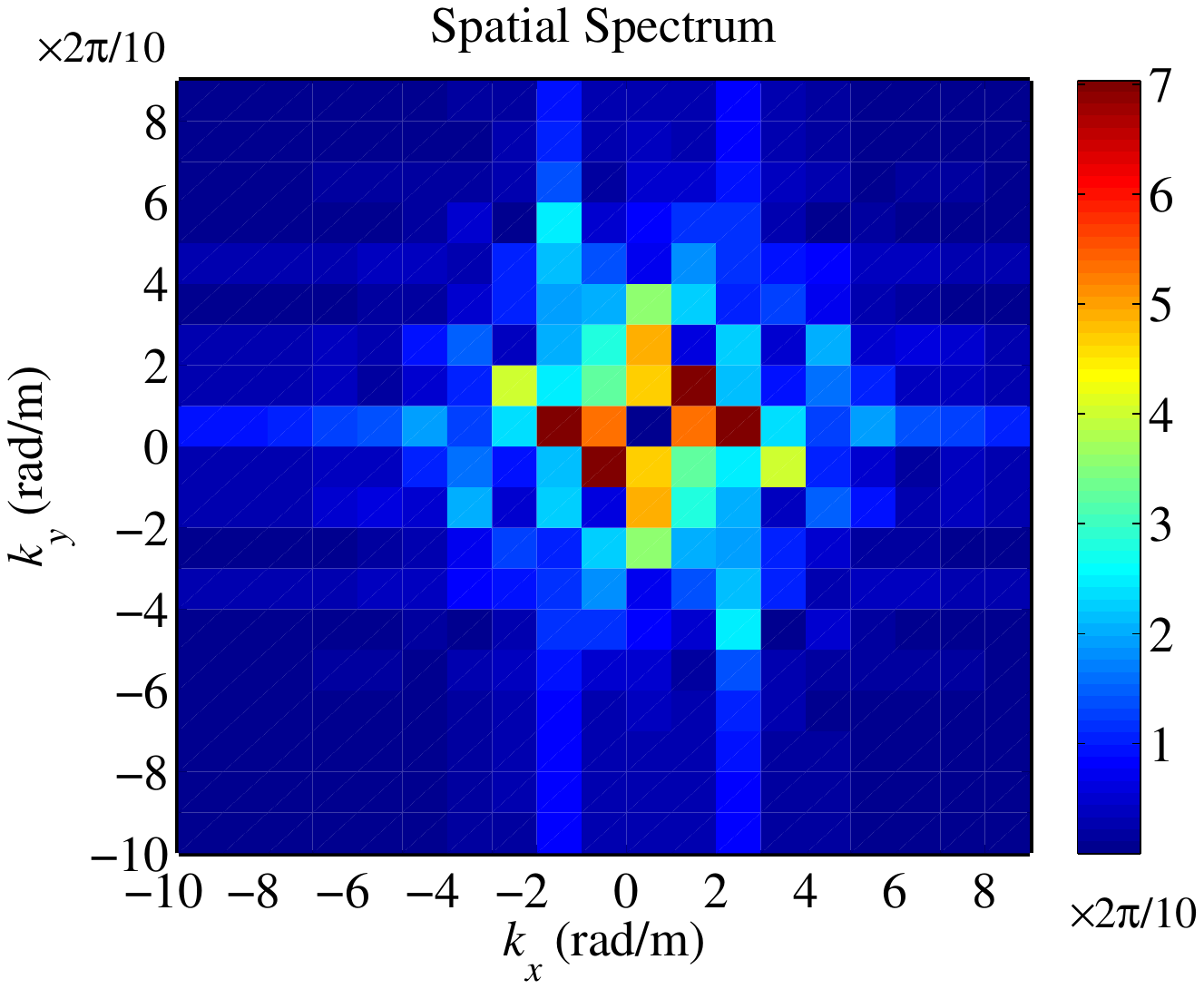}}
\caption{Example of the permittivity profile of a continuous random medium. (a) Spatial distribution. (b) Spatial spectrum.}
\label{FigCH6_0}
\end{figure}

An example of the permittivity contrast of a continuous random medium is shown in Fig. \ref{FigCH6_0}(a). This distribution is a realization of a Gaussian random process with zero-mean, standard deviation of 0.064, and a Gaussian correlation function with correlation length $l_c$=1.25 m. The spatial spectrum is shown in Fig. \ref{FigCH6_0}(b). For simplicity, we consider 2-D models throughout this work, but evidently the same analysis can be easily extended to 3-D cases.
The background medium is assumed to have mean permittivity of 3 and mean conductivity of 0.15 mS/m. This example may correspond to underground imaging of dry soil \cite{MelesGPR1}. The interrogating frequency band ranges from 5-250 MHz with 50 samples. Note that the maximum wavenumber of the interrogating signal $k_{max}$=1.48$\times$2$\pi$ rad/m is larger than the maximum spatial harmonic of the medium (=1.1$\times$2$\pi$ rad/m computed across the diagonal). Forward problem simulations are carried out using the finite-difference time-domain method~\cite{TafloveBook}. 

\begin{figure}[!t]
\centering
\subfigure[]{\includegraphics[width=2.0in,trim= 100 180 100 180,clip=true]{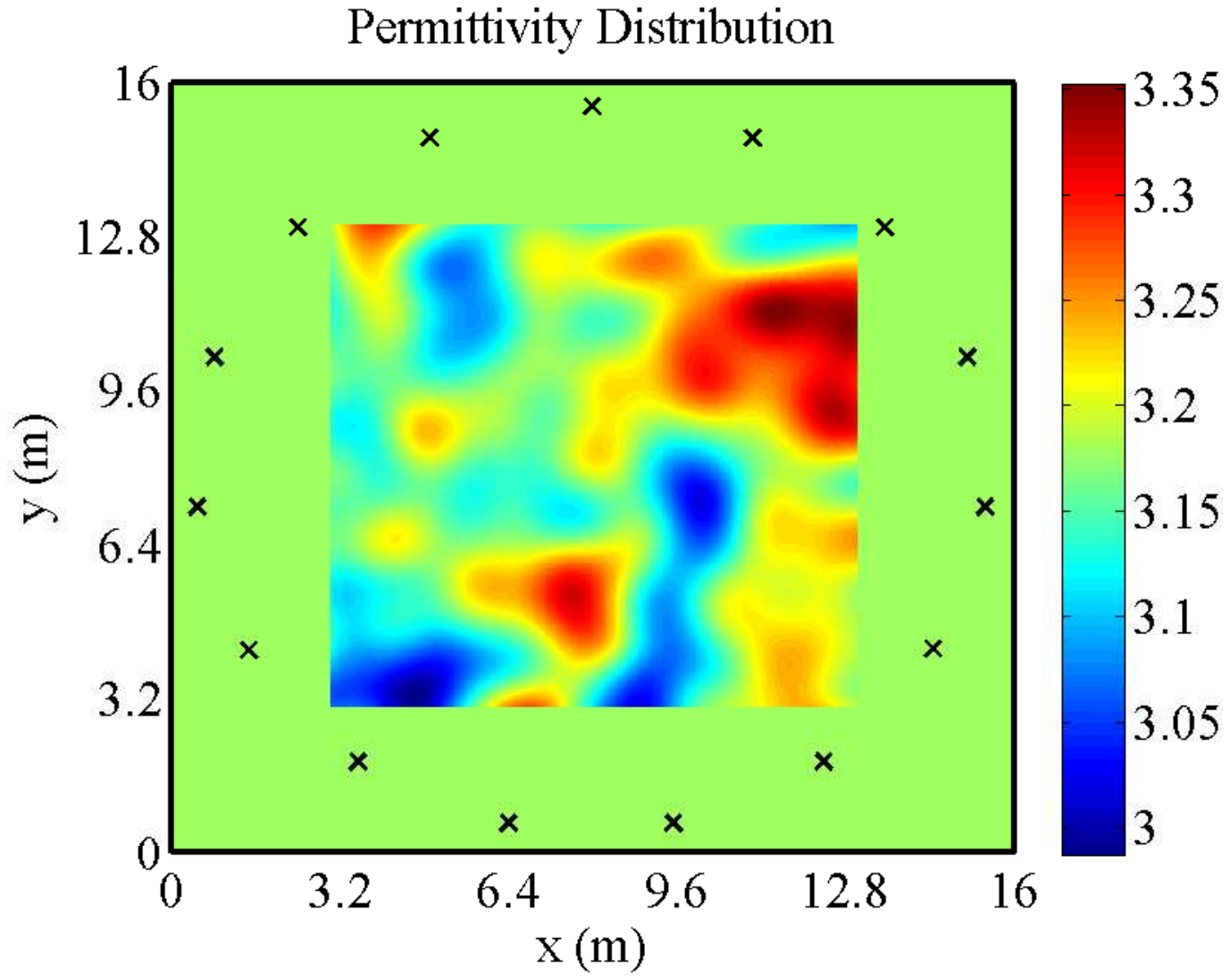}}
\subfigure[]{\includegraphics[width=2.0in,trim= 100 180 100 180,clip=true]{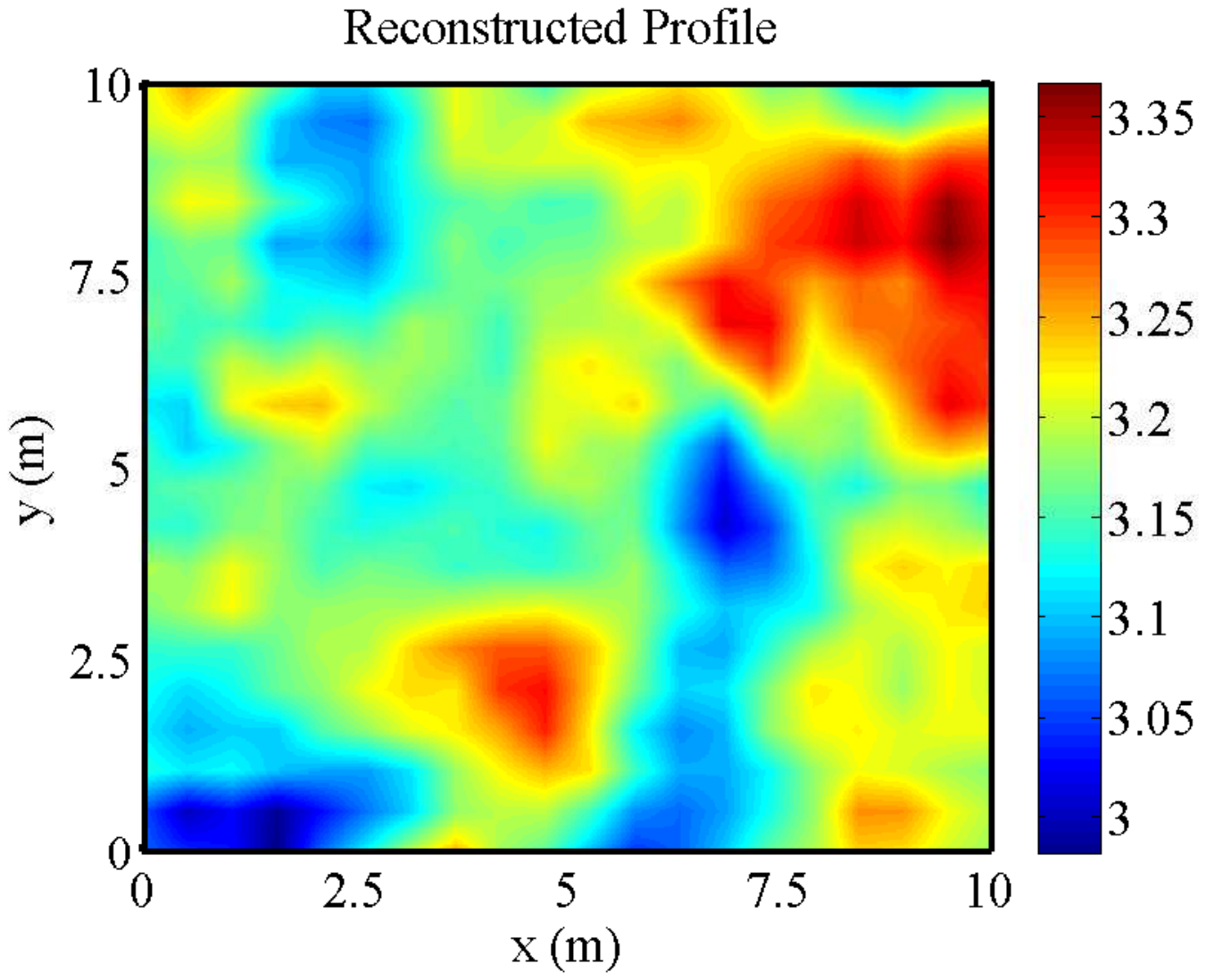}}
\subfigure[]{\includegraphics[width=2.0in,trim= 100 180 100 180,clip=true]{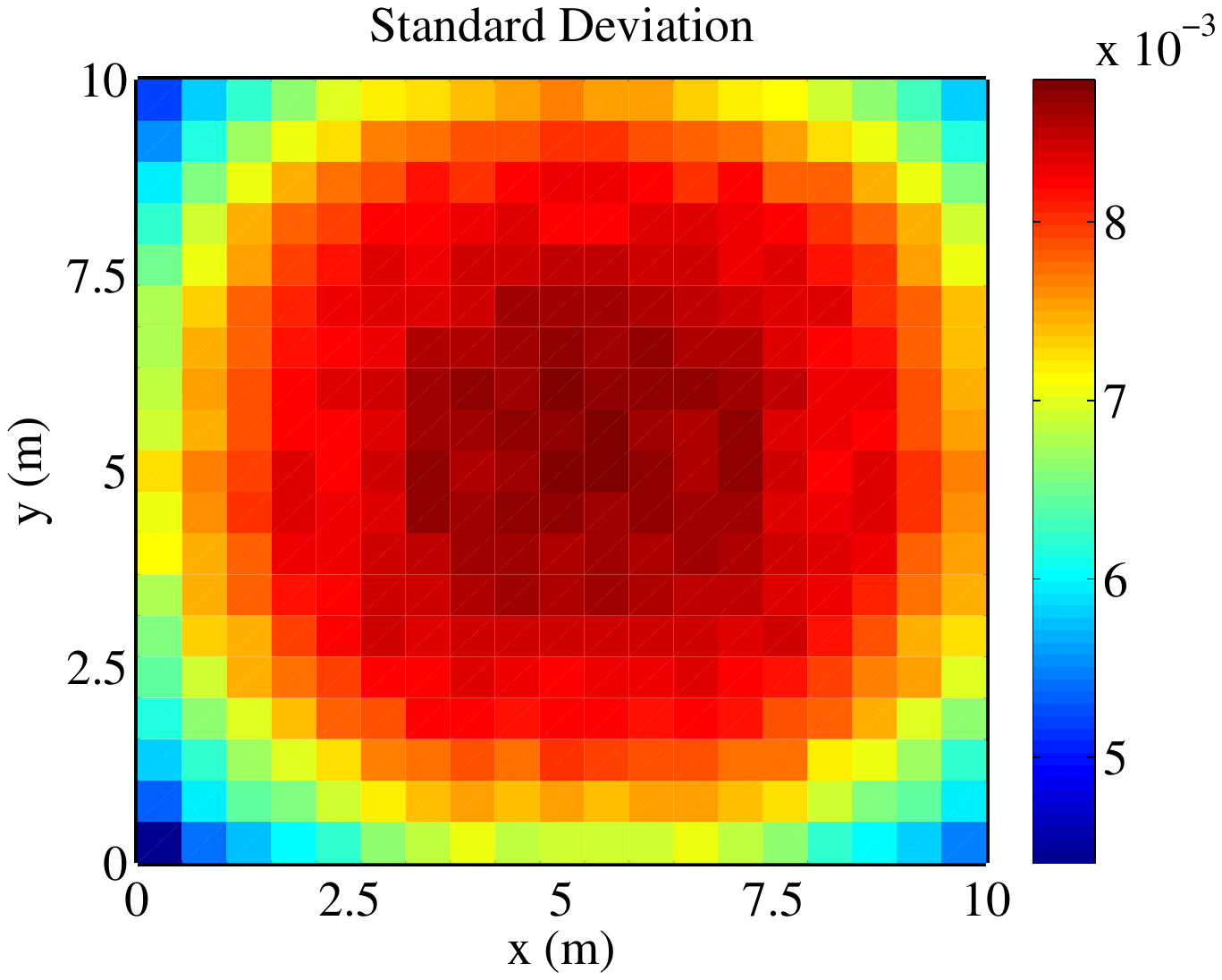}}
\subfigure[]{\includegraphics[width=2.0in,trim= 100 180 100 180,clip=true]{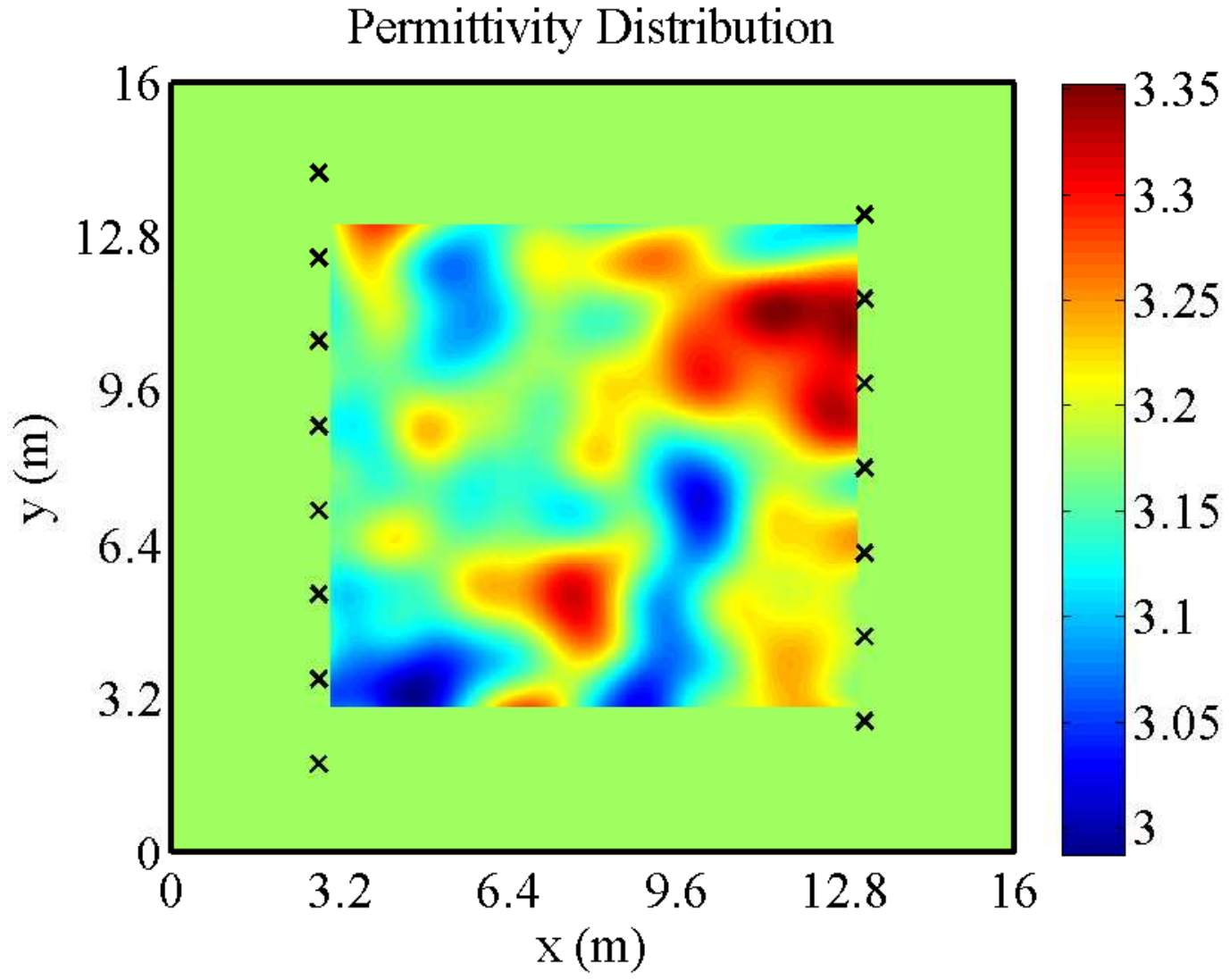}}
\subfigure[]{\includegraphics[width=2.0in,trim= 100 180 100 180,clip=true]{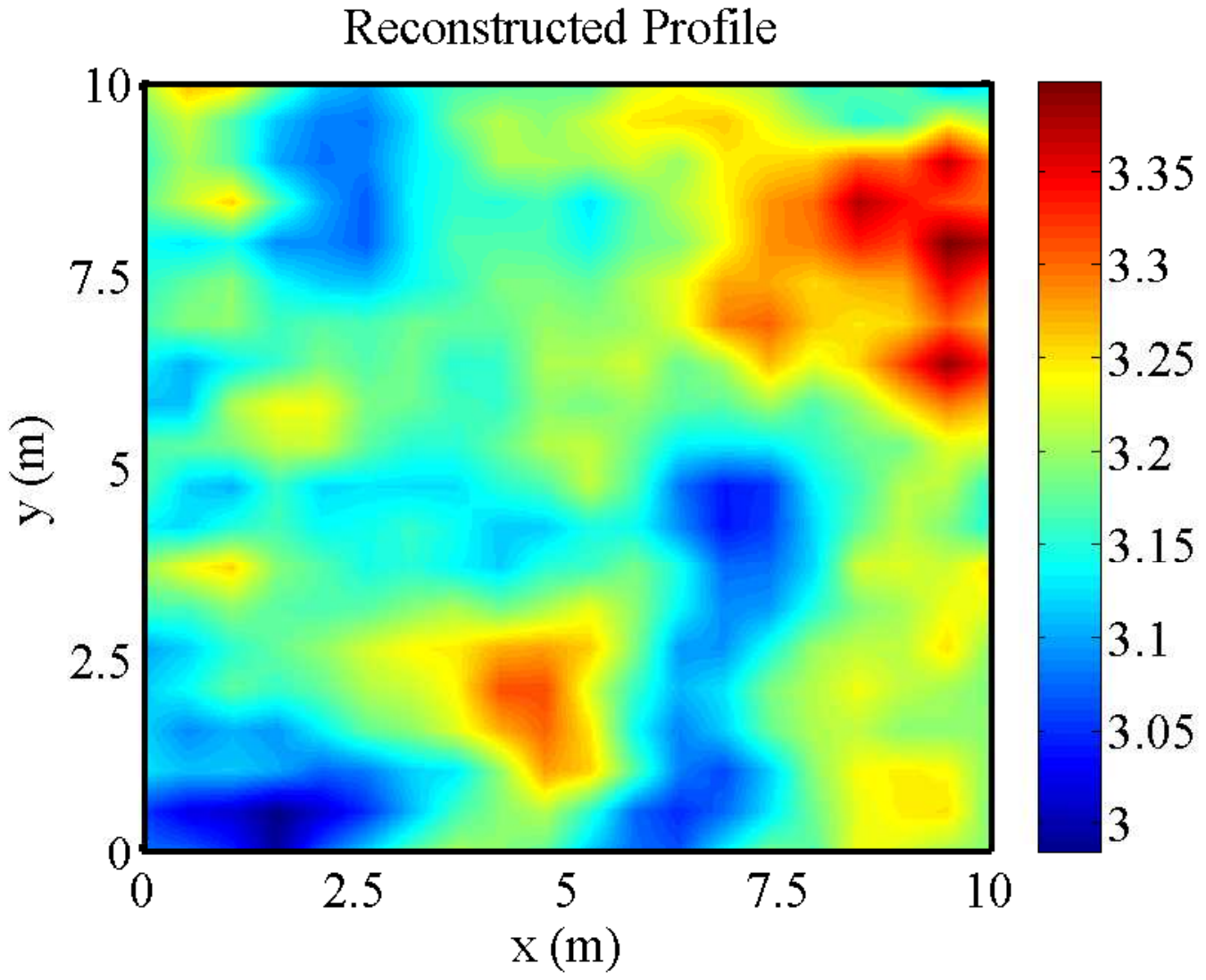}}
\subfigure[]{\includegraphics[width=2.0in,trim= 100 180 100 180,clip=true]{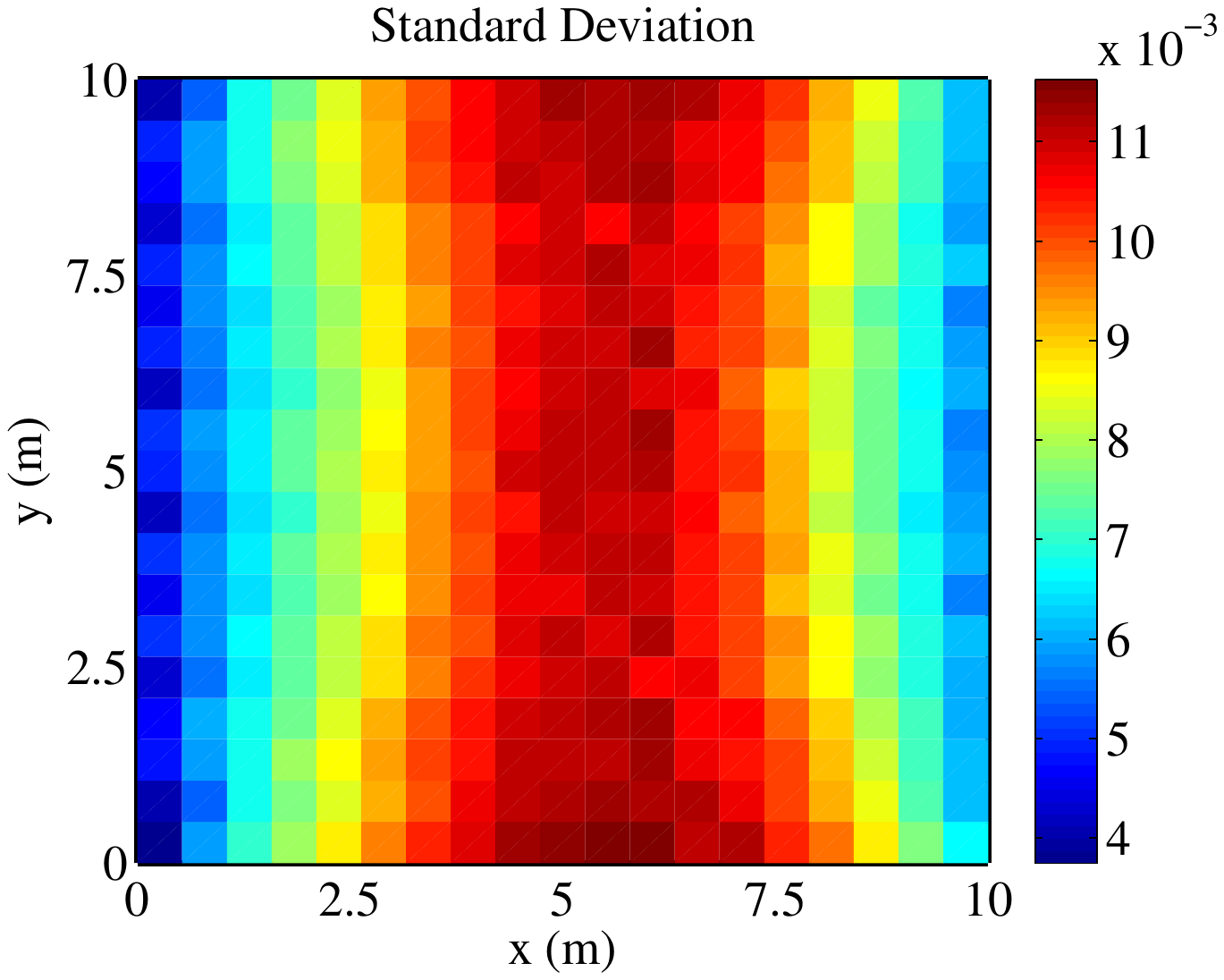}}
\subfigure[]{\includegraphics[width=2.0in,trim= 100 180 100 180,clip=true]{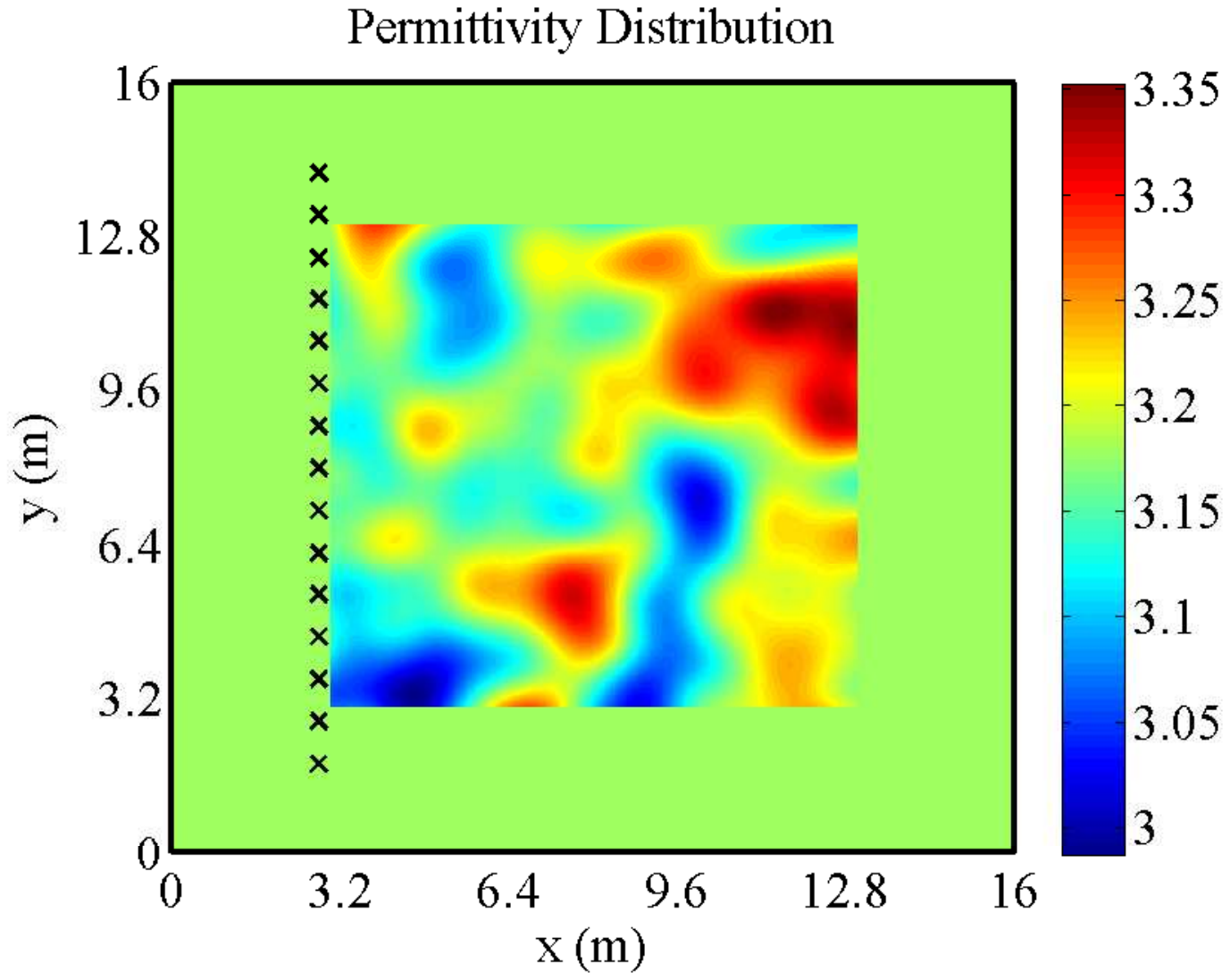}}
\subfigure[]{\includegraphics[width=2.0in,trim= 100 180 100 180,clip=true]{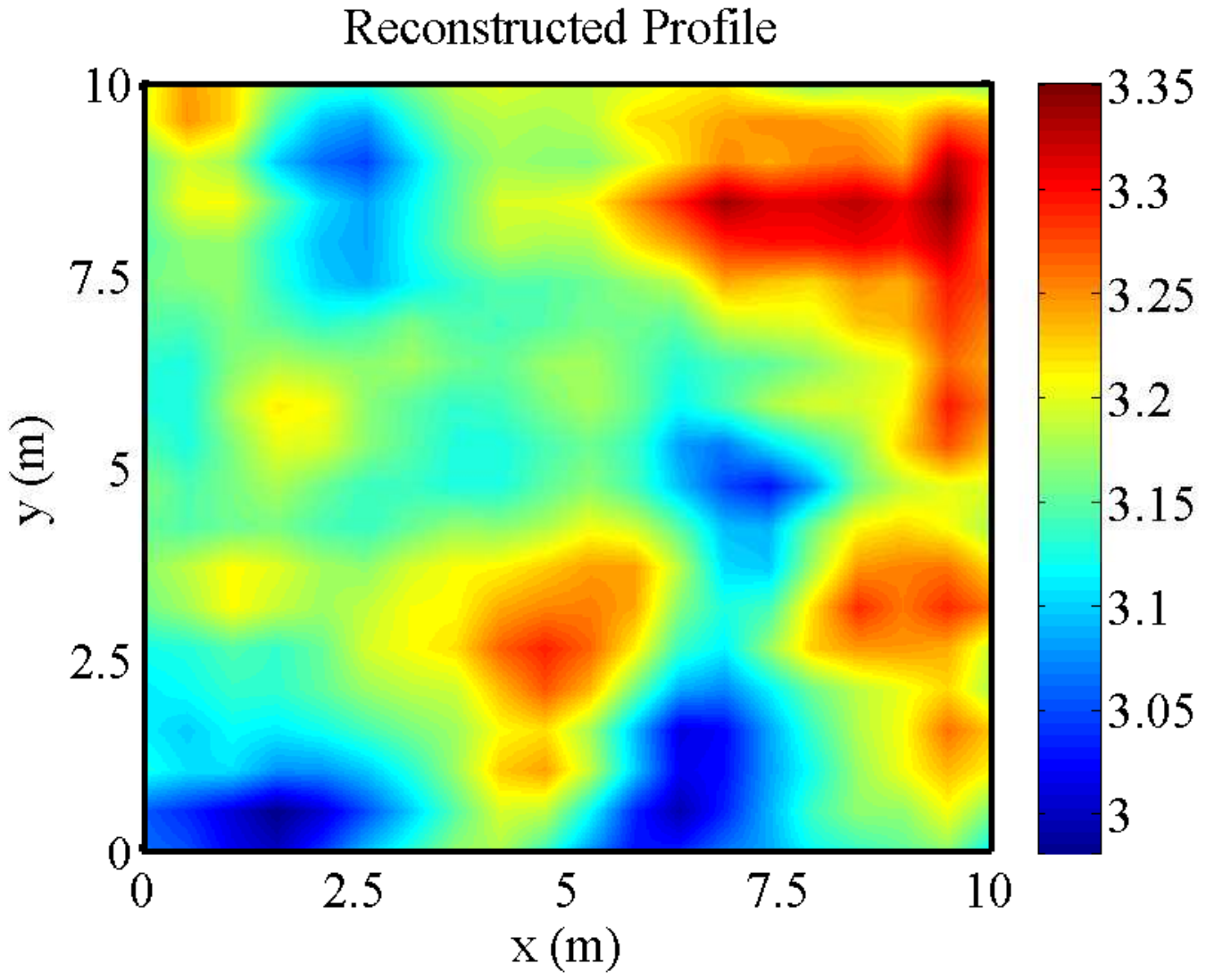}}
\subfigure[]{\includegraphics[width=2.0in,trim= 100 180 100 180,clip=true]{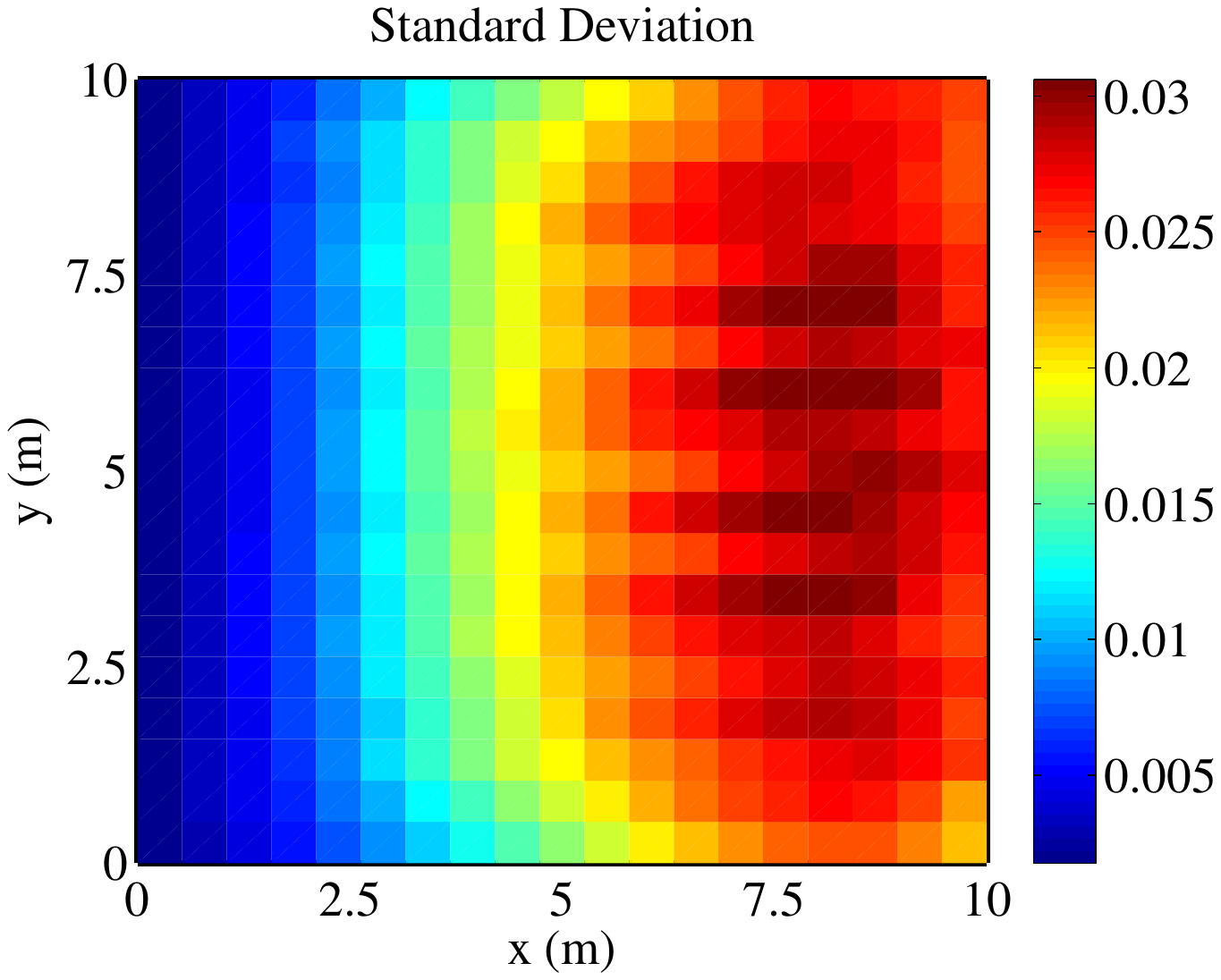}}
\caption{Forward problem permittivity distribution, reconstructed profiles and estimated standard deviation for different sensor array geometries. Sensors are indicated with `x's. $N_s$=15 and SNR=10 dB. (a)-(c) Full-aspect. (d)-(f) Crosshole. (g)-(i) Borehole.}
\label{FigCH6_1}
\end{figure}

In the inverse problem, the DOI is discretized uniformly into 20$\times$20 pixels, and the 2-D Green's function is computed analytically assuming known average medium properties. 
We use $N_s$=15 multistatic sensors deployed either in full-aspect (FA) circular geometry as shown in Fig. \ref{FigCH6_1}(a), crosshole (CH) geometry as in Fig \ref{FigCH6_1}(d), or borehole (BH) geometry as in Fig. \ref{FigCH6_1}(g). Transmitters are point sources in 2-D (infinite line source) radiating TM$_{z}$ polarization. For the particular application of underground imaging, the $x-y$ plane in the FA case can be perceived as the horizontal plane, with the shown distribution being a horizontal cross-section in the formation, and the sensors are deployed in circularly distributed wells. For the BH and CH cases, the shown distribution is a vertical cross-section, and the sensors are deployed in one or two wells, respectively.  
The SNR is assumed to be 10 dB for all measurements performed using different sensors and frequencies.
Reconstructed profiles for the three geometries are shown in Fig. \ref{FigCH6_1}(b), (e) and (h), and the estimated standard deviations (which determine the confidence level of the inversions) are shown in Fig. \ref{FigCH6_1}(c), (f) and (i), respectively. Reconstructed images are interpolated to a finer grid for the sake of visualization. Comparing actual and reconstructed profiles, we note that FA and CH outperform BH. Also the estimated standard deviation provides a reasonably good measure for the inversion accuracy, this more obvious in the BH case, where the reduced-accuracy inversion in the right half of the investigation domain (farther from the array) is associated with higher standard deviation. Roughly speaking, the inversion accuracy and the reciprocal of standard deviation at a certain point are proportional to the spatial resolution offered by the sensors array at that point.  
%

\subsection{Performance analysis}

\begin{figure}[!t]
\centering
\subfigure[]{\includegraphics[width=2.8in,trim= 90 120 90 120,clip=true]{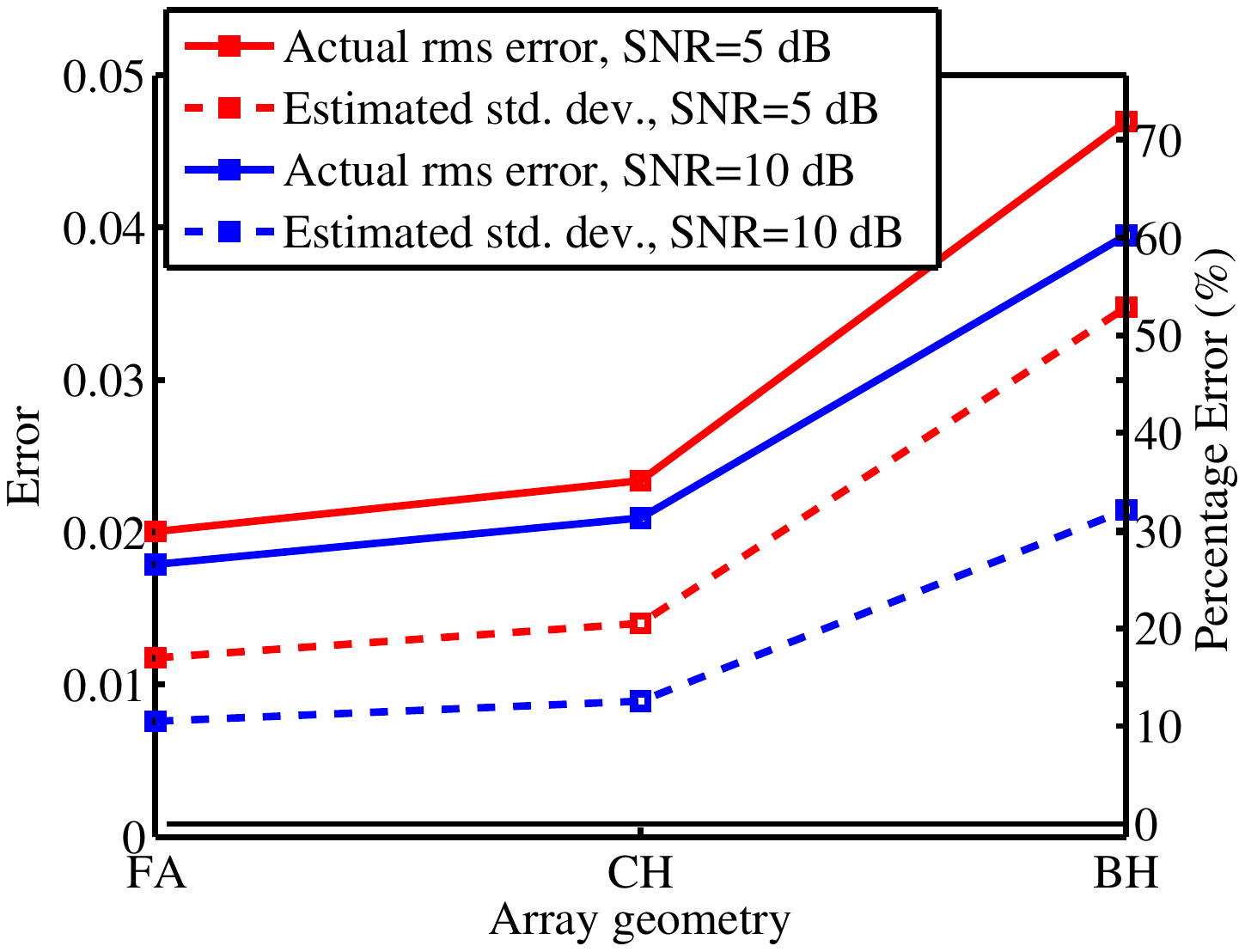}}
\subfigure[]{\includegraphics[width=2.8in,trim= 90 120 90 120,clip=true]{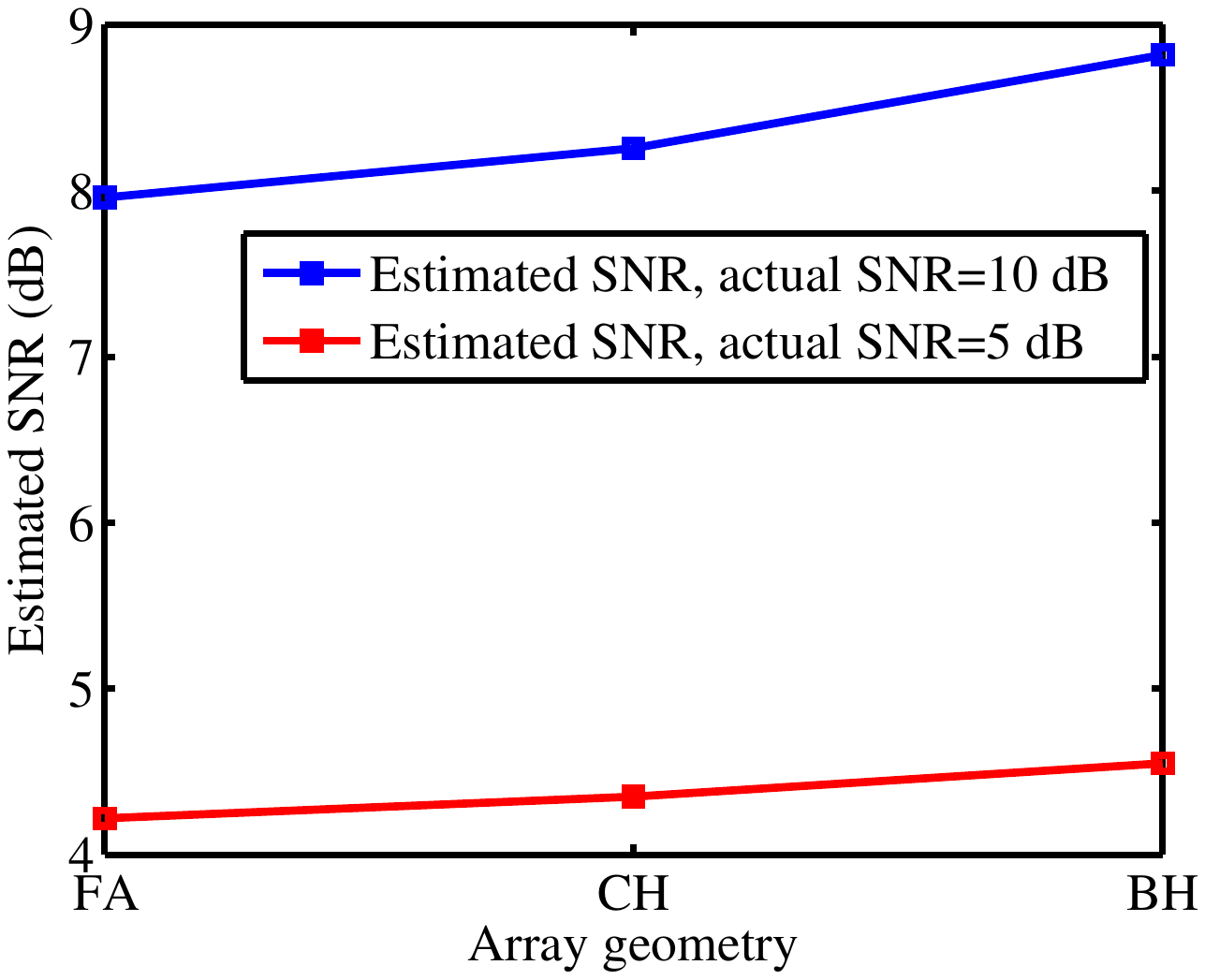}}
\caption{(a) Actual and estimated errors for different array geometries and SNRs. Percentile error is the ratio of the absolute error to the r.m.s. of the actual contrast function. (b) Estimated SNR. }
\label{FigCH6_2}
\end{figure}

In this section, we provide some qualitative measures for assessing inversion accuracy and efficiency. We first define the actual r.m.s. error as 
\begin{equation}
\label{equnCH6_29} 
\textrm{Actual r.m.s. error}=\sqrt{\textrm{avg}_{D}\left|\hat{\epsilon}_{r}(\textbf{r})-\epsilon_{r}(\textbf{r})\right|^2}
\end{equation}
where $\hat{\epsilon}_{r}$ is the estimated permittivity. The (average) estimated standard deviation can be defined as 
\begin{equation}
\label{equnCH6_30} 
\textrm{Estimated std. deviation}=\sqrt{\textrm{avg}\left[\textrm{diag}(\textrm{cov}(\textbf{F}^{-1} \textbf{t}))\right]}
\end{equation}
Both actual and estimated errors are plotted in Fig. \ref{FigCH6_2}(a) for the three previously discussed geometries and two SNRs. Percentile error is defined as the ratio of the absolute error to the r.m.s. of the actual contrast function ($\sqrt{\textrm{avg}_{D}\left|\epsilon_{r}(\textbf{r})\right|^2}$). This plot shows that estimated error follows pretty well the actual error, with the latter being always larger. This makes perfect sense, since the estimated error only accounts for errors due to additive noise, whereas actual error encloses, in addition to noise, errors due to the adopted Born approximation and discretization error.
Estimated SNRs are shown in Fig. \ref{FigCH6_2}(b). They are below their actual values by 1-2 dB, which indicates that the noise variance was over-estimated by the RVM solver.     

Errors and processing times for a FA array with uniformly distributed increasing number of sensors are tabulated in Table \ref{TableCH6_1}.  Errors decrease monotonically with increasing the number of sensors at the expense of increasing the processing time, as expected. Listed times are those required for solving the fast RVM, using non-optimized Matlab code, running on a machine with average CPU speed of 2.7 G.cycle/s. They are very short times (almost real-times) w.r.t. the size and the number of measurements of the considered problem. Note that there are costs associated with computing the Green's function and constructing the projection matrix, but those are considered as pre-processing costs. 

\begin{table*}[!t]
\renewcommand{\arraystretch}{1.5}
\caption{Actual error, estimated standard deviation, and processing time using a FA array with increasing number of sensors and SNR=5 dB. }
\label{TableCH6_1}
\centering
\begin{tabular}{|c|c|c|c|}
\hline
Number of sensors & Actual r.m.s. error & Estimated std. dev. & Processing time (sec.)\\
\hline
10 & 0.0267 & 0.0196 & 5.4 \\
\hline
15 & 0.02 & 0.01176 & 10.6 \\
\hline
30 & 0.0114 & 0.005 & 36 \\
\hline
\end{tabular}
\end{table*}
Another measure for quantifying the confidence level of the inversion is the differential entropy (DE) \cite{CoverITBook}. Referring to (\ref{equnCH6_1}), the DE of the posterior multivariate Gaussian PDF is given by
\begin{equation}
\label{equnCH6_31} 
h(p)=-\int p(\textbf{w}\left|\textbf{y}\right. ) \ln \left(p(\textbf{w}\left|\textbf{y}\right. )\right)d\textbf{w}\\
=\frac{1}{2} \ln \left[\left(2\pi e\right)^{M} \left|\bm{\Sigma} \right|\right]
\end{equation}
The DE given by the above equation is in information units (nats). It can be divided by ln(2) to give the DE in bits. DE measures randomness -random variables with PDF concentrated on  a small interval yields smaller DE. For continuous random variables, DE can be negative (as opposed to the entropy of discrete random variables which is always positive). Differential entropies for the setups of Fig. \ref{FigCH6_2}(a) are summarized in Table \ref{TableCH6_2}. Individual values of DE do not give much information about the randomness of the PDF; however, comparing DEs of two setups gives an idea about the accuracy gained or lost (measured in units of information) on going from one setup to the other. This measure agrees well with the behavior described in Fig. \ref{FigCH6_2}(a).
\begin{table*}[!t]
\renewcommand{\arraystretch}{1.5}
\caption{Differential entropy (in kb) for the setups in Fig. \ref{FigCH6_1}.}
\label{TableCH6_2}
\centering
\begin{tabular}{|c|c|c|}
\hline
Array geometry & SNR=10 dB & SNR=5 dB \\
\hline
FA & -3.34 & -3.09 \\
\hline
CH & -3.33 & -3.03 \\
\hline
BH & -3.32 & -2.8 \\
\hline
\end{tabular}
\end{table*}

\subsection{Adaptive sensing}
Our goal in this section is to develop a systematic procedure for optimizing the location(s) of subsequent measurement(s), such that the information gain from each measurement is maximized \cite{BayesExpDes, BCSCarin}. The DE, after adding the $(N+1)^{th}$ measurement, can be written in terms of the DE of $N$ measurements as follows \cite{BCSCarin}   
\begin{equation}
\label{equnCH6_32} 
h(p_{new})=h(p)-\frac{1}{2} \ln\left(1+\sigma _{n}^{-2}\textbf{r}_{\textbf{B},N+1}^T\bm{\Sigma}\textbf{r}_{\textbf{B},N+1}\right) 
\end{equation}
where $\textbf{r}_{\textbf{B},N+1}^T$ is the new row added to the projection matrix $\textbf{B}$ associated with the $(N+1)^{th}$ measurement. To maximize information gain, the absolute value of the second term in the r.h.s. of (\ref{equnCH6_32}) should be maximized, which implies that  $\textbf{r}_{\textbf{B},N+1}^T$ should be chosen such that
\begin{equation}
\label{equnCH6_33} 
\textbf{r}_{\textbf{B},N+1}^T\bm{\Sigma}\textbf{r}_{\textbf{B},N+1}=\textrm{var}\left(y_{N+1}\right)
\end{equation}
is maximized. In other words, we choose to place the next sensor where we expect highest uncertainty in the measurement, in this way, the information gain is maximized \cite{BCSCarin}. The above equation is maximized by choosing $\textbf{r}_{\textbf{B},N+1}$ to be the eigenvector of $\bm{\Sigma}$ corresponding to the largest eigenvalue \cite{BCSCarin}. 

\begin{figure}[!t]
\centering
{\includegraphics[width=1.8in,trim= 100 235 100 235,clip=true]{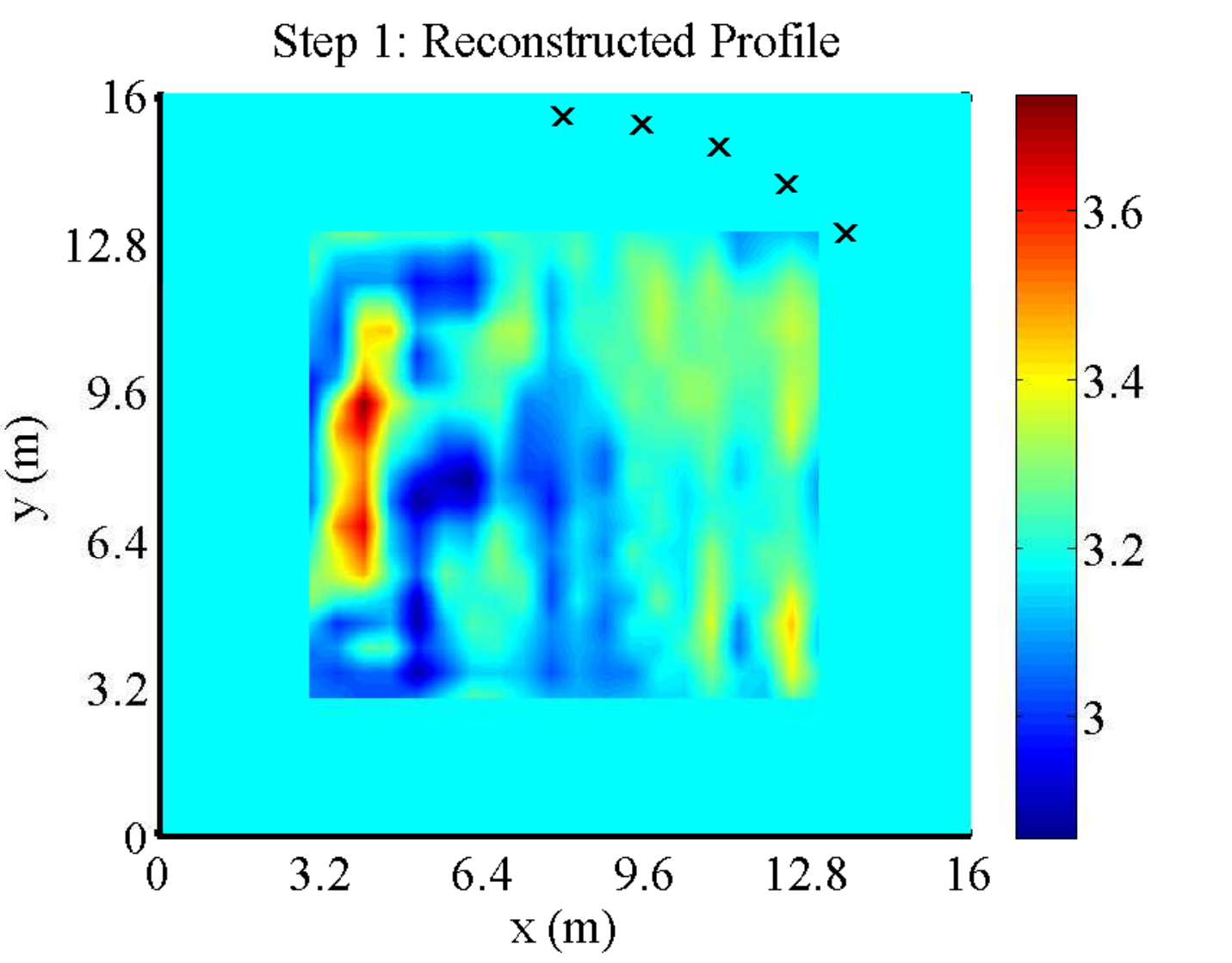}}
{\includegraphics[width=1.8in,trim= 100 235 100 235,clip=true]{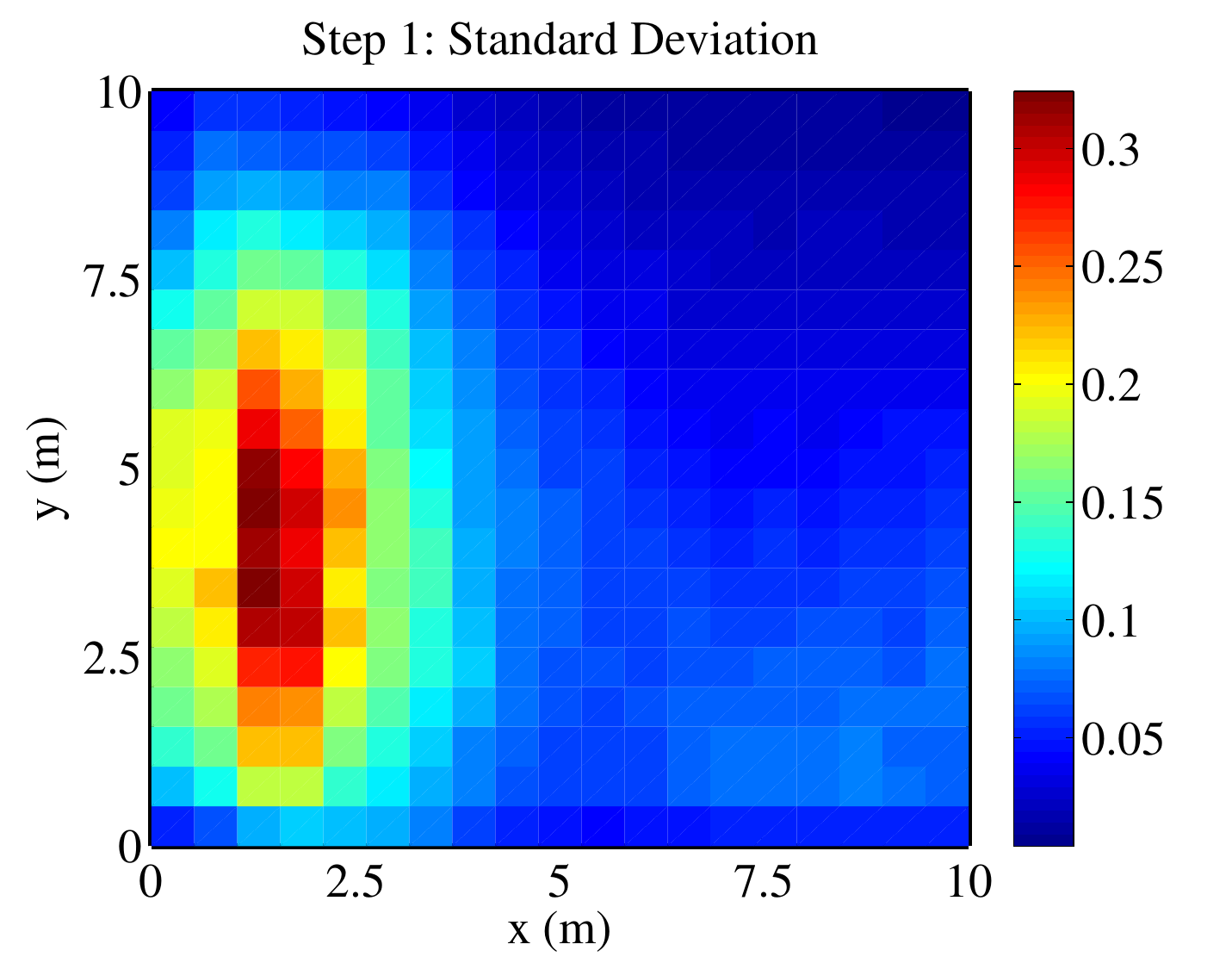}}
{\includegraphics[width=1.8in,trim= 100 235 100 235,clip=true]{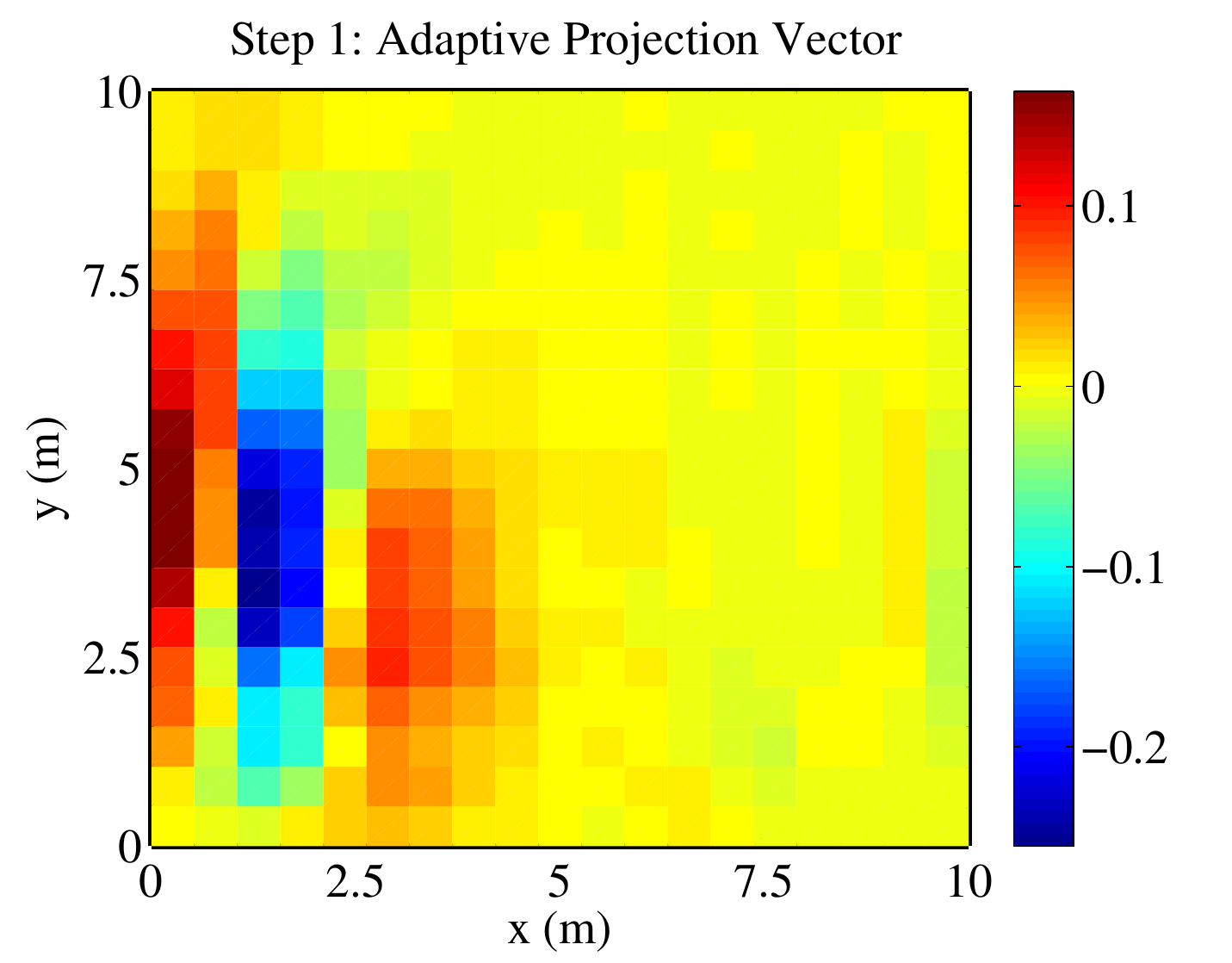}}
{\includegraphics[width=1.8in,trim= 100 235 100 235,clip=true]{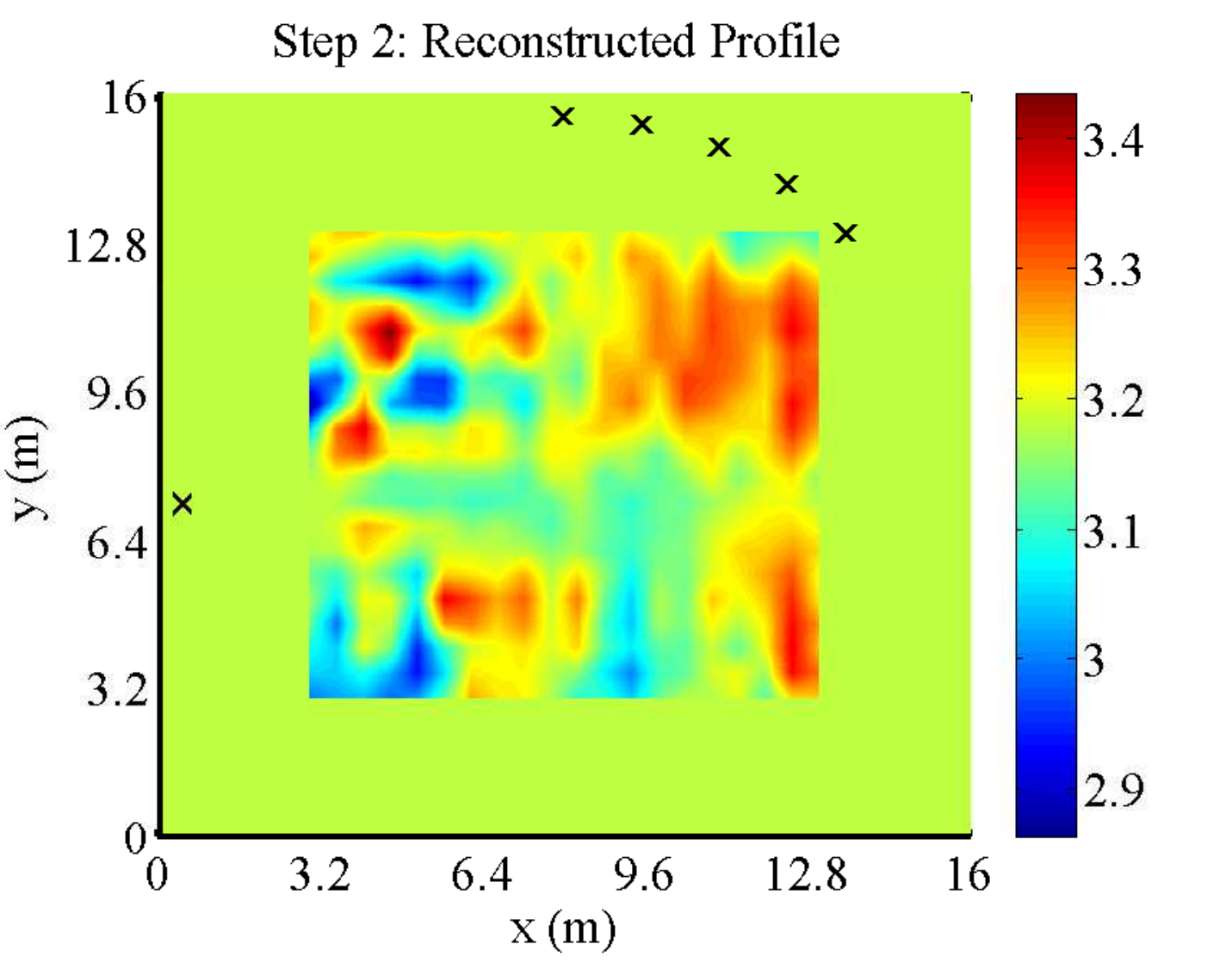}}
{\includegraphics[width=1.8in,trim= 100 235 100 235,clip=true]{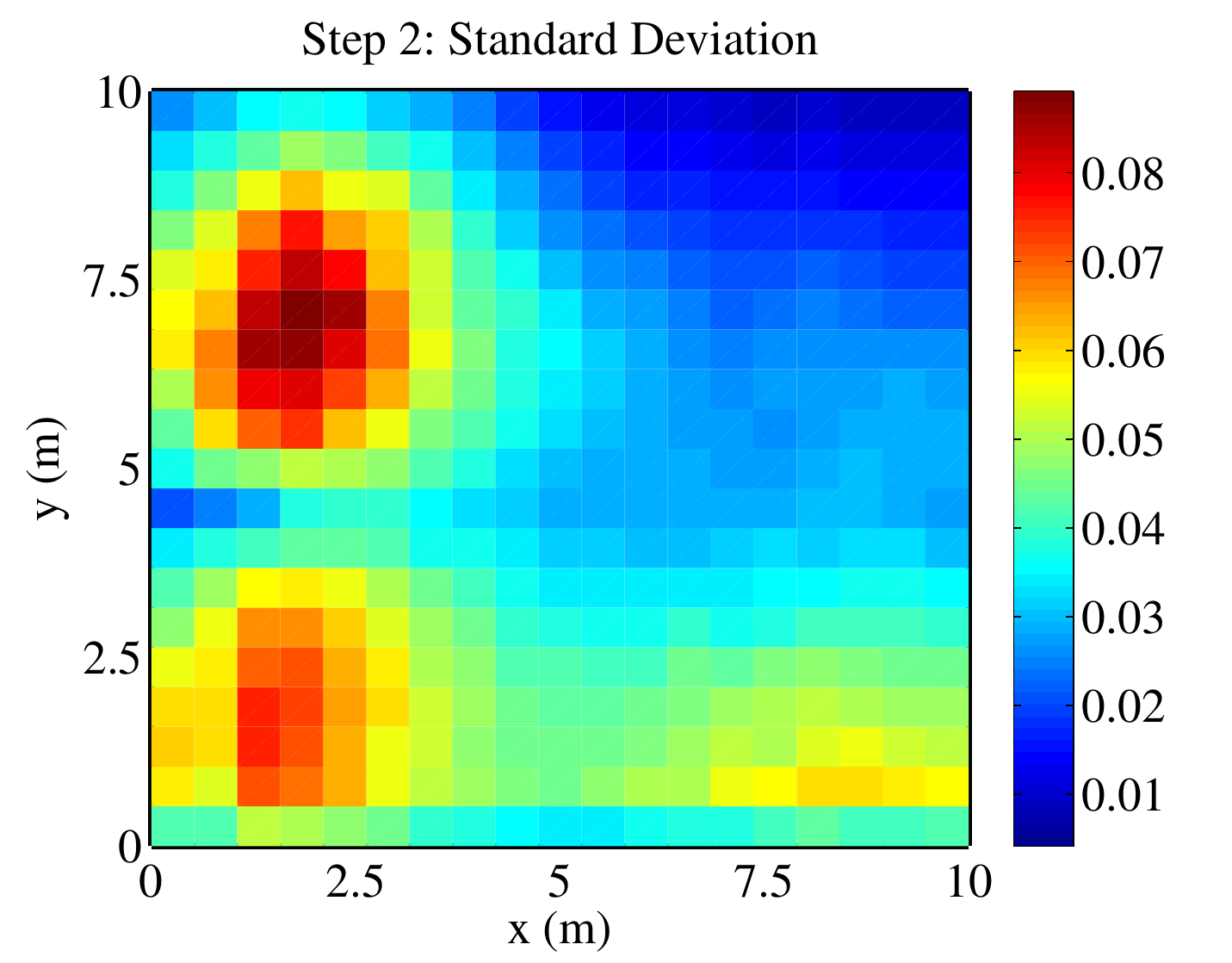}}
{\includegraphics[width=1.8in,trim= 100 235 100 235,clip=true]{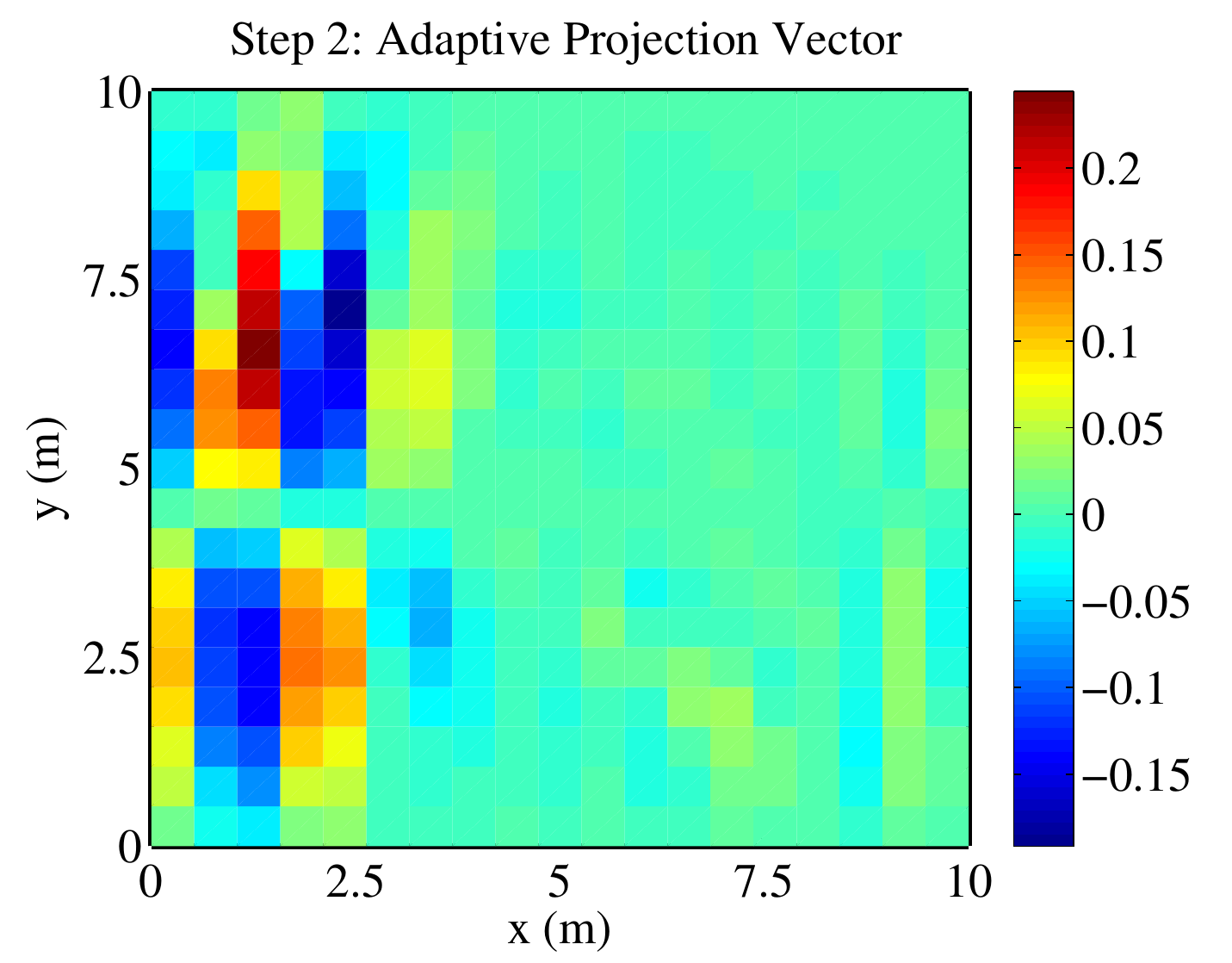}}
{\includegraphics[width=1.8in,trim= 100 235 100 235,clip=true]{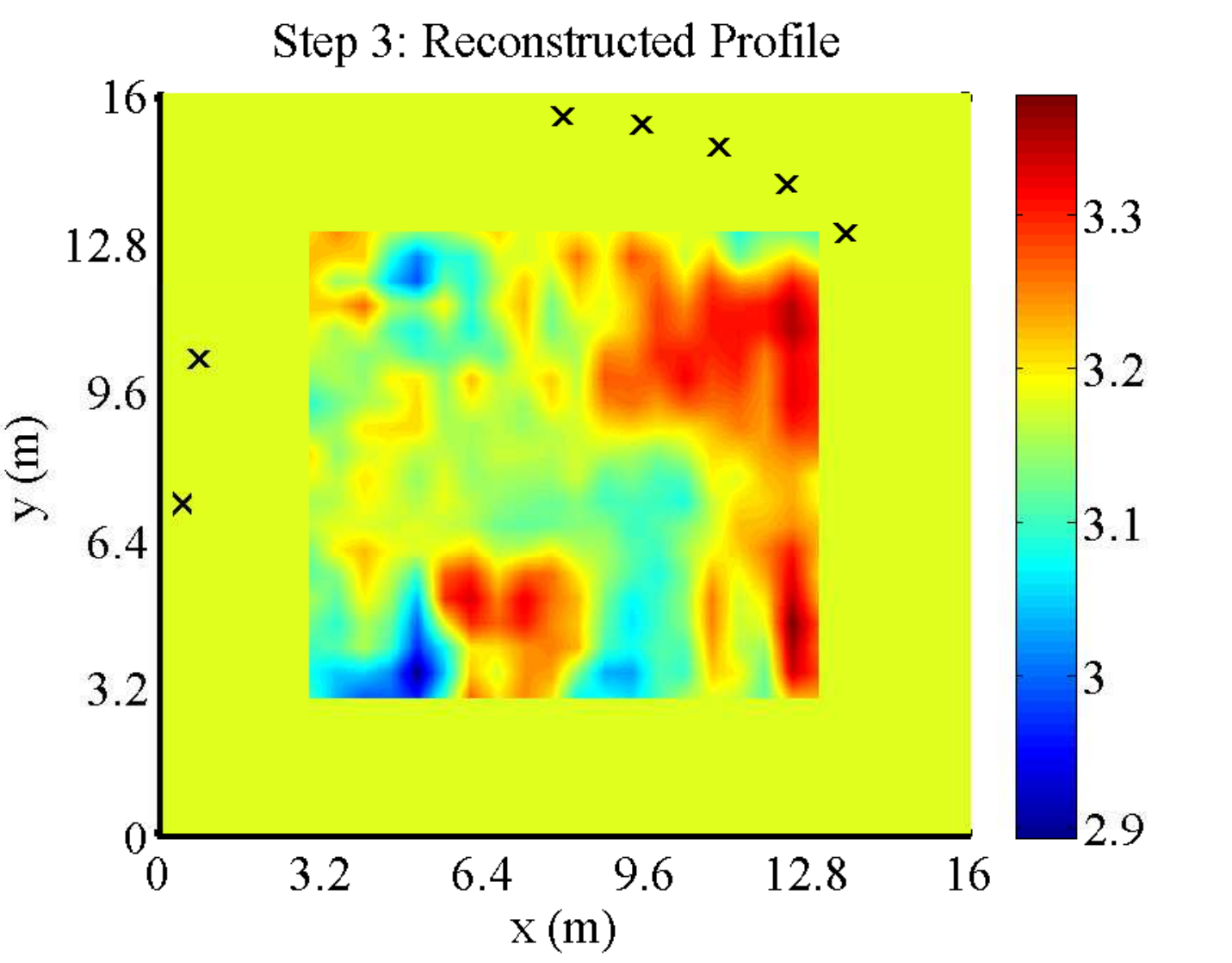}}
{\includegraphics[width=1.8in,trim= 100 235 100 235,clip=true]{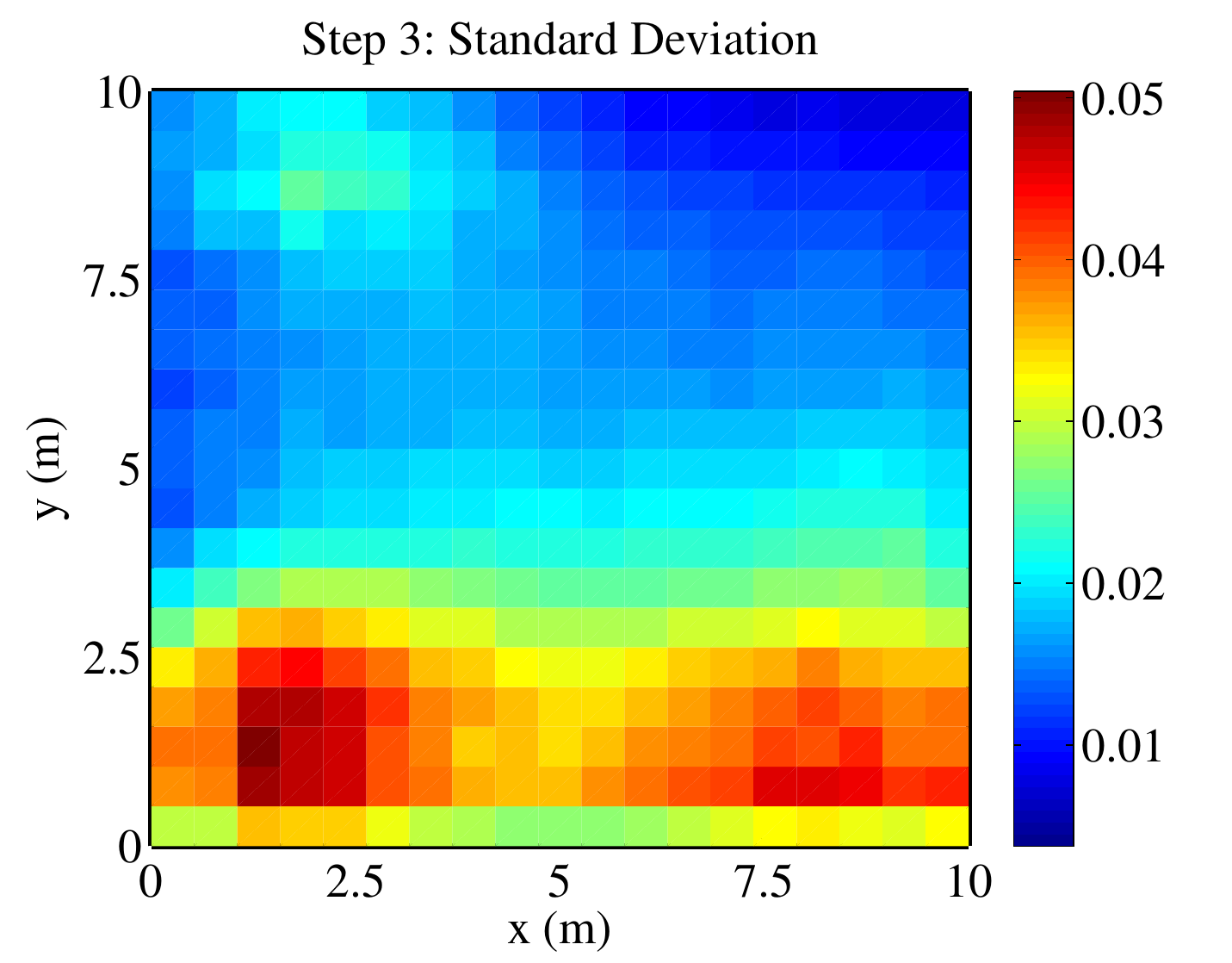}}
{\includegraphics[width=1.8in,trim= 100 235 100 235,clip=true]{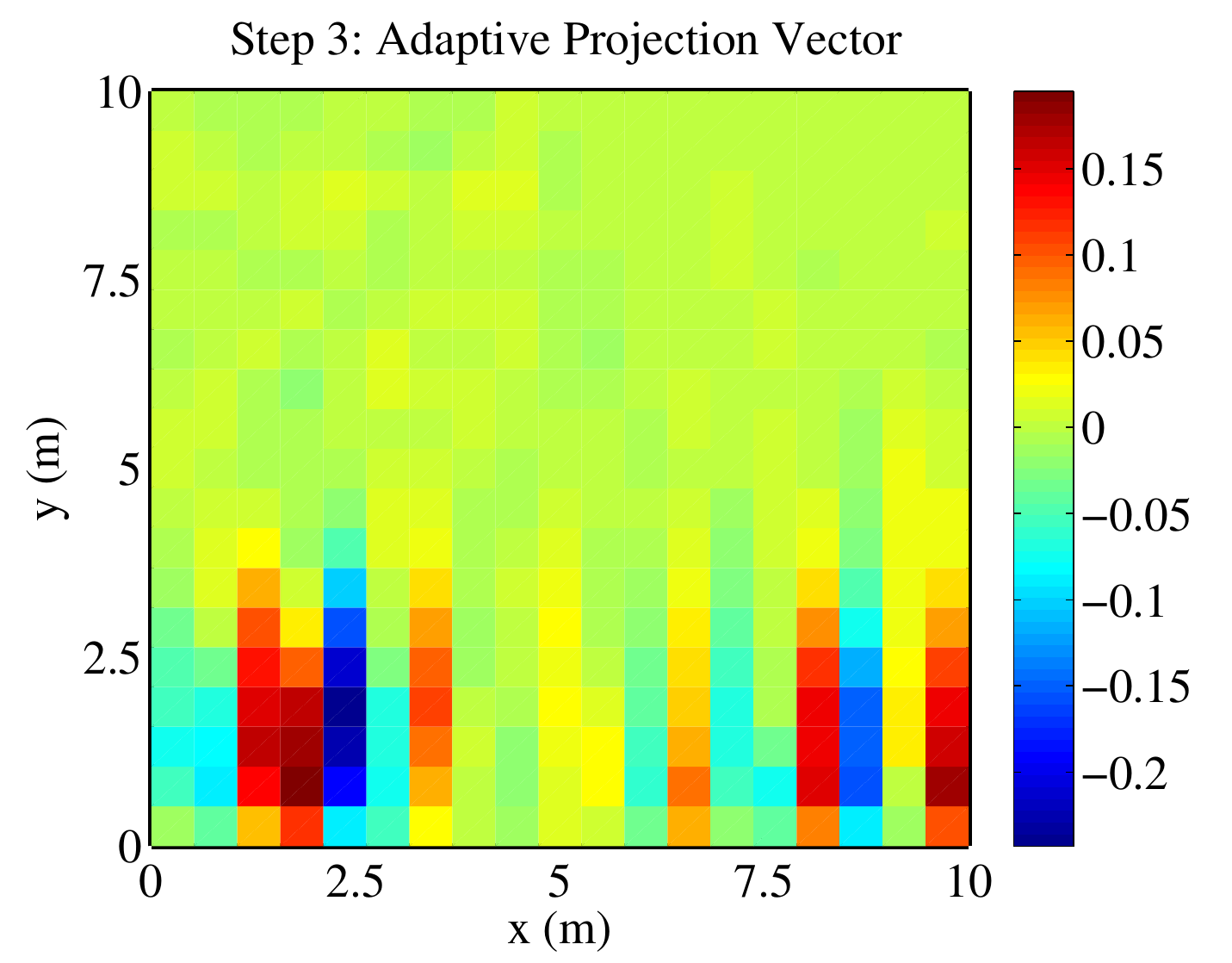}}
{\includegraphics[width=1.8in,trim= 100 235 100 235,clip=true]{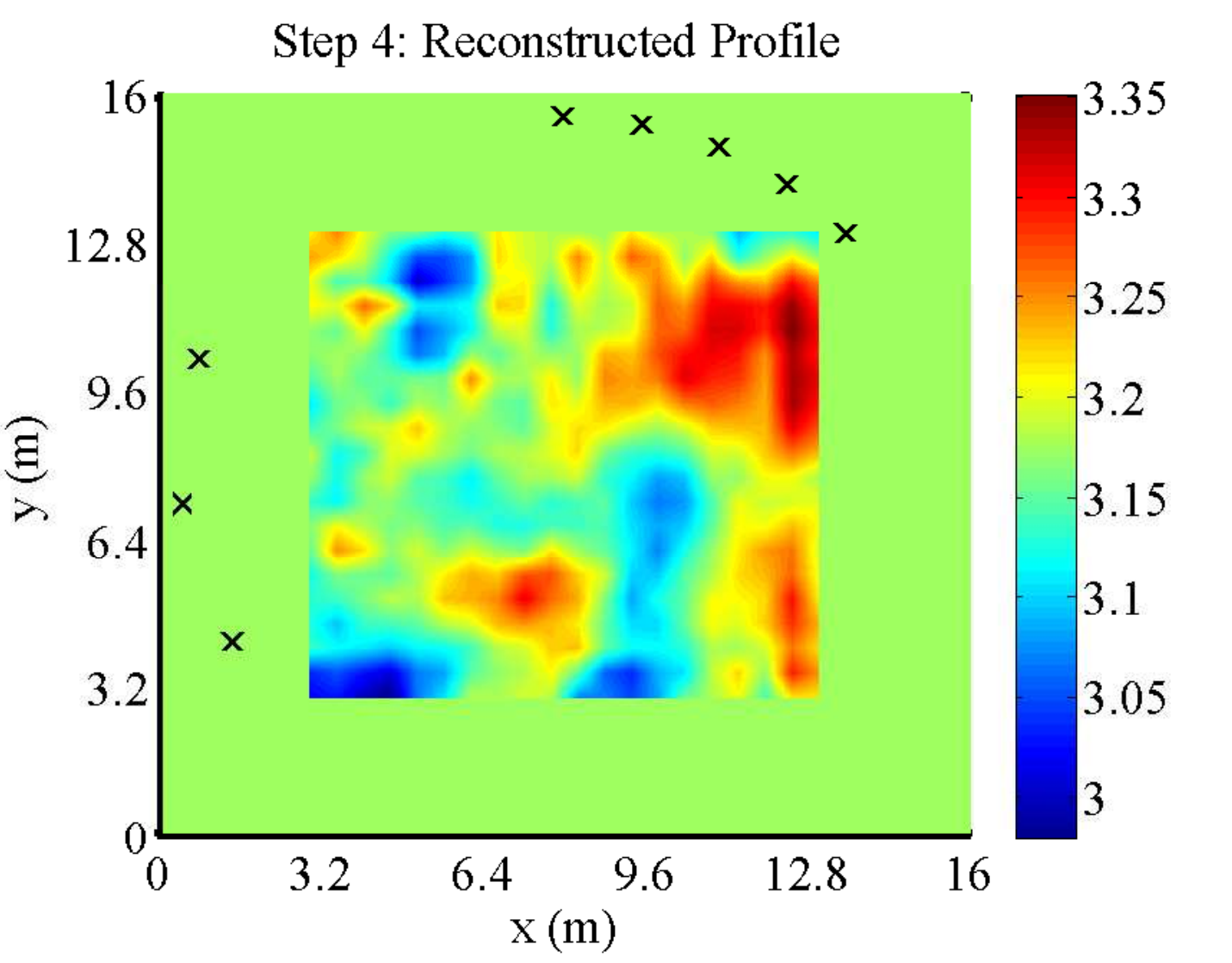}}
{\includegraphics[width=1.8in,trim= 100 235 100 235,clip=true]{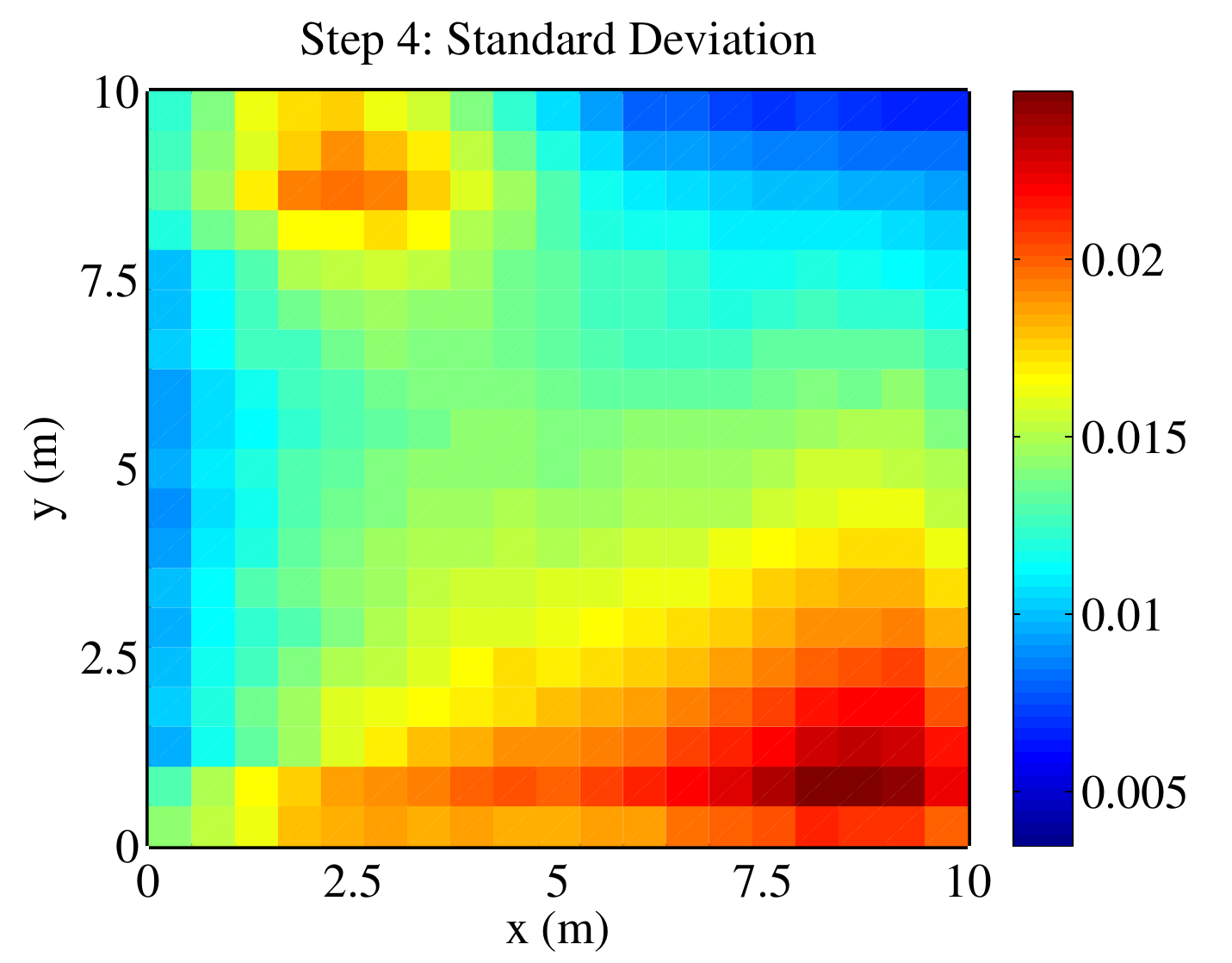}}
{\includegraphics[width=1.8in,trim= 100 235 100 235,clip=true]{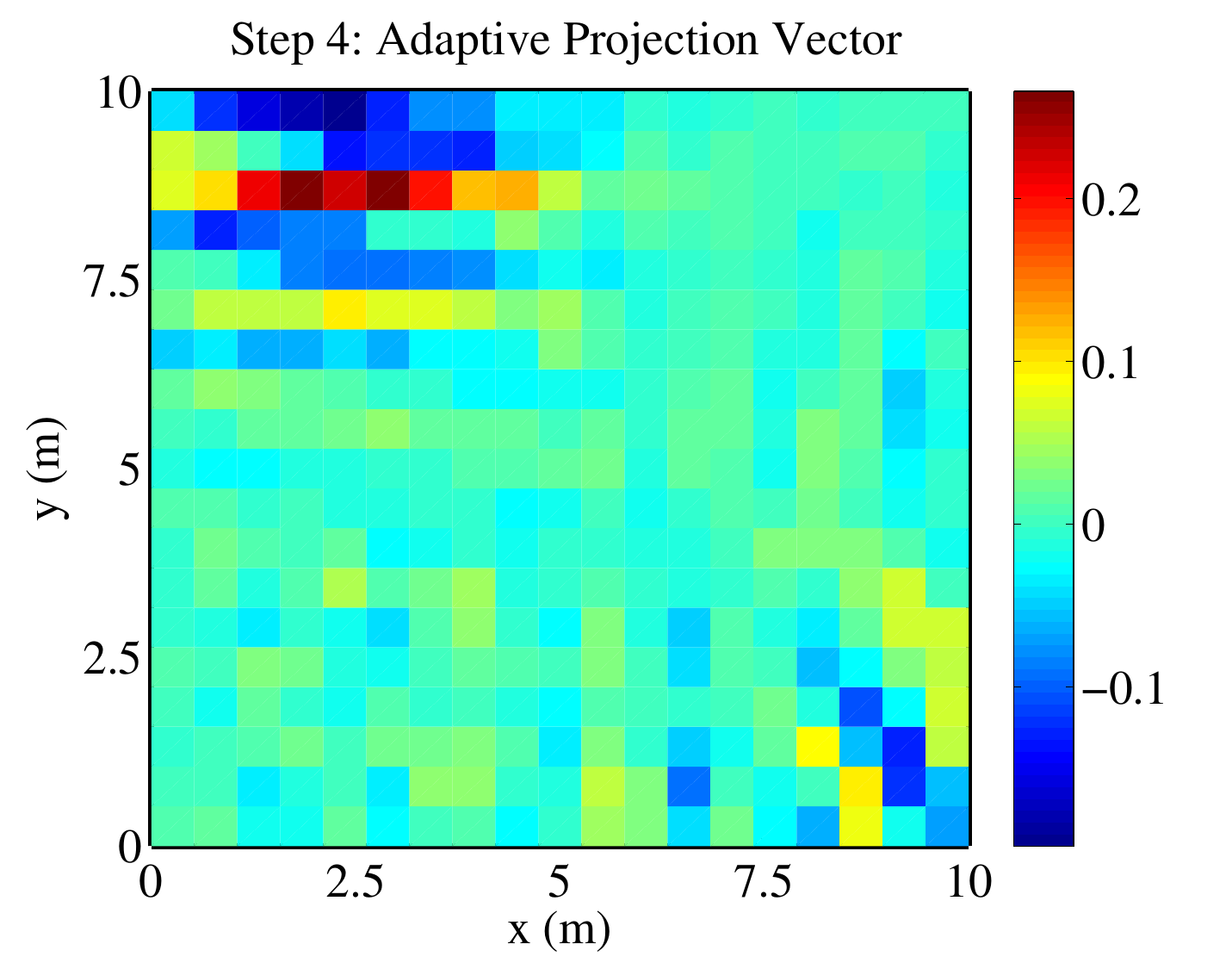}}
{\includegraphics[width=1.8in,trim= 100 235 100 235,clip=true]{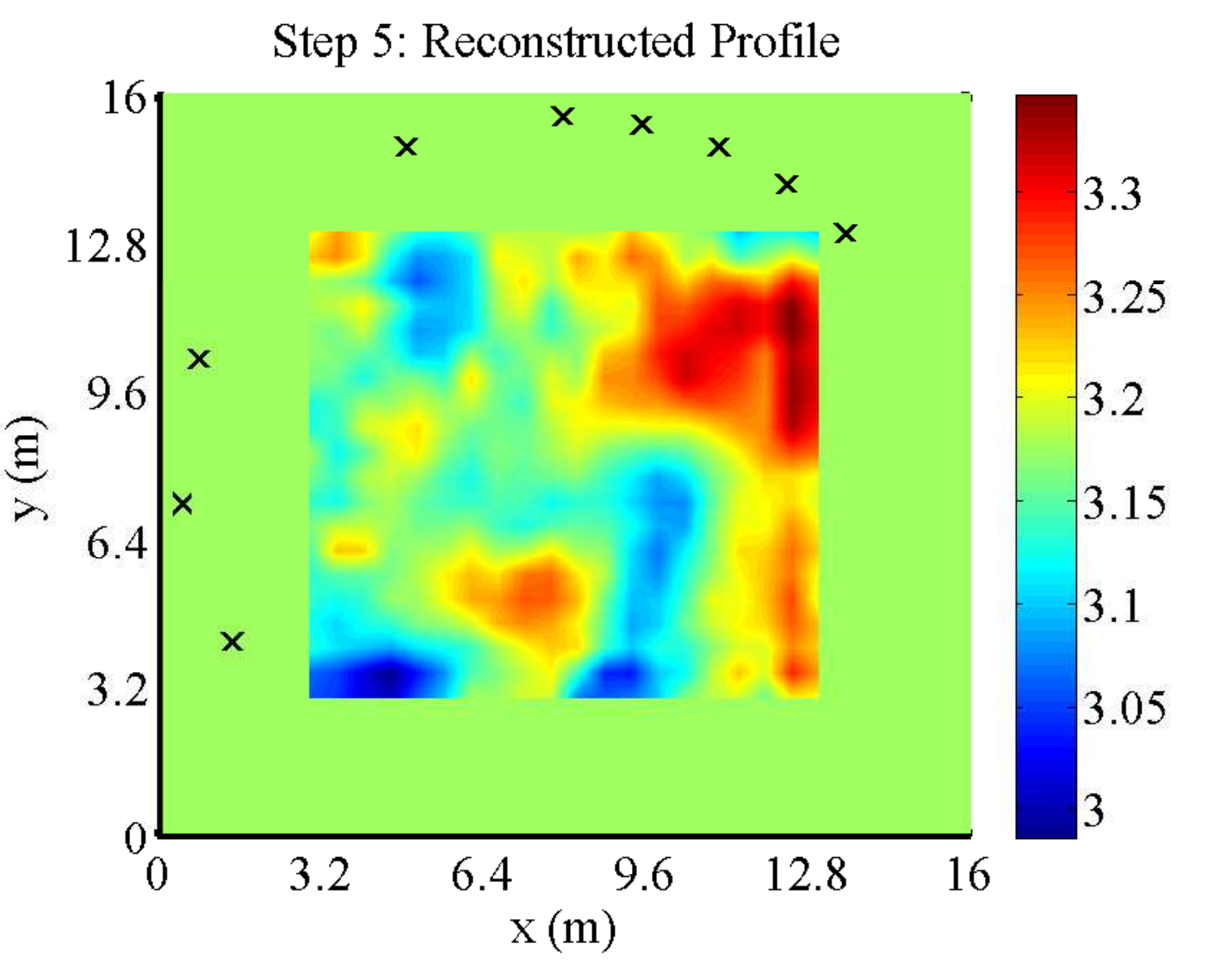}}
{\includegraphics[width=1.8in,trim= 100 235 100 235,clip=true]{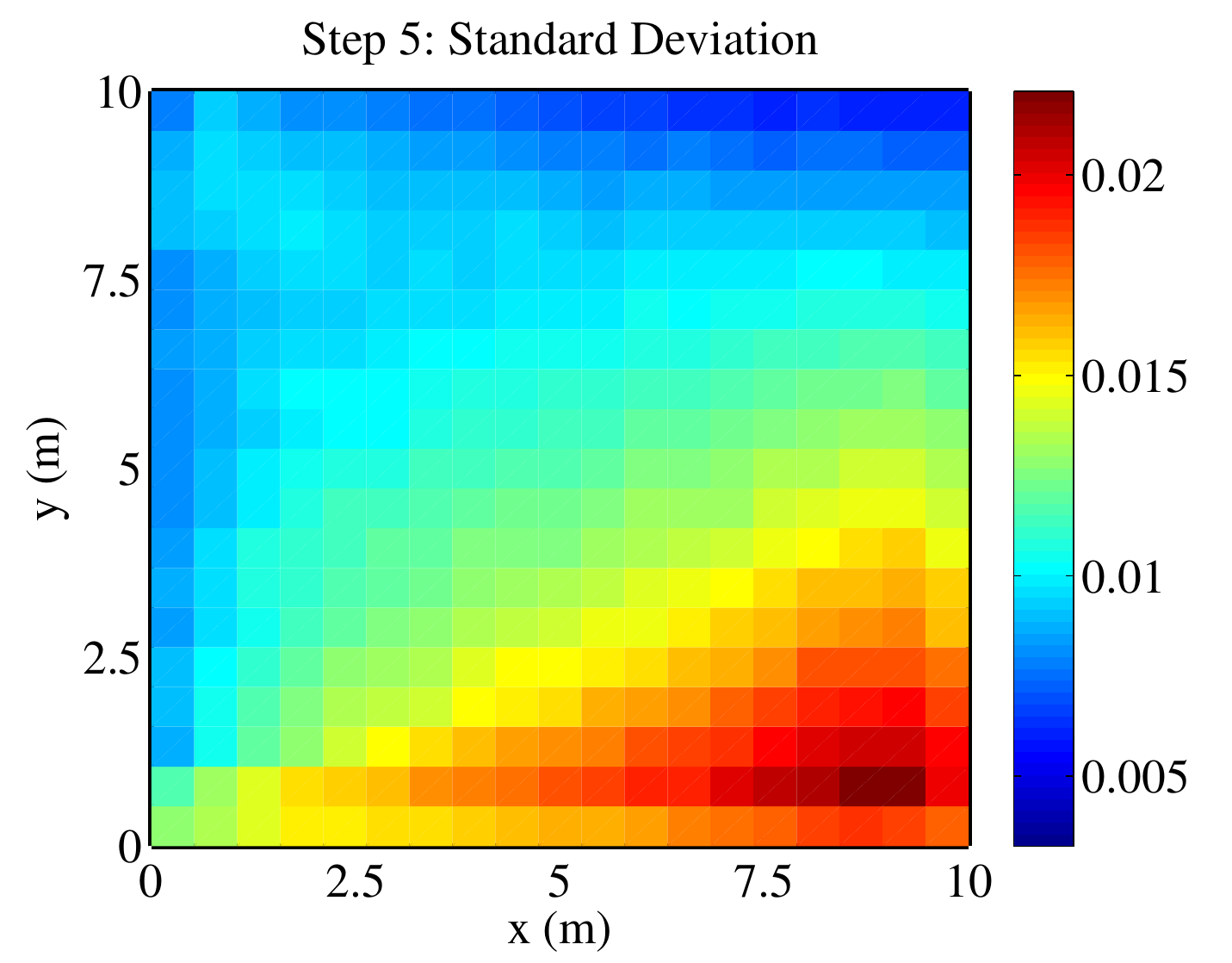}}
{\includegraphics[width=1.8in,trim= 100 235 100 235,clip=true]{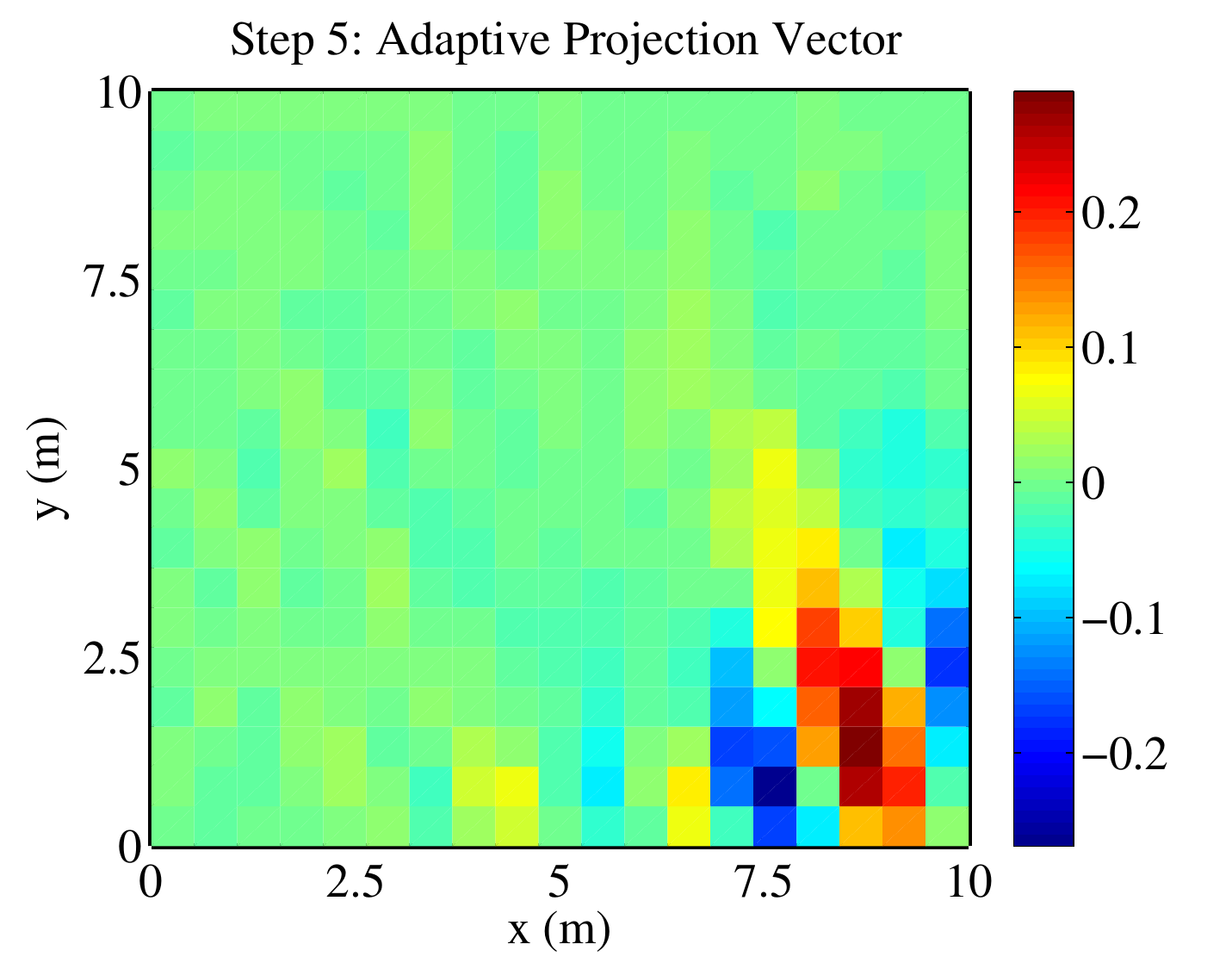}}
{\includegraphics[width=1.8in,trim= 100 235 100 235,clip=true]{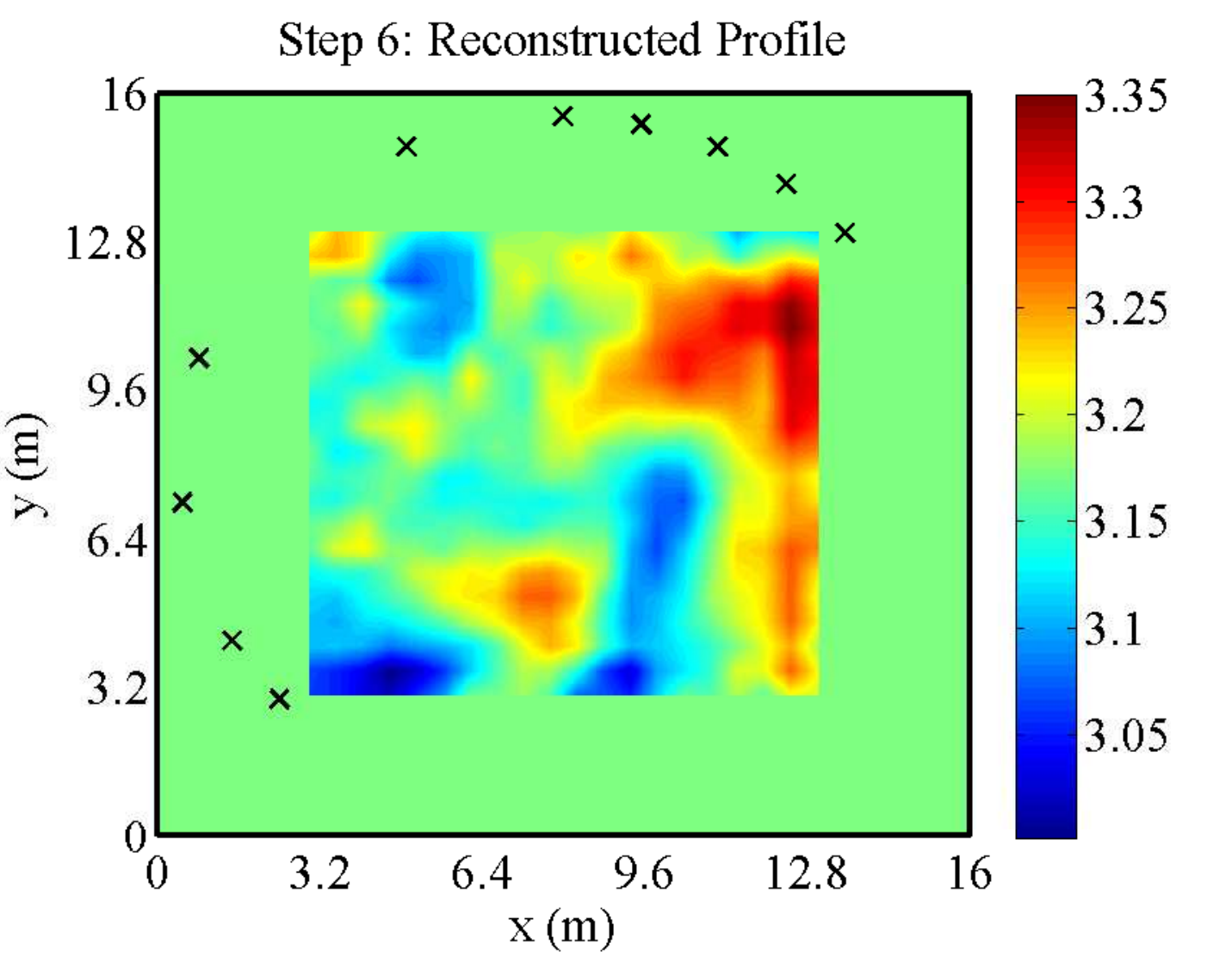}}
{\includegraphics[width=1.8in,trim= 100 235 100 235,clip=true]{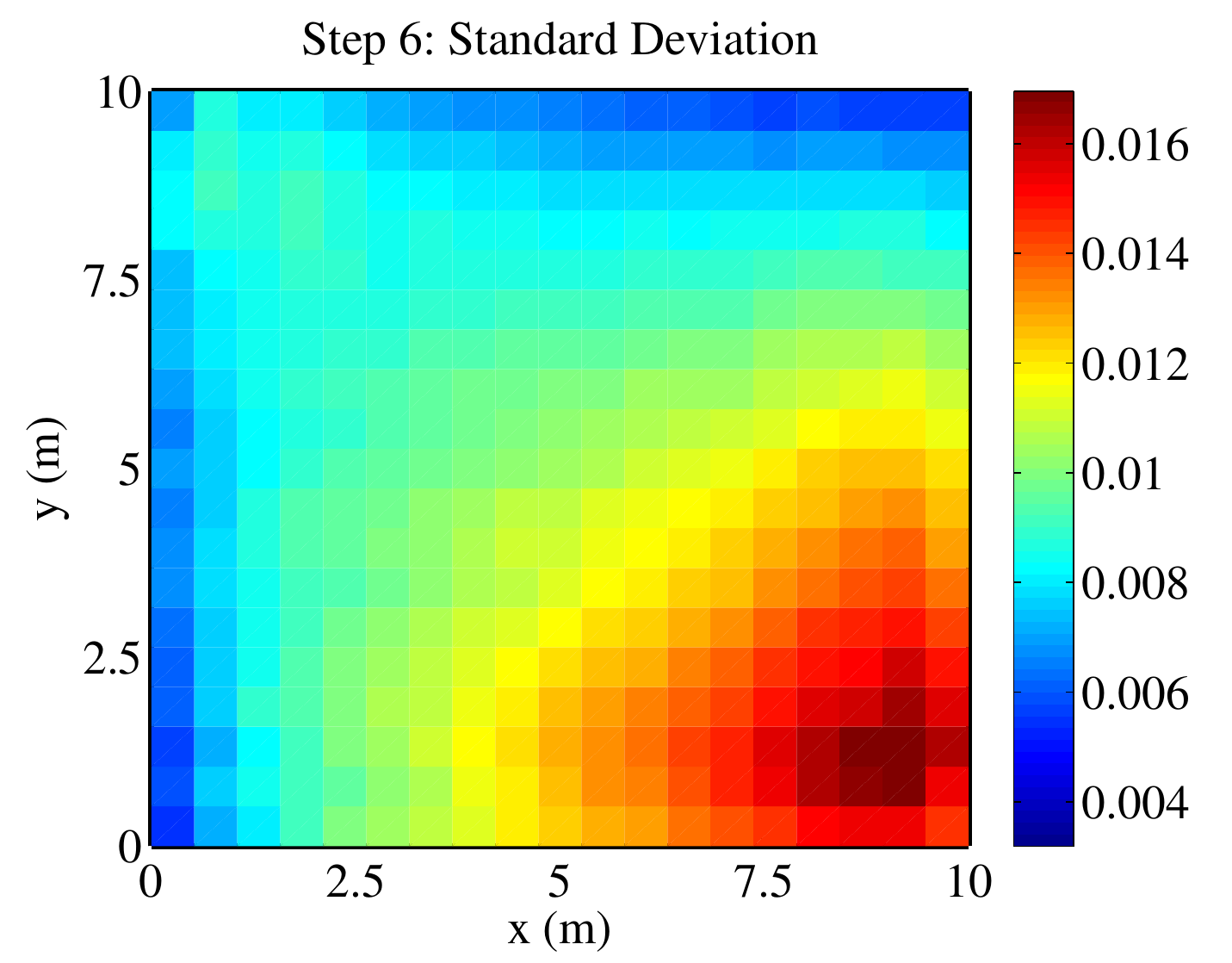}}
{\includegraphics[width=1.8in,trim= 100 235 100 235,clip=true]{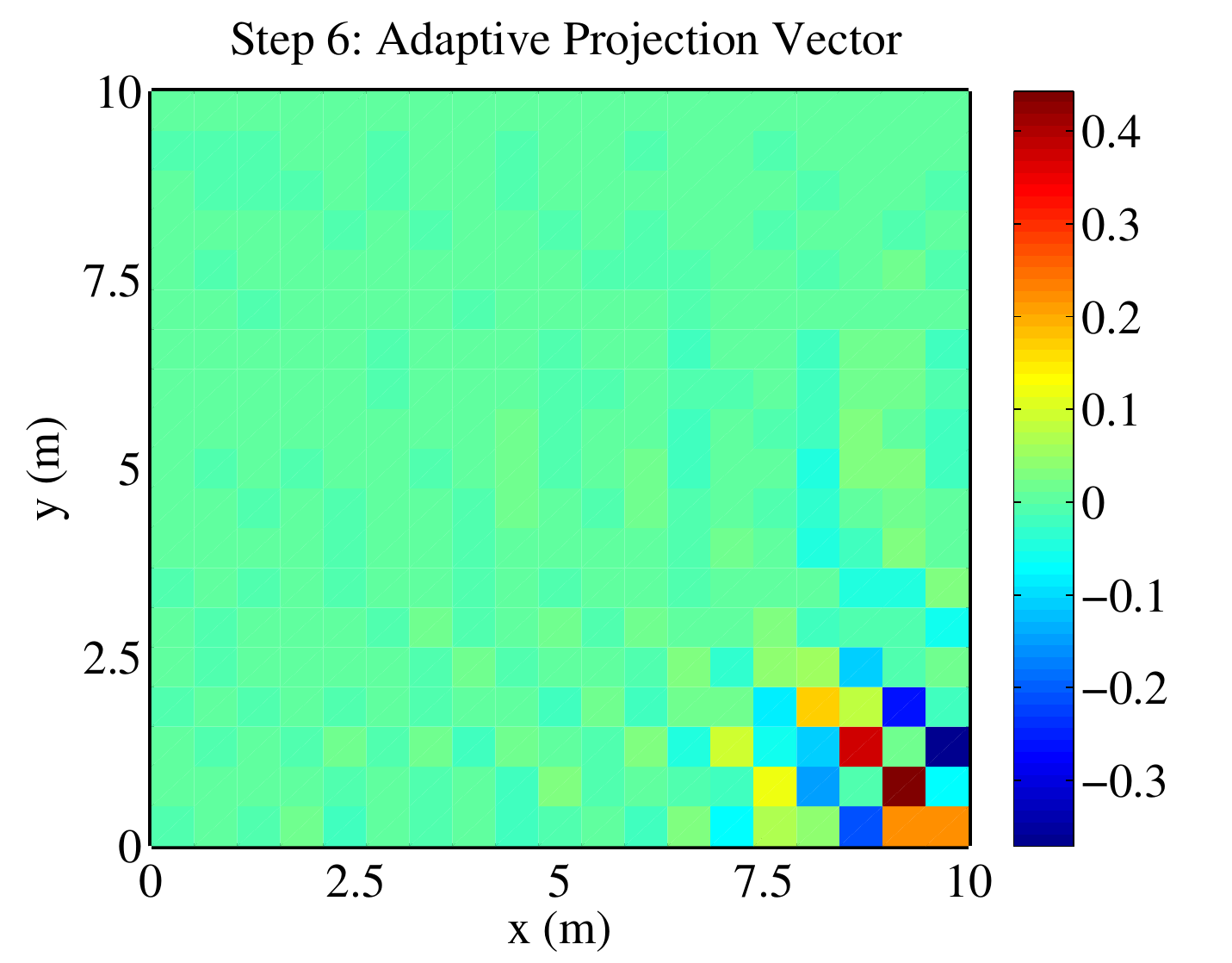}}
\caption{Reconstructed profiles, standard deviations, and projection vectors for six steps of an adaptive sensing scenario. Sensors used in each step are indicated with `x's. SNR=10 dB.}
\label{FigCH6_3}
\end{figure}
Referring to our case study, we apply the adaptive scheme to place new sensors in a `myopic' sense (one sensor in each step) as shown in Fig. \ref{FigCH6_3}. Suppose that the locations of five sensors in step 1 are pre-determined, the goal is to optimally place five more sensors. Also, suppose that sensors can only be deployed on a circle with 7.5 m radius. The figure shows the reconstructed profile from each step, the location of the utilized sensors, the estimated standard deviation, and the optimized projection vector. The new sensor has to be placed such that the field pattern produced from it best matches the optimized projection vector. Note that the standard deviation distribution is not enough to determine the location of the new sensor without computing its eigenvalue decomposition.     
\begin{figure}[!t]
\centering
\subfigure[]{\includegraphics[width=3.0in,trim= 100 120 100 120,clip=true]{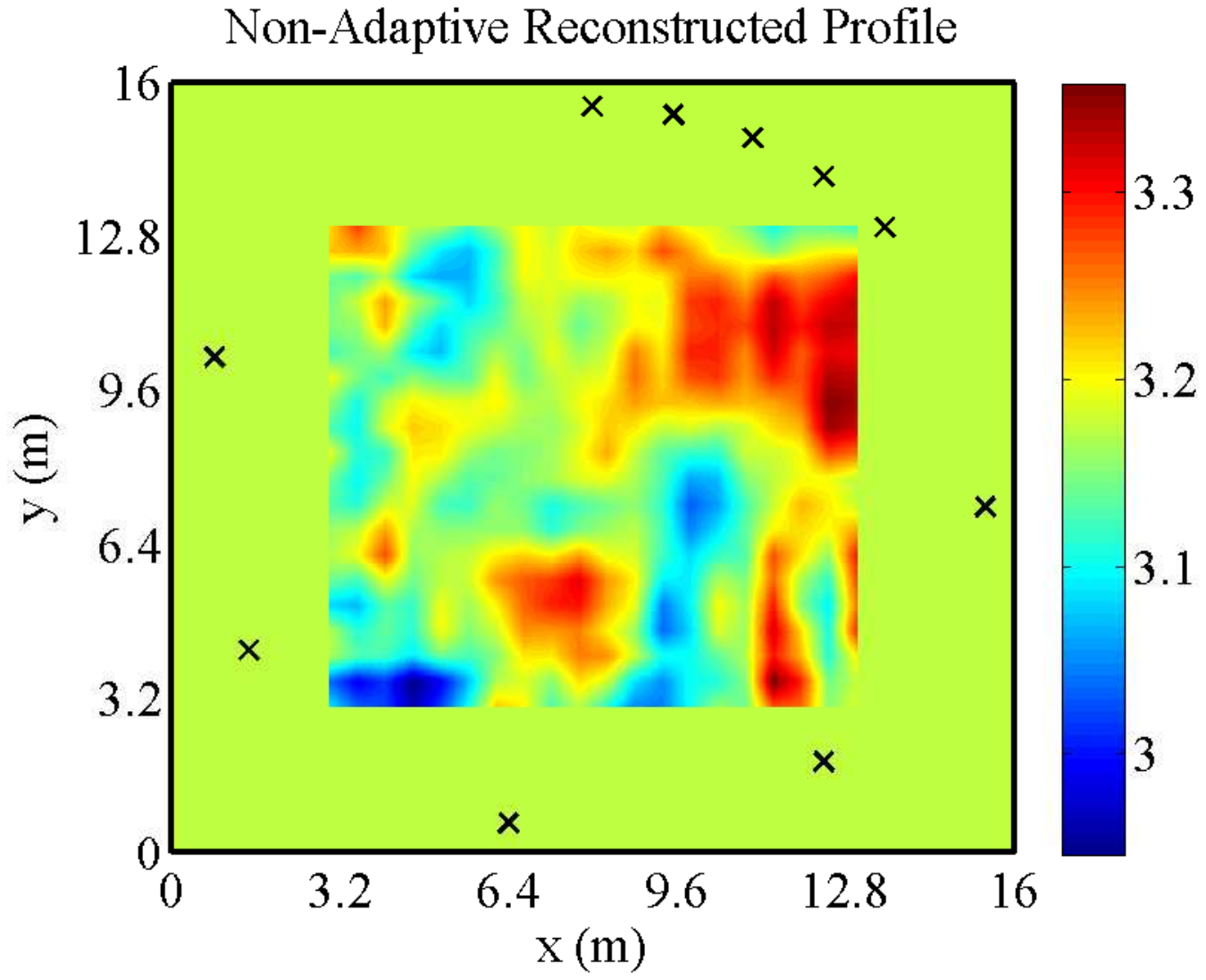}}
\subfigure[]{\includegraphics[width=3.0in,trim= 100 120 100 120,clip=true]{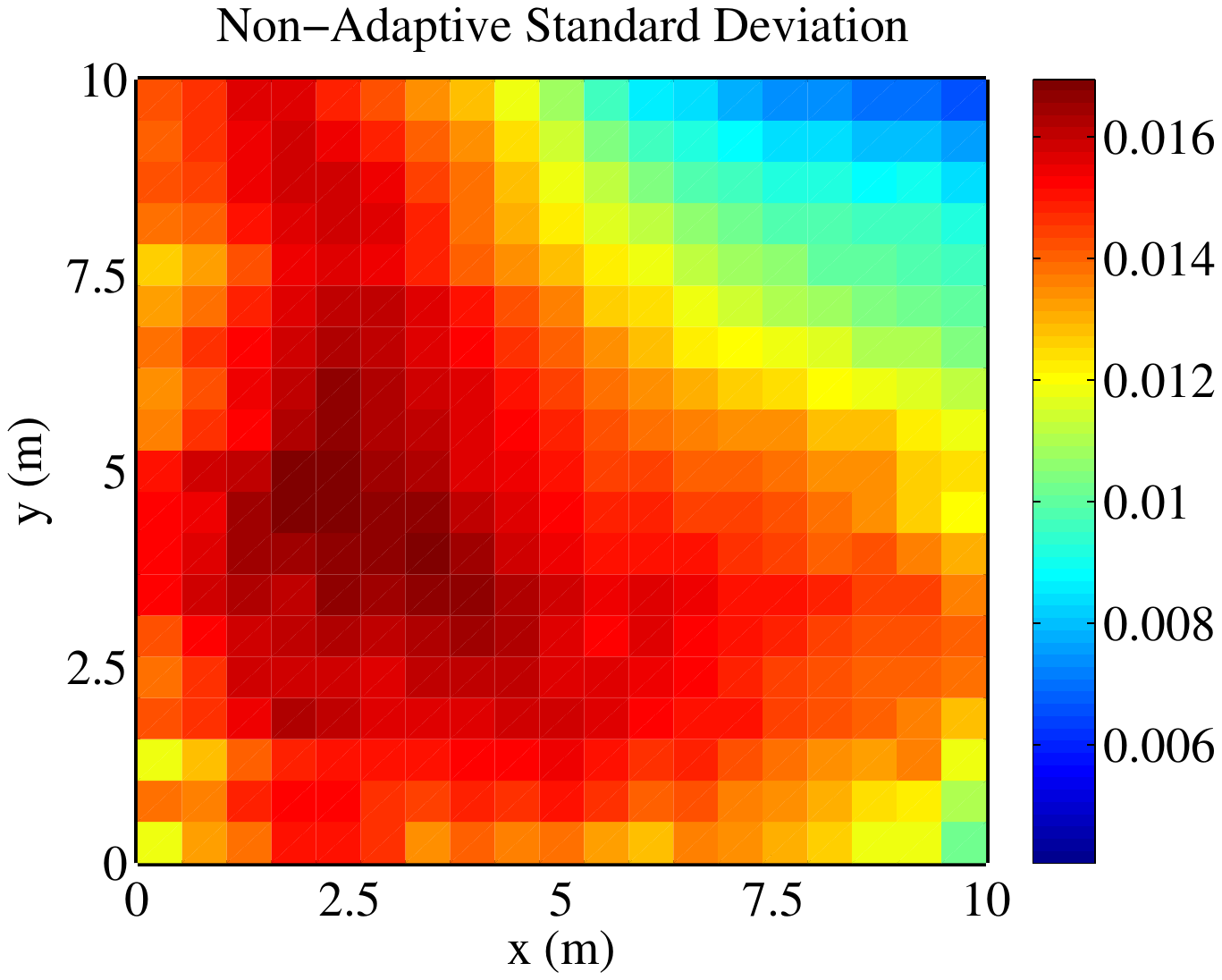}}
\caption{Non-adaptive sensing scenario. $N_s$=10 and SNR=10 dB.}
\label{FigCH6_4}
\end{figure}
Actual and estimated errors as well as the DE of each step are summarized in Table \ref{TableCH6_3}. For comparison, a non-adaptive scenario is shown in Fig. \ref{FigCH6_4}, where the same five sensors are pre-determined and the other five sensors are uniformly distributed as shown. Corresponding performance parameters are shown in Table \ref{TableCH6_3}, as well. From this comparison, it is obvious how adaptive optimized sensing yields more accurate inversion given the same number of sensors, or in other words, adaptive sensing can achieve a given inversion accuracy with less number of sensors.  
\begin{table*}[!t]
\renewcommand{\arraystretch}{1.5}
\caption{Summary of the performance parameters of adaptive and non-adaptive sensing scenarios.}
\label{TableCH6_3}
\centering
\begin{tabular}{|c|c|c|c|c|}\hline 
 & Step & Actual Error & Estimated Error & Differential Entropy (kb) \\ \hline 
\multirow{6}{*}{Adaptive Sensing} & 1 & 0.117 & 0.108 & -2.32 \\ 
 & 2 & 0.067 & 0.042 & -2.56 \\ 
 & 3 & 0.041 & 0.025 & -2.82 \\ 
 & 4 & 0.028 & 0.015 & -3.03 \\ 
 & 5 & 0.023 & 0.012 & -3.12 \\ 
 & 6 & 0.021 & 0.01 & -3.18 \\ \hline 
\multicolumn{2}{|p{1in}|}{Non-Adaptive Sensing} & 0.031 & 0.014 & -3.01 \\ \hline 
\end{tabular}
\end{table*}
\section{Time-Reversal-Assisted Localized-Inversion}

\begin{figure}[!t]
\centering
\subfigure[]{\includegraphics[width=2.0in,trim= 100 180 100 180,clip=true]{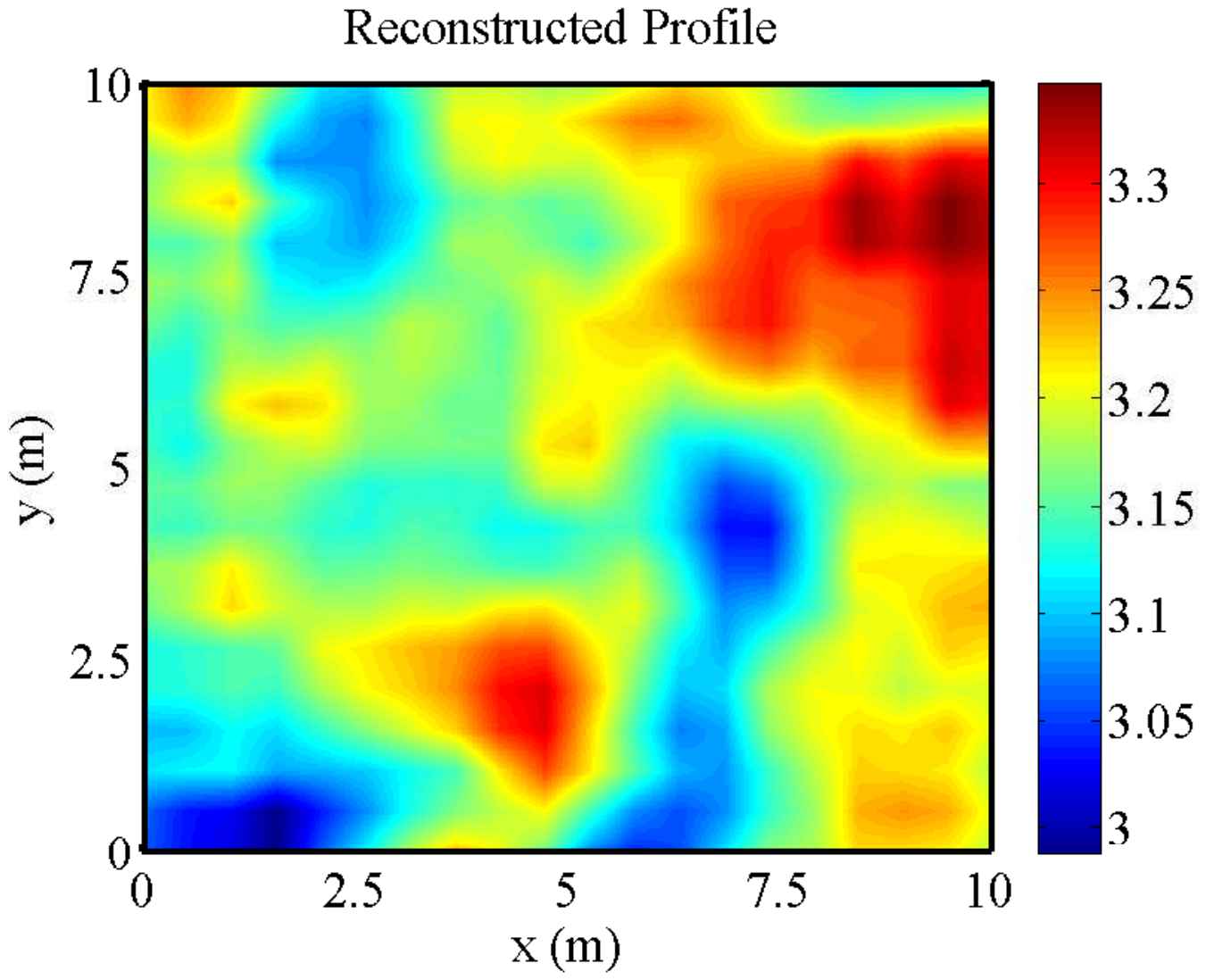}}
\subfigure[]{\includegraphics[width=2.0in,trim= 100 180 100 180,clip=true]{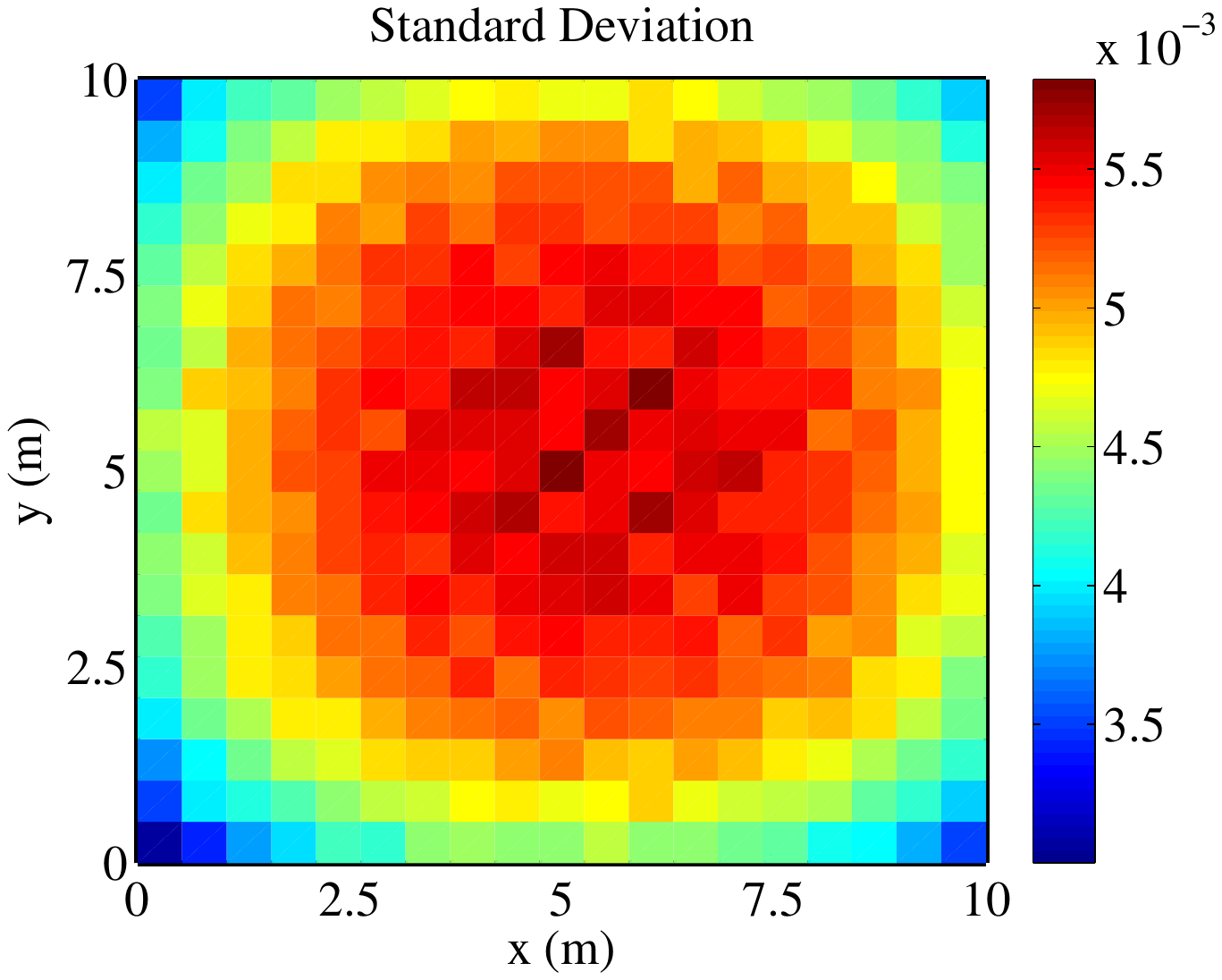}}
\subfigure[]{\includegraphics[width=2.0in,trim= 100 180 100 180,clip=true]{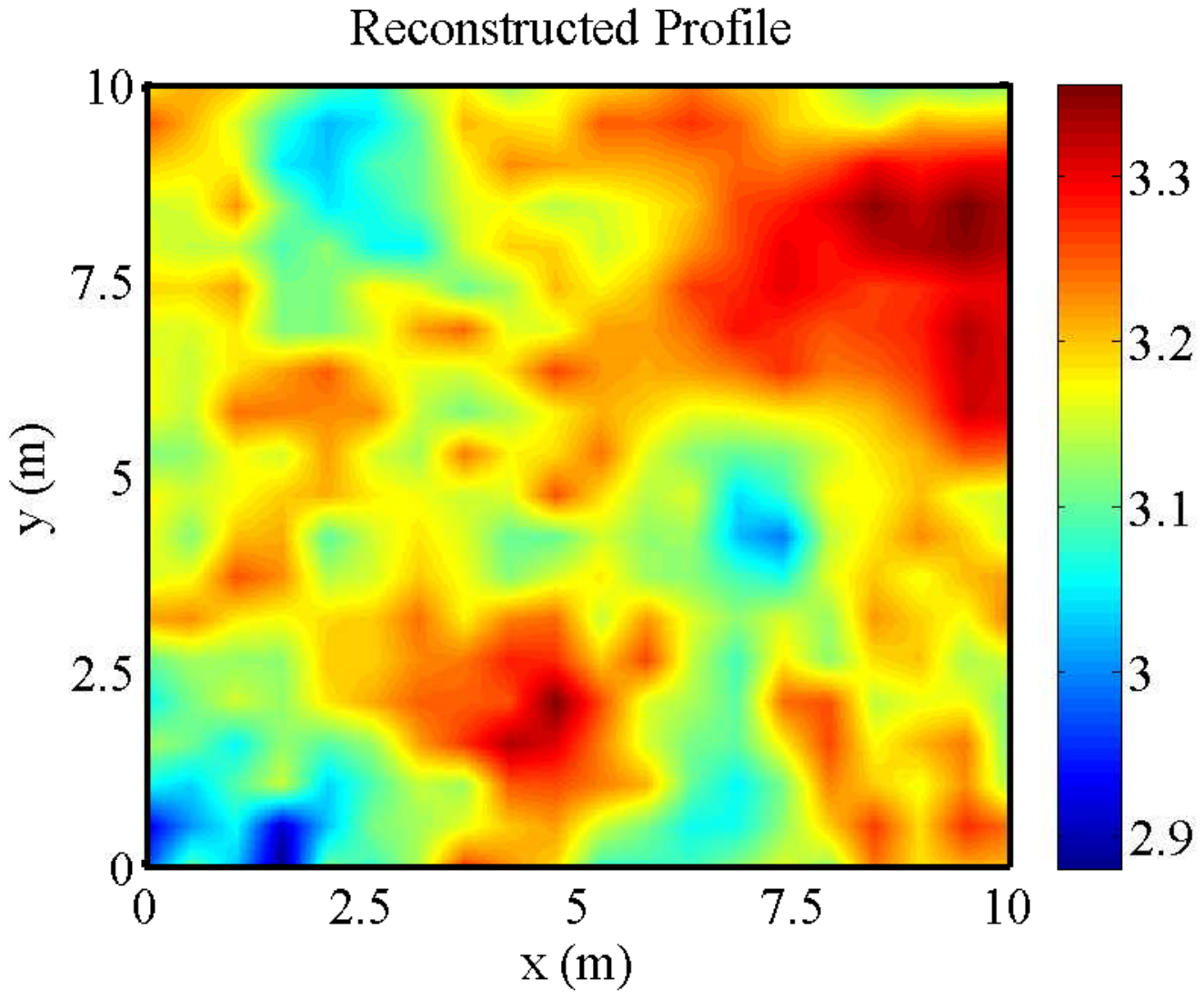}}
\subfigure[]{\includegraphics[width=2.0in,trim= 100 180 100 180,clip=true]{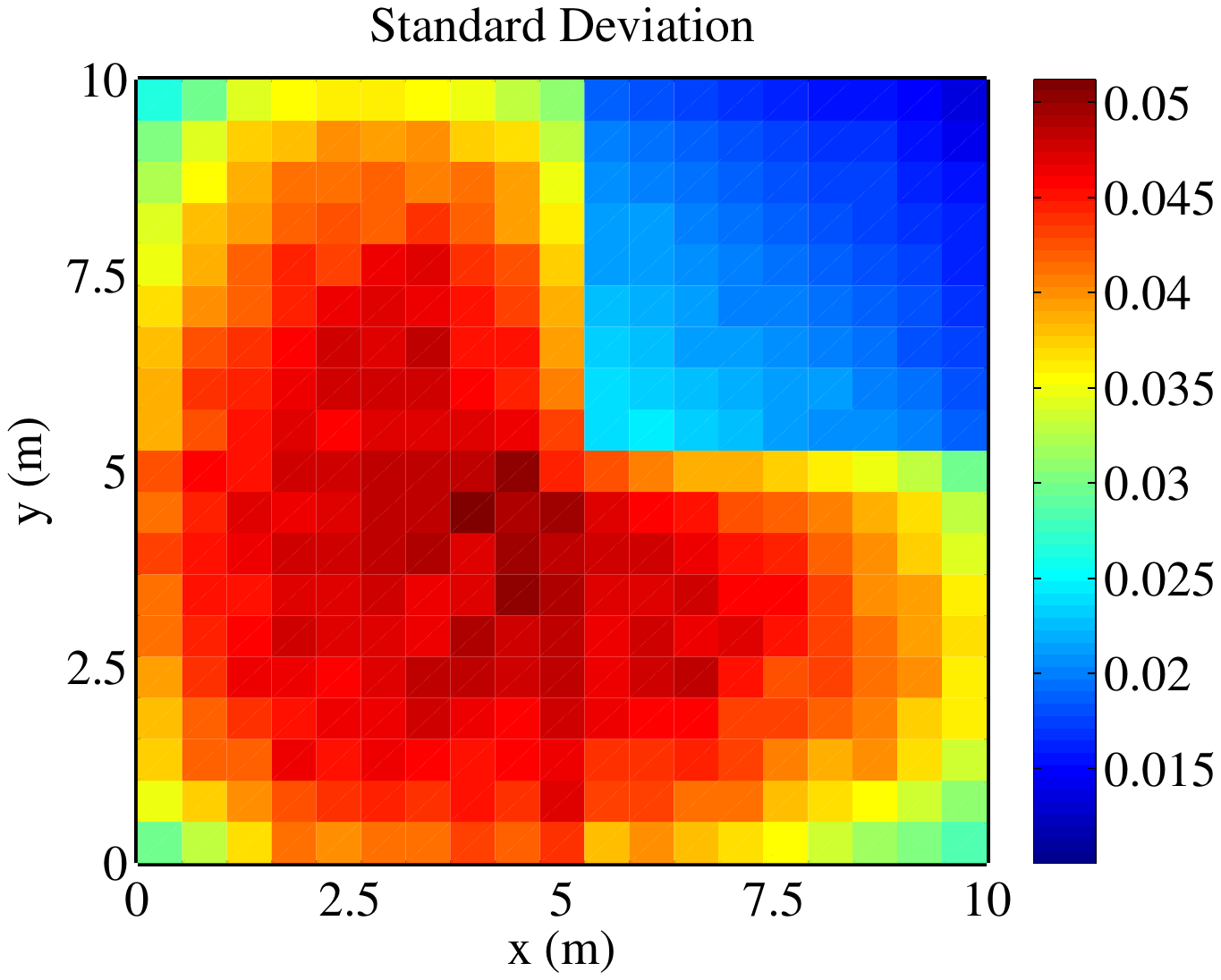}}
\subfigure[]{\includegraphics[width=2.0in,trim= 100 180 100 180,clip=true]{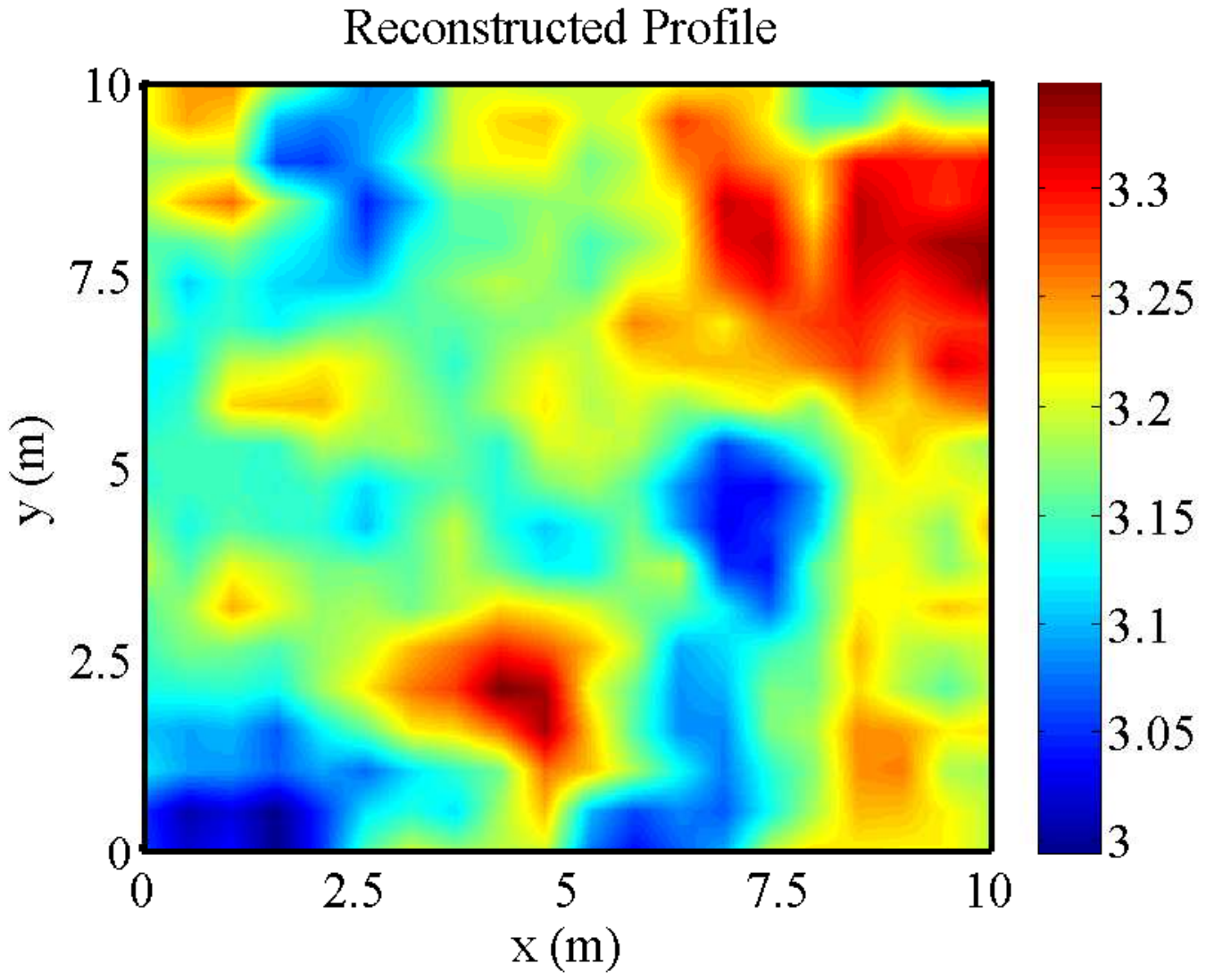}}
\subfigure[]{\includegraphics[width=2.0in,trim= 100 180 100 180,clip=true]{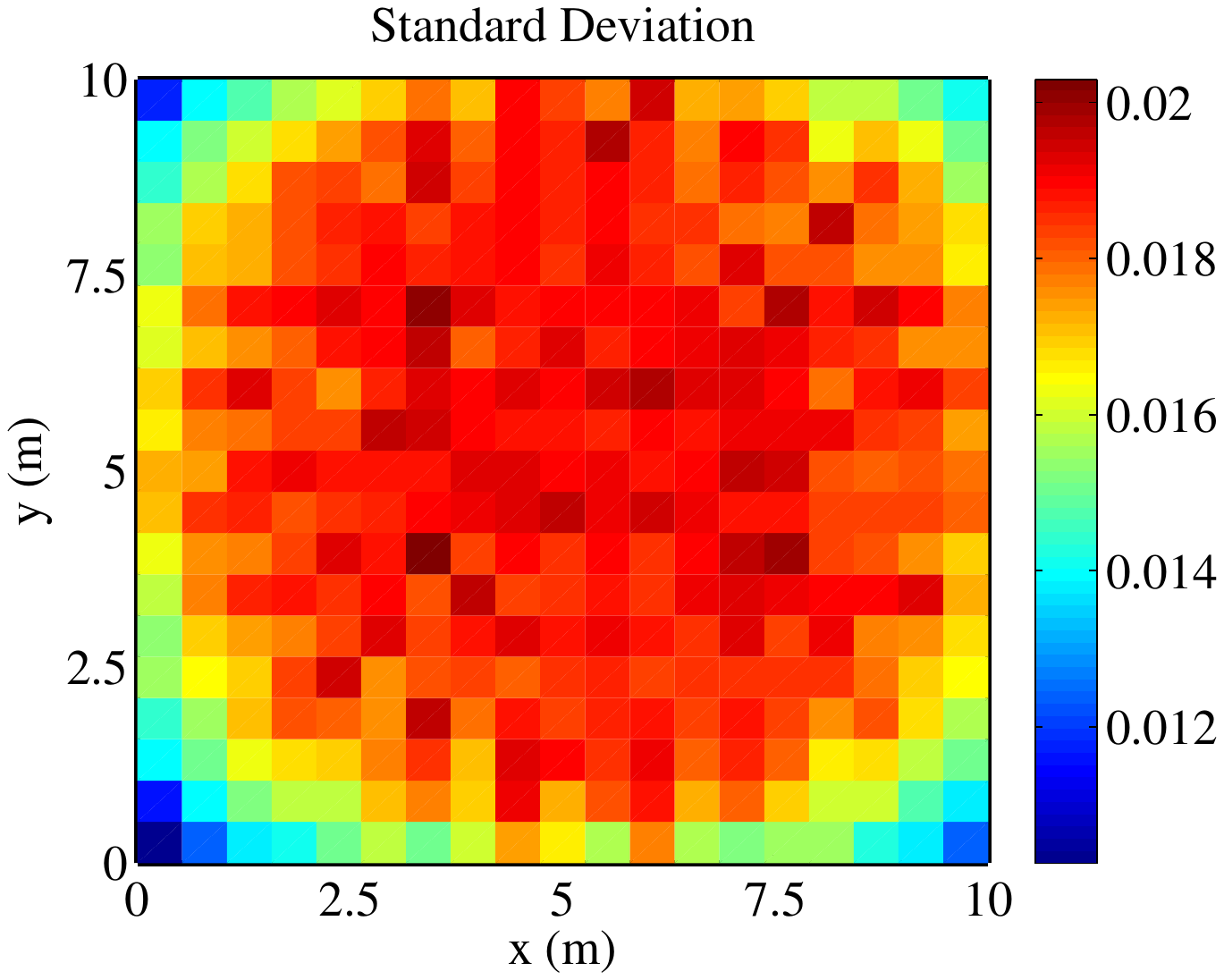}}
\caption{Time-reversal-assisted localized-inversion. (a)-(b) $N_s$=30 without TR. (c)-(d) $N_s$=30 with TR focused on one hundred pixels within the upper right quarter of the investigation domain. (e)-(f)$N_s$=10 without TR. SNR=5 dB.}
\label{FigCH6_5}
\end{figure}

If our interest is to reconstruct only a localized region of the investigation domain that can change dynamically, time-reversal (TR) focusing \cite{FinkTRmirrors, YavuzSenstivPerturb, Fouda1, TRinchangingmedia} can be used to achieve accurate localized inversion with significantly shorter processing time. We call this technique `Time-Reversal-Assisted Localized-Inversion' (TRALI). In TRALI, measurements from different sensors are linearly combined as follows
\begin{equation}
\label{equnCH6_34} 
E^{TR}(\textbf{r}_p,\omega_k)=\sum_{r=1}^{N_s}\sum_{t=1}^{N_s} G^*(\textbf{r}_p,\textbf{r}_r,\omega_k)G^*(\textbf{r}_p,\textbf{r}_t,\omega_k)E^{s}_t(\textbf{r}_r,\omega_k)
\end{equation}
where $G^*(\textbf{r}_p,\textbf{r},\omega_k)$ is the complex conjugated Green's function between pixel $p$, in the region of interest, and location $\textbf{r}$. Note that complex conjugation in the frequency domain is equivalent to TR. Assuming multistatic acquisition, the above equation is equivalent to simultaneously firing all transmitters to illuminate the DOI by a beam focused at location $\textbf{r}_p$, backscattering is then recorded by all receivers, time-reversed and projected on pixel $p$. In this way, $E^{TR}(\textbf{r}_p,\omega_k)$ will be most sensitive to the contrast of pixel $p$, and consequently, using $E^{TR}(\textbf{r}_p,\omega_k)$ in place of $E^{s}_{t} (\textbf{r},\omega_k)$ in the linear regression model (\ref{equnCH6_27}), yields accurate localized inversion. Of course, the rows of the projection matrix need to undergo the same linear combination in (\ref{equnCH6_34}). 
An example is shown in Fig. \ref{FigCH6_5}. A thirty-transceivers FA array is used to obtain very accurate inversion of the entire DOI as shown in Fig. \ref{FigCH6_5}(a) and (b). Using the same array, TRALI is applied to obtain localized inversion of one hundred pixels in the upper right quarter of the DOI, as shown in Fig. \ref{FigCH6_5}(c) and (d). Note that the local inversion sub-domain does not need to be static or contiguous, also it can be extended to encompass the entire DOI.  To further assess the performance of TRALI, a ten-transceivers FA array (which has the same data points and requires the same processing time as TRALI) is used in Fig. \ref{FigCH6_5}(e) and (f). Corresponding total error, local error (of the upper right quarter), and processing time are summarized in Table \ref{TableCH6_4}. 
TRALI is shown to produce local inversion with almost the same accuracy as the full multistatic acquisition, but with much less processing time. This comes at the expense of sacrificing the accuracy elsewhere outside the local domain of interest. Using the same number of multistatic acquisitions as the TR focusing pixels results in a larger local error, but less overall error.  
%
%
\begin{table*}[!t]
\renewcommand{\arraystretch}{1.5}
\caption{Summary of the performance parameters for the setups in Fig. \ref{FigCH6_5}.}
\label{TableCH6_4}
\centering
\begin{tabular}{|c|c|c|c|}
\hline
Setup & Total r.m.s. error & Local r.m.s. error & Processing time (sec.)\\
\hline
$N_s$=30 w/o TR& 0.01146 & 0.01122 & 36 \\
\hline
$N_s$=30 w/ TR & 0.0306 & 0.0135 & 2 \\
\hline
$N_s$=10 w/o TR& 0.0267 & 0.0315 & 5.4 \\
\hline
\end{tabular}
\end{table*}

\section{Bayesian Distorted-Born Iterative Method}
So far, we considered the application of the proposed UWB BCS inversion to low contrast media obeying the first order Born approximation. In this section, we extend the applicability of the method to high contrast continuous media. The proposed Bayesian inversion scheme can be applied iteratively, yielding what we call `Bayesian Distorted-Born Iterative Method' (BDBIM). In conventional DBIM \cite{Moghadam1, Weedon1, MoraInv, WangDBIM, MoghadamDBIM, WangDBIM2}, a cost function is defined, usually as the $L_2$ norm between measured scattered field and synthetic scattered field computed from the reconstructed profile, and the method proceeds iteratively to minimize that cost function. Reconstructed profile from each iteration is used to compute the synthetic scattered field as well as the Green's function used in the next iteration. The method converges when the cost function gets below a certain pre-determined threshold. BDBIM proceeds the same way; however, instead of explicitly defining a cost function on the scattered fields, the estimated standard deviation from the Bayesian solver can be used as stopping criterion. 
\begin{figure}[!t]
\centering
\subfigure[]{\includegraphics[width=2.0in,trim= 100 180 100 180,clip=true]{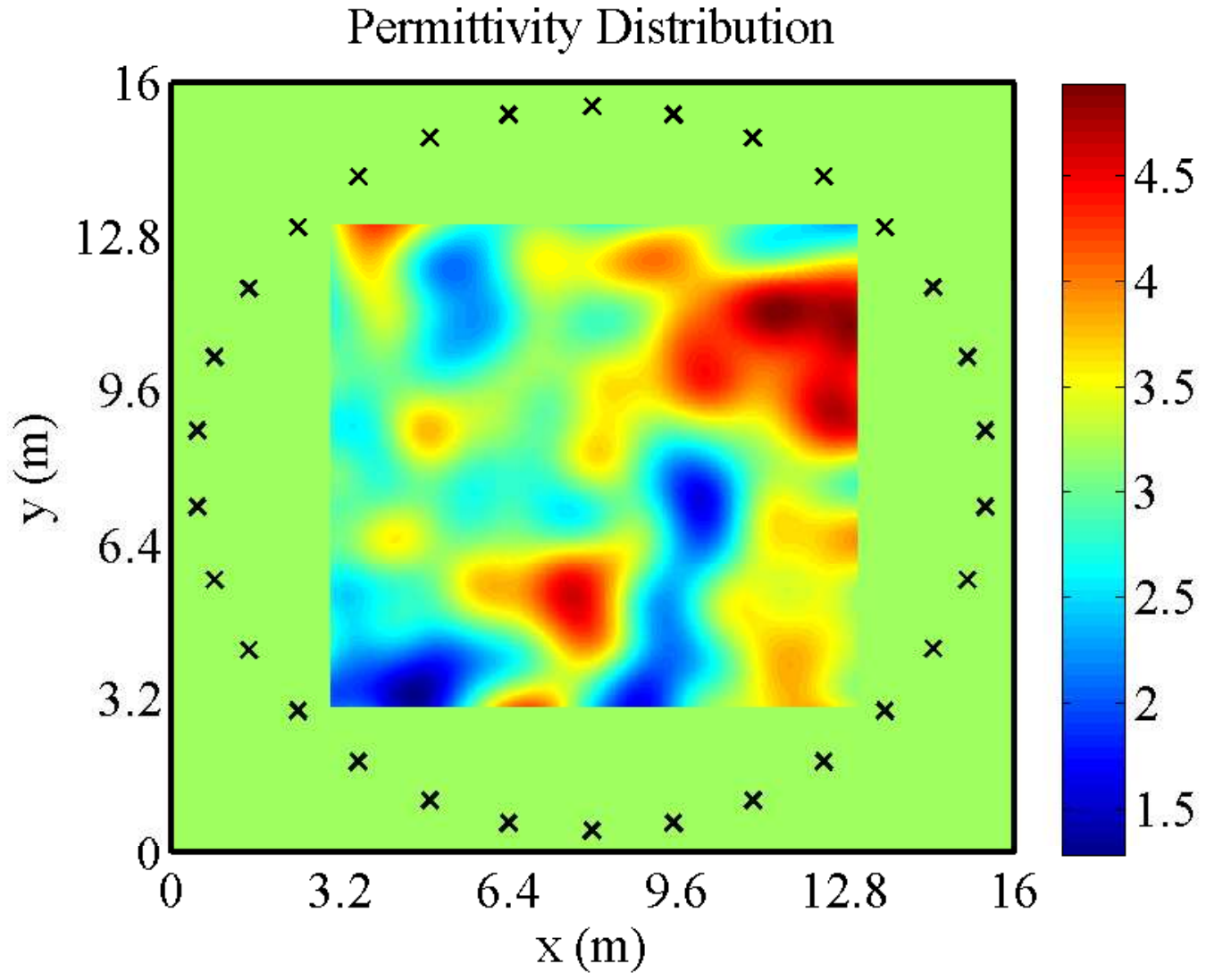}}
\subfigure[]{\includegraphics[width=2.0in,trim= 100 180 100 180,clip=true]{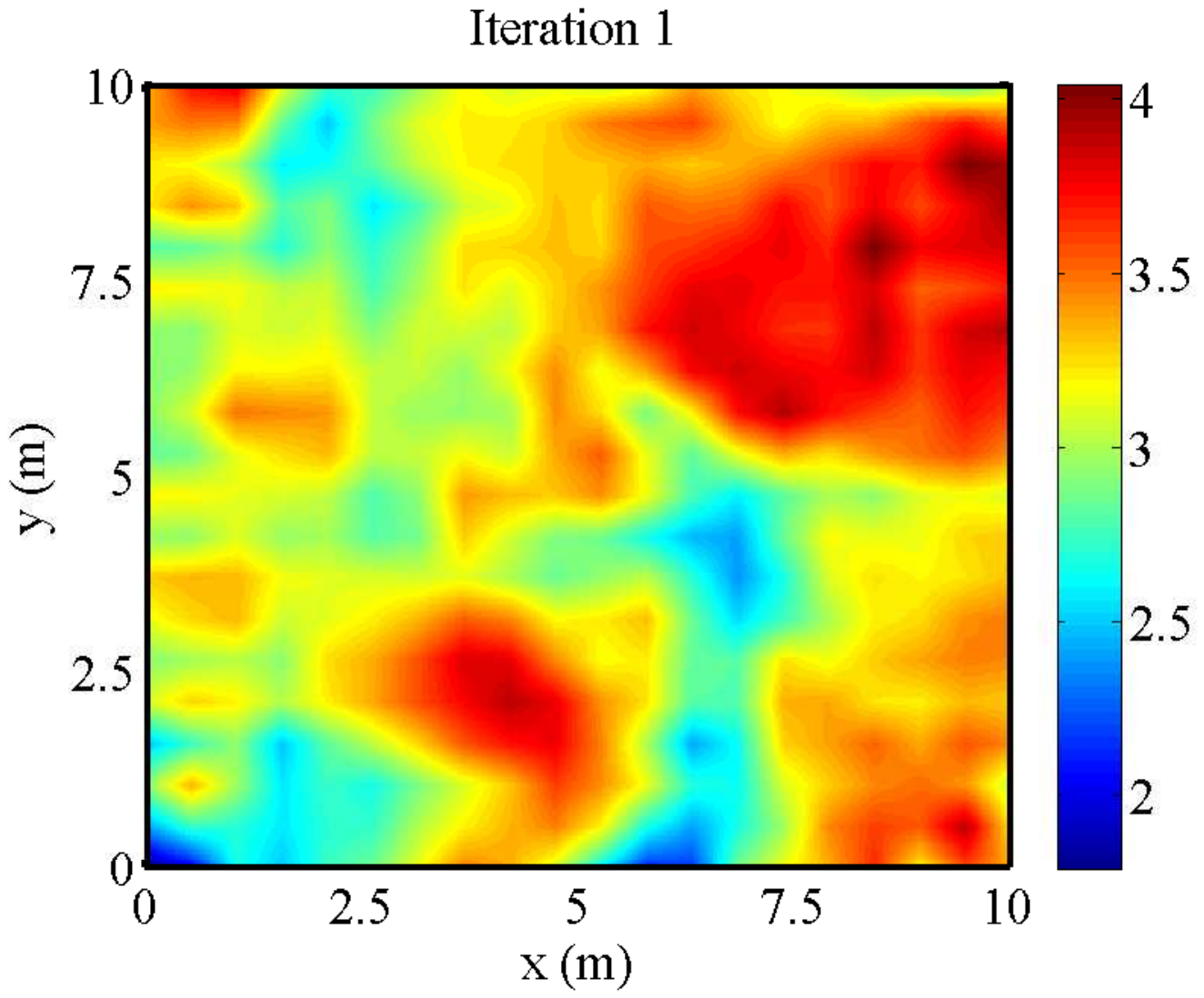}}
\subfigure[]{\includegraphics[width=2.0in,trim= 100 180 100 180,clip=true]{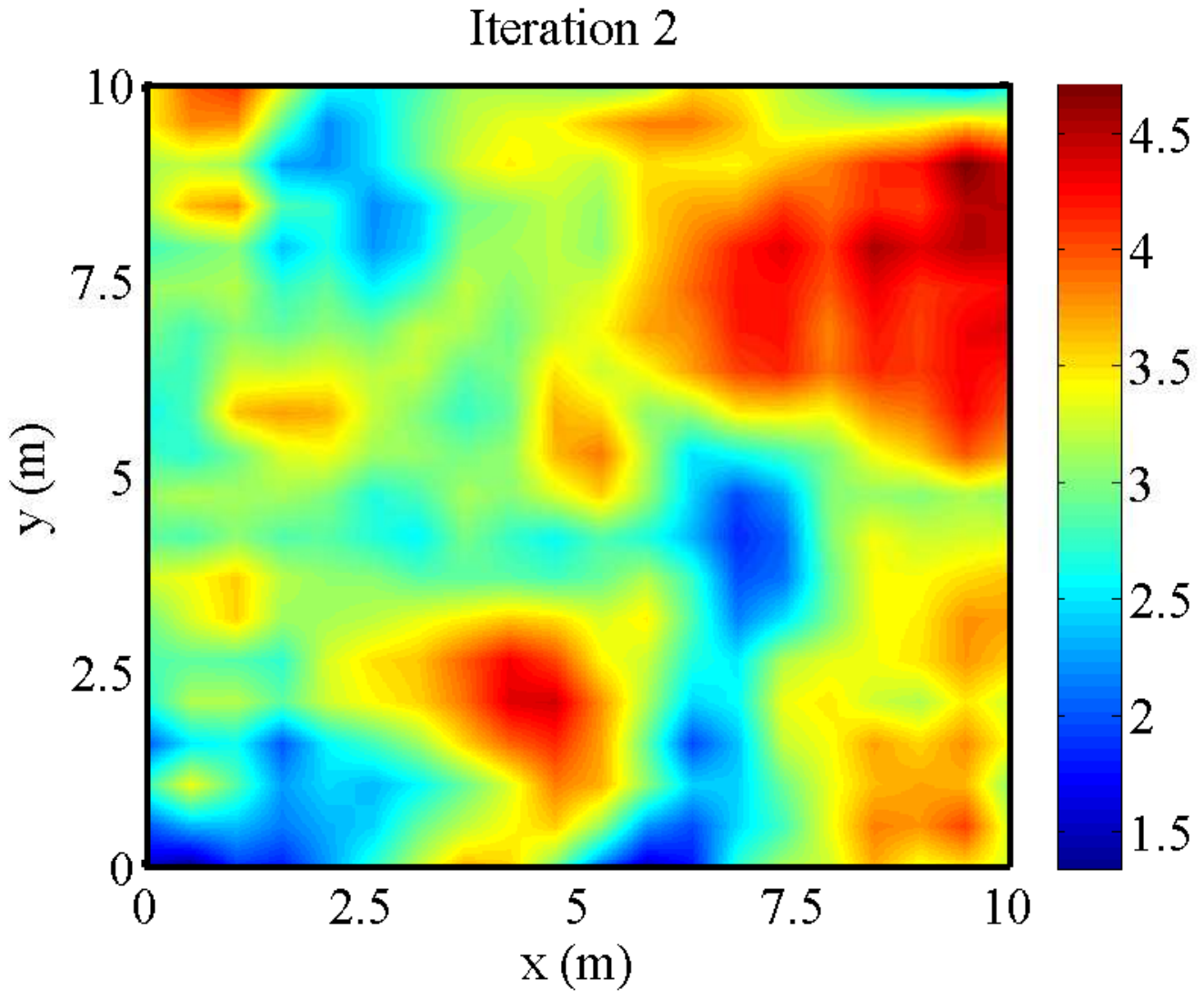}}
\subfigure[]{\includegraphics[width=2.0in,trim= 100 180 100 180,clip=true]{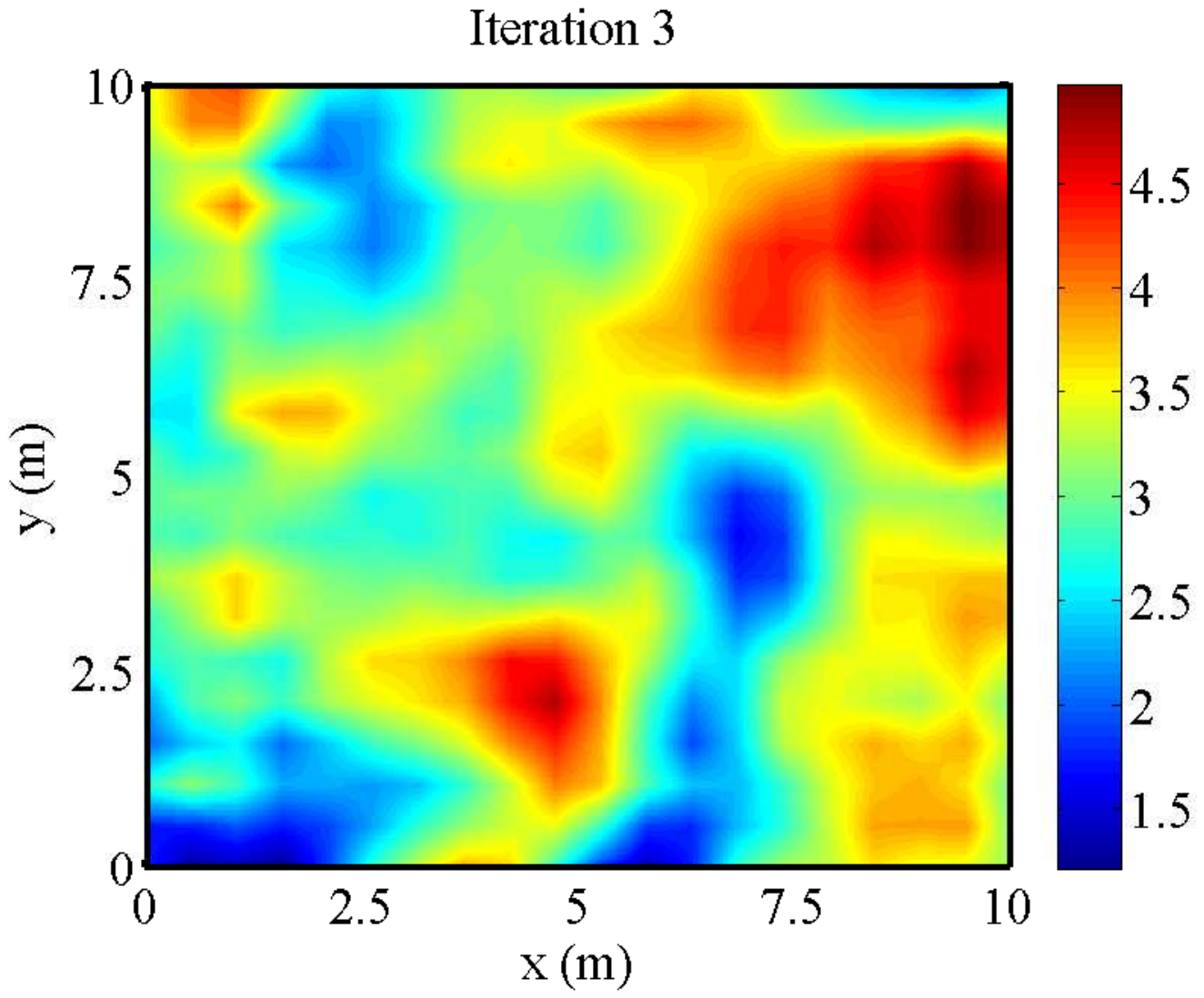}}
\subfigure[]{\includegraphics[width=2.0in,trim= 100 180 100 180,clip=true]{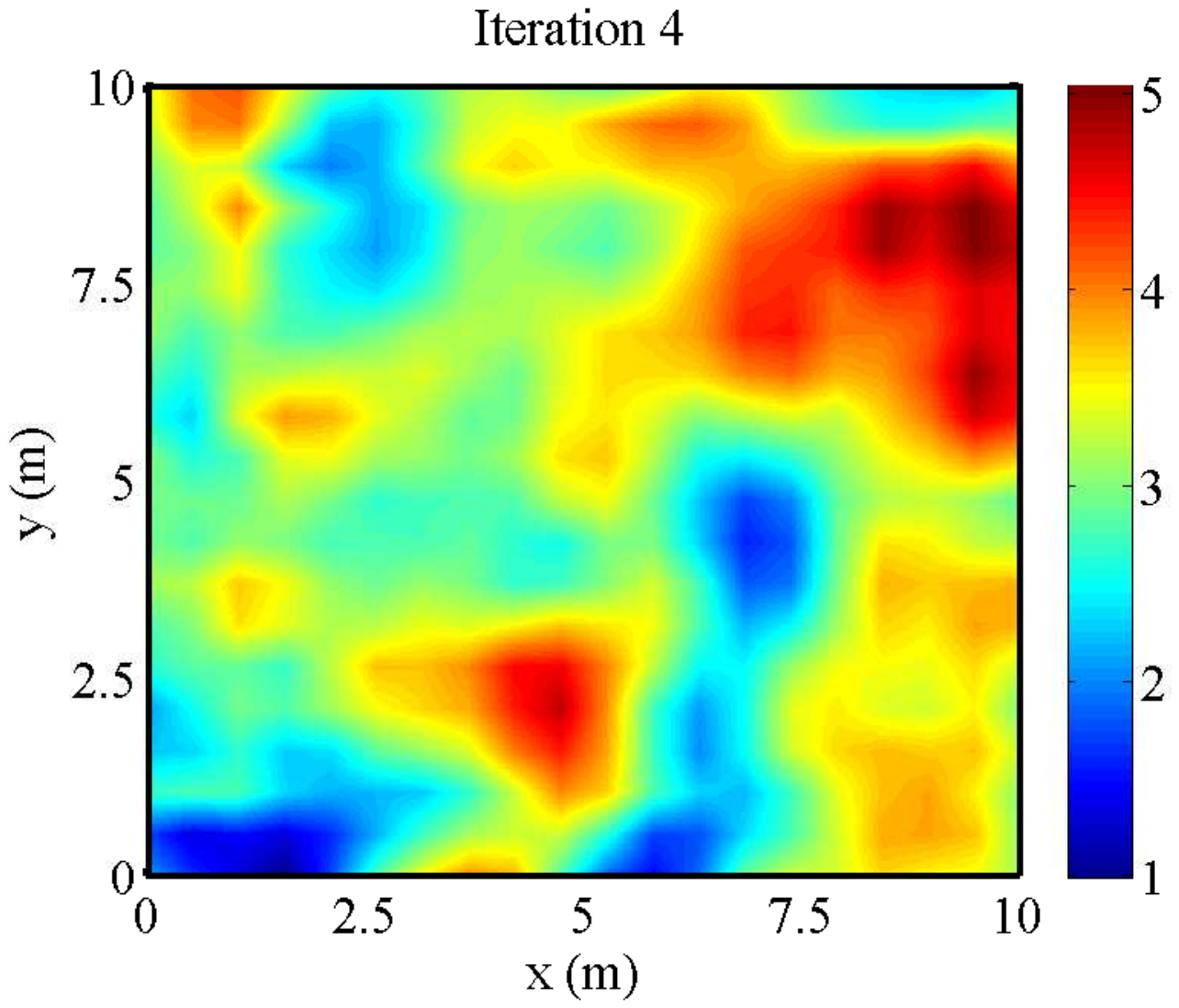}}
\subfigure[]{\includegraphics[width=2.0in,trim= 100 180 100 180,clip=true]{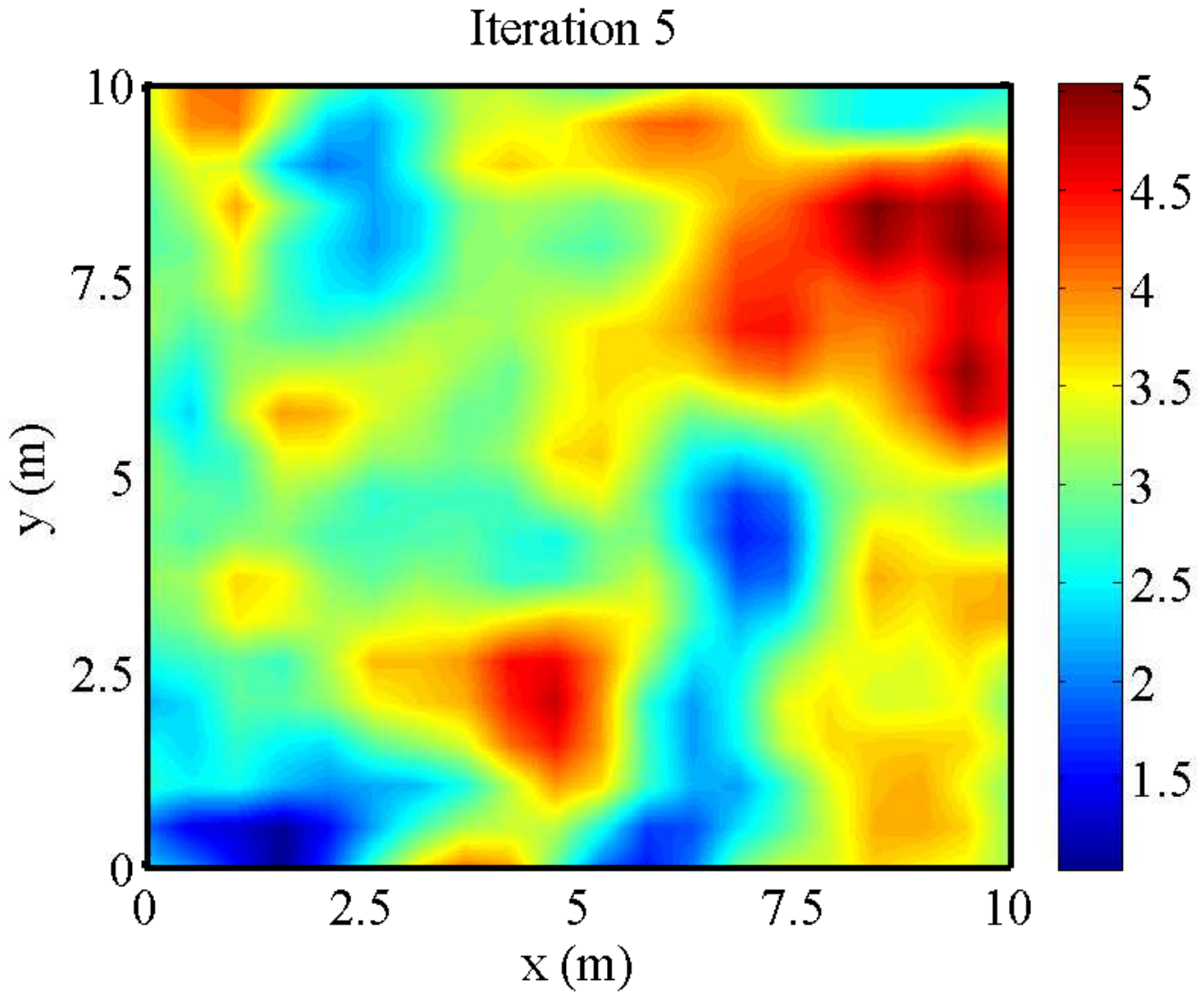}}
\caption{Bayesian DBIM. (a) Forward permittivity distribution. (b)-(f) Reconstructed profiles from five iterations. $N_s$=30 and SNR=10 dB.}
\label{FigCH6_6}
\end{figure}
\begin{figure}[!t]
\centering
\subfigure[]{\includegraphics[width=3.0in,trim= 90 100 90 120,clip=true]{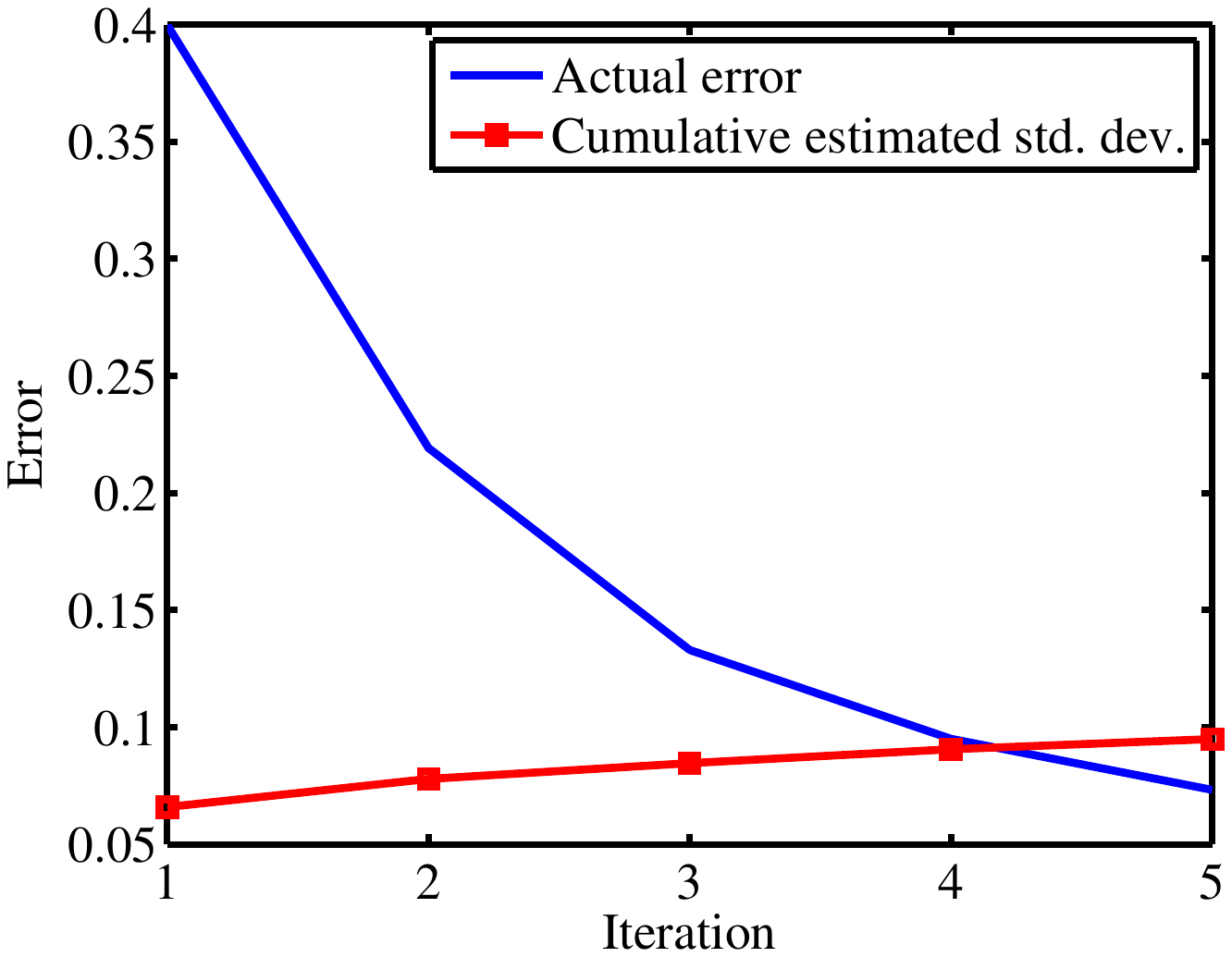}}
\subfigure[]{\includegraphics[width=4.1in,angle=90,trim= 90 100 90 0,clip=true]{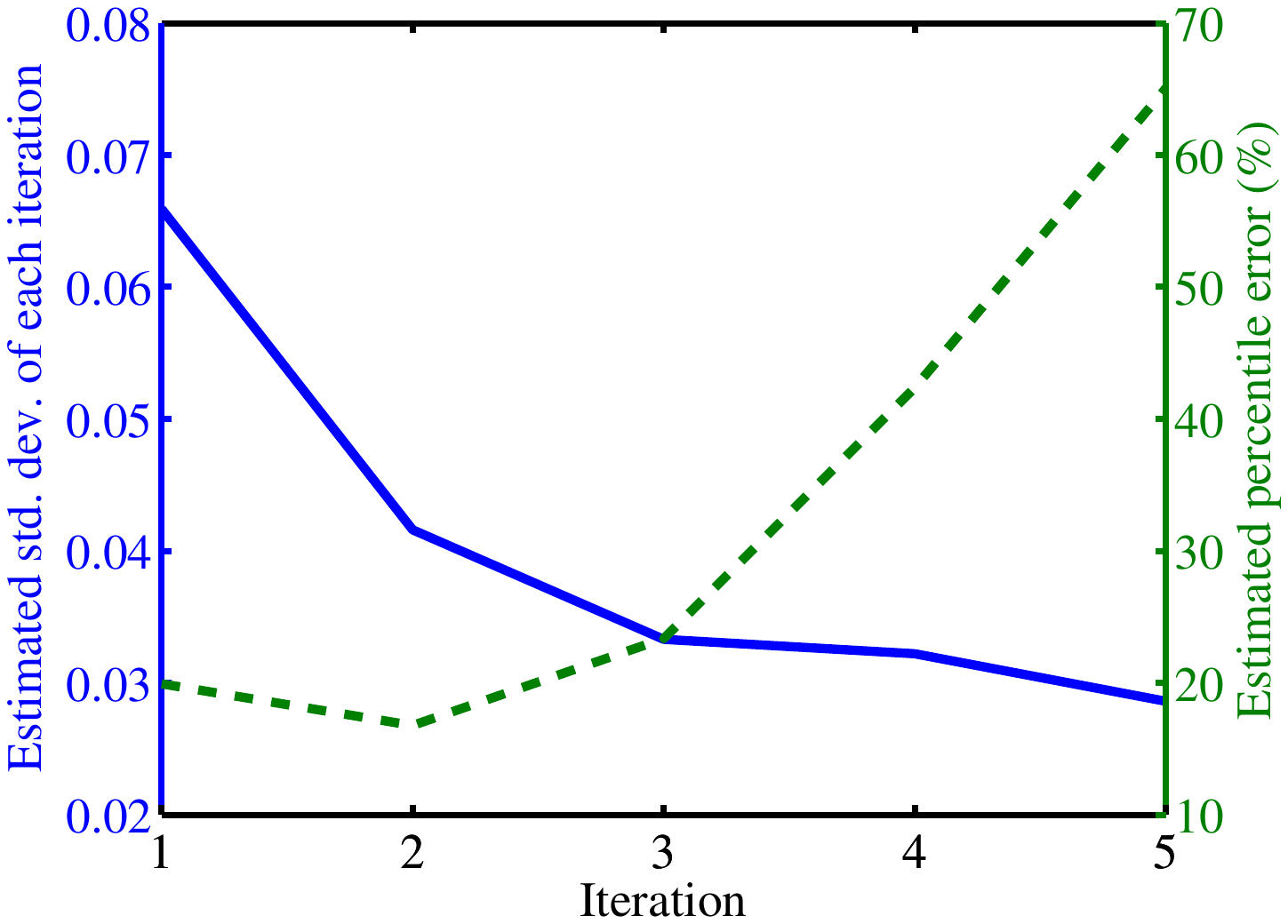}}
\caption{Error analysis of Bayesian DBIM. (a) Actual total error and cumulative standard deviation after each iteration. (b) Estimated standard deviation of individual contributions from each iteration. Percentile error is the ratio of the estimated standard deviation to the r.m.s. of the differential contrast function contributed from each iteration.}
\label{FigCH6_7}
\end{figure}
To illustrate that, consider the example in Fig. \ref{FigCH6_6}. In the first iteration, a uniform homogeneous background is used in BCS inversion. The reconstructed profile is shown in Fig. \ref{FigCH6_6}(b). Actual error and estimated standard deviations are shown in Fig. \ref{FigCH6_7}(a) and (b), respectively. The percentile error shown in Fig. \ref{FigCH6_7}(b) is the ratio of the estimated standard deviation to the r.m.s. of the contrast function contributed from each iteration. The reconstructed profile from the first iteration is plugged into a forward problem numerical solver, and used to compute the synthetic scattered field and the Green's function to be used in the following iteration. The synthetic scattered field is subtracted from the (noisy) measurements, and that differential signal is used as the measurements vector in the second iteration. The reconstructed profile from the second iteration (refereed to as iteration 2 contribution) is added to the reconstructed profile from the first iteration to yield the overall profile of iteration 2 shown in Fig. \ref{FigCH6_6}(c). The process is then repeated.  
Assuming that reconstructed contributions from different iterations are independent random variables, covariance matrices form all iterations can be added up, yielding the cumulative estimated standard deviation plotted in Fig. \ref{FigCH6_7}(a). There are several interesting points to note here. In the early iterations, the cumulative estimated error is not an accurate measure for the actual error; this is because of the deficiency of the underlying Born approximation to precisely model the scattered field as these stages. With increasing iterations, the discrepancy between actual and estimated errors gets smaller. 
As the method proceeds, reconstructed contribution gets smaller and smaller, and so does the associated estimated standard deviation. However, the standard deviation decreases at a slower rate, because the SNR of each inversion also decreases, this explains the increase in the percentile error shown in Fig. \ref{FigCH6_7}(b) with iterations. The percentile error is inversely proportional to the confidence level, therefore, a maximum threshold can be set on the former to determine when to stop. Intuitively, the higher SNR we have, the further we can go on with iterations, and the more accurate the inversion will be for a given (desired) confidence level.

\section{Layered Media Examples}

\begin{figure}[!t]
\centering
\subfigure[]{\includegraphics[width=2.9in,trim= 90 180 90 180,clip=true]{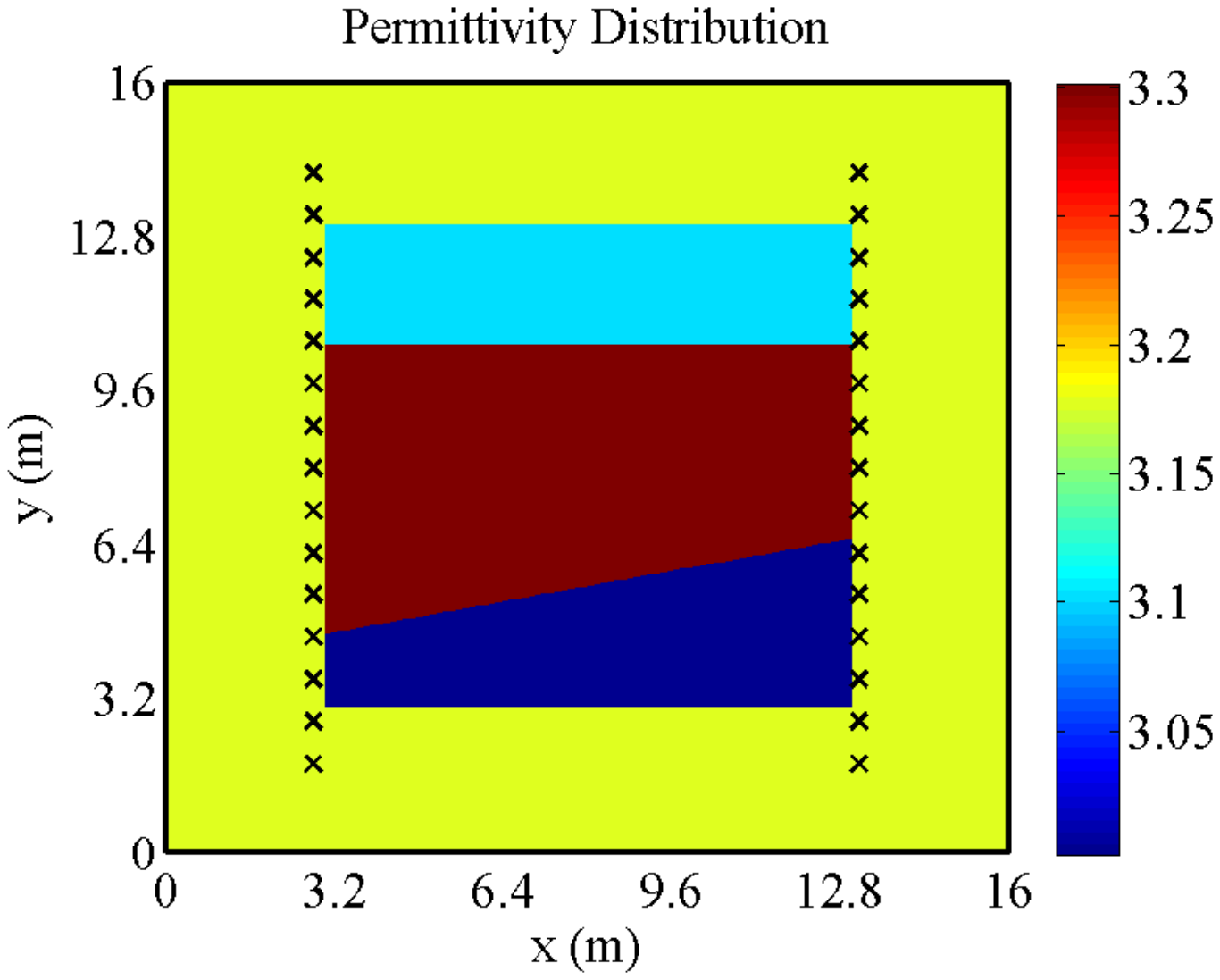}}
\subfigure[]{\includegraphics[width=2.9in,trim= 90 180 90 180,clip=true]{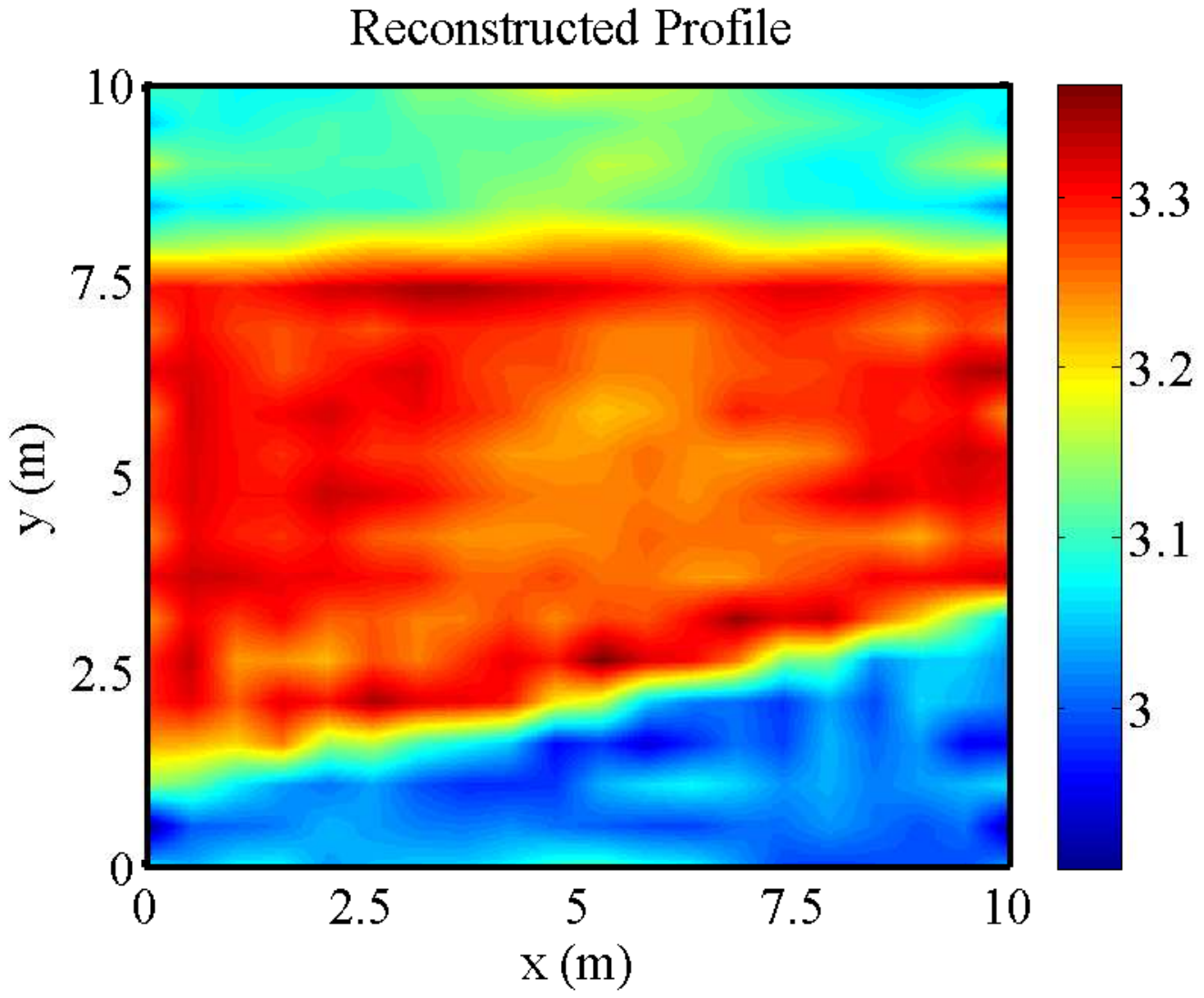}}
\subfigure[]{\includegraphics[width=2.9in,trim= 90 180 90 180,clip=true]{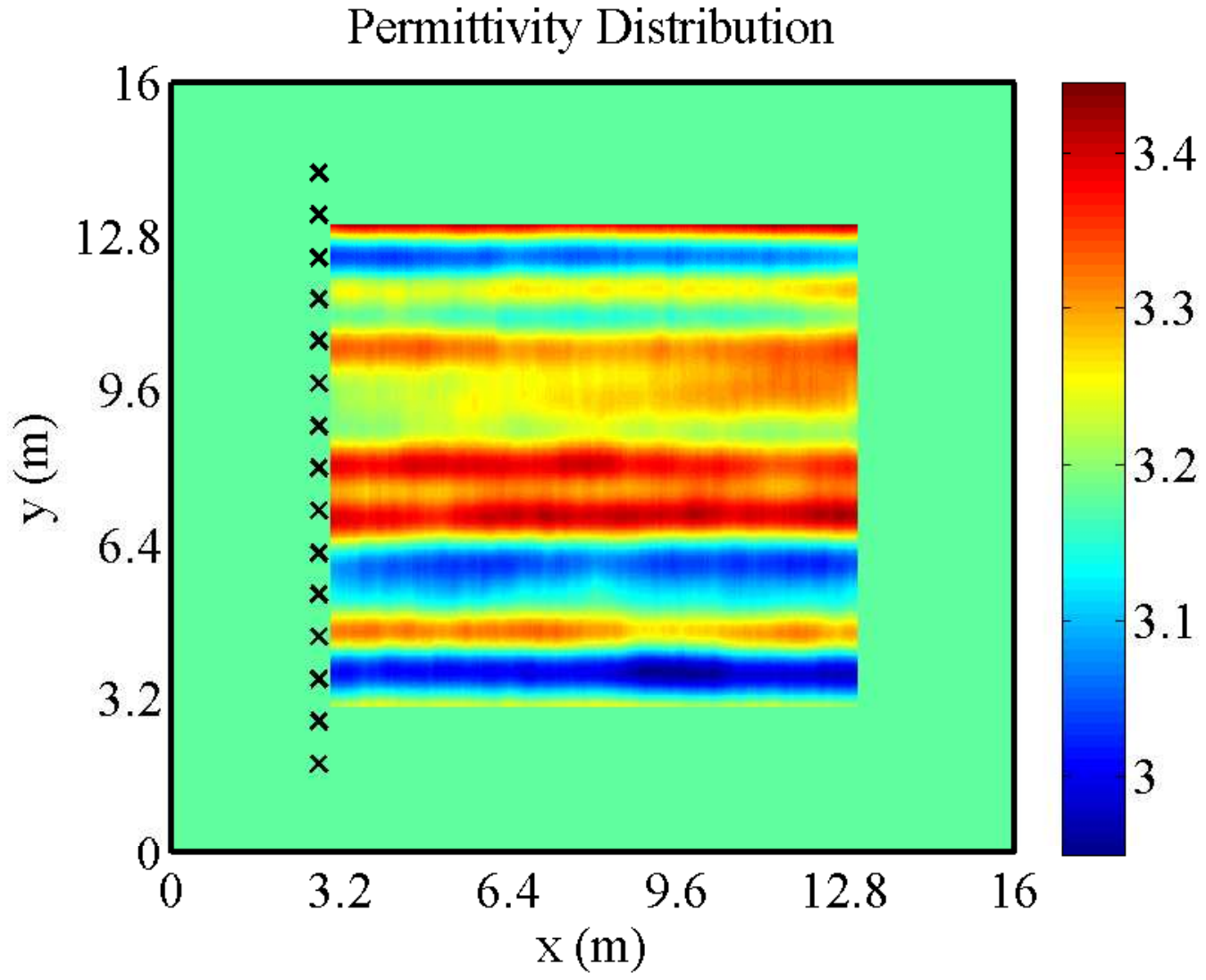}}
\subfigure[]{\includegraphics[width=2.9in,trim= 90 180 90 180,clip=true]{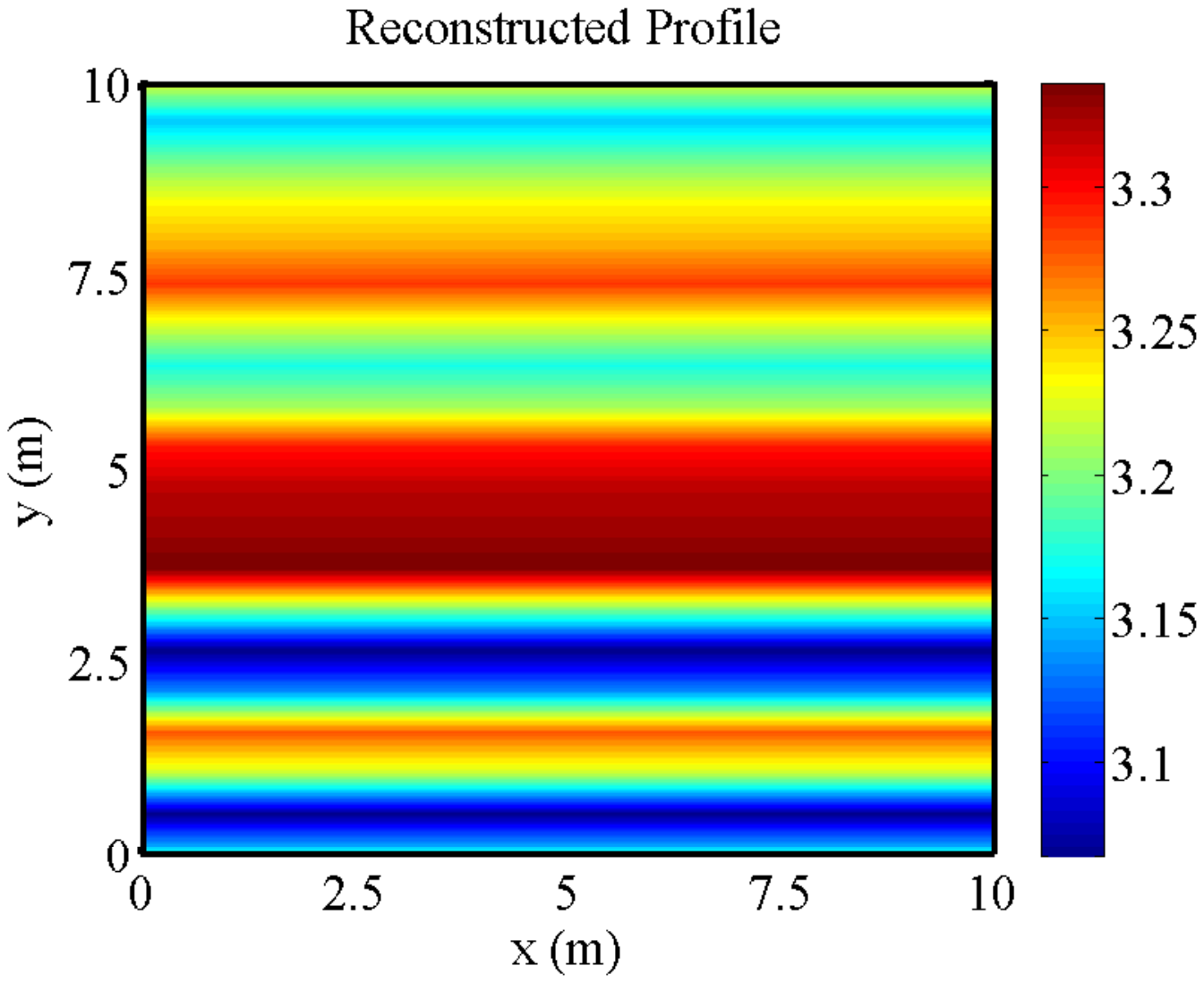}}
\caption{Forward problem and reconstructed permittivity profiles for layered media. (a)-(b) Slanted layers imaged by $N_s$=30 CH sensors. (c)-(d) (Quasi)-horizontally layered medium imaged by $N_s$=15 BH sensors. This setup is efficiently solved as 1-D inverse problem. SNR=10 dB.}
\label{FigCH6_8}
\end{figure}
Two examples of layered media are shown in Fig. \ref{FigCH6_8}. In Fig \ref{FigCH6_8}(a), CH sensors are used to reconstruct a slanted layered medium with abrupt changes in permittivity. The problem is solved as a 2-D problem and the reconstructed profile is shown in Fig. \ref{FigCH6_8}(b). Fig. \ref{FigCH6_8}(c) shows a quasi-horizontally layered medium. It is a realization of an anisotropic continuous random Gaussian medium with $l_{cx}$=125 m along $x$-direction and $l_{cy}$=1.25 m along $y$-direction. This is a good model for layered Earth formations encountered in geophysical exploration \cite{teixeirLWD1, teixeirLWD2, teixeirLWD3}. Prior knowledge of the layered nature of problem can simplify the inversion significantly by solving the problem as 1-D inversion problem (i.e. restricting the unknowns to spatial harmonics along $y$-direction), as shown in Fig. \ref{FigCH6_8}(d) for a BH scenario. The linear array shown in Fig. \ref{FigCH6_8}(c) can be deployed horizontally along $x$-direction as a surface controlled source electromagnetic (CSEM) array. In that case, the problem becomes 1-D along a direction normal to the array. Shown results are for a single iteration inversion. For high contrast media, BDBIM can be used.

\section{Conclusion}
An approach based on Bayesian compressive sensing (in a broad sense of the term) was applied for ultrawideband multistatic inverse scattering problems in continuous random media. It was shown that UWB BCS not only provides accurate reconstructions in different scenarios but also provides means for estimating the accuracy of the inversion. In addition, it allows for a systematic way of determining optimal locations so that the information gain is maximized using sequential measurements.  Furthermore, time-reversal-based focusing was combined with UWB BCS to achieve localized, adaptive inversion and reduce overall inversion costs.  The proposed methodology was successfully applied to a number of canonical geophysical imaging problems.


\section*{Acknowledgments}
This work has been supported by the National Science Foundation (NSF) under grant ECCS-0925272 and by the Ohio Supercomputing Center (OSC) under grant PAS-0110.

\ifCLASSOPTIONcaptionsoff
  \newpage
\fi






\bibliographystyle{IEEEtran}
\bibliography{RefsDissrtationAll}

\begin{thebibliography}{10}
\providecommand{\url}[1]{#1}
\csname url@samestyle\endcsname
\providecommand{\newblock}{\relax}
\providecommand{\bibinfo}[2]{#2}
\providecommand{\BIBentrySTDinterwordspacing}{\spaceskip=0pt\relax}
\providecommand{\BIBentryALTinterwordstretchfactor}{4}
\providecommand{\BIBentryALTinterwordspacing}{\spaceskip=\fontdimen2\font plus
\BIBentryALTinterwordstretchfactor\fontdimen3\font minus
  \fontdimen4\font\relax}
\providecommand{\BIBforeignlanguage}[2]{{%
\expandafter\ifx\csname l@#1\endcsname\relax
\typeout{** WARNING: IEEEtran.bst: No hyphenation pattern has been}%
\typeout{** loaded for the language `#1'. Using the pattern for}%
\typeout{** the default language instead.}%
\else
\language=\csname l@#1\endcsname
\fi
#2}}
\providecommand{\BIBdecl}{\relax}
\BIBdecl

\bibitem{OceanAcustInv}
P.~Gerstoft and C.~F. Mecklenbrauker, ``Ocean acoustic inversion with
  estimation of a posteriori probability distributions,'' \emph{J. Acoust. Soc.
  Amer.}, vol. 104, no.~2, pp. 808--819, 1998.

\bibitem{EMMarine}
M.~Birsan, ``A \uppercase{B}ayesian approach to electromagnetic sounding in a
  marine environment,'' \emph{IEEE Trans. Geosci. Remote Sensing}, vol.~41,
  no.~6, pp. 1455--1460, 2003.

\bibitem{MelesGPR1}
J.~R. Ernst, H.~Maurer, A.~G. Green, and K.~Holliger, ``Full-waveform inversion
  of crosshole radar data based on 2-\uppercase{D} finite-difference time
  domain solutions of \uppercase{M}axwell’s equations,'' \emph{IEEE Trans.
  Geosci. Remote Sensing}, vol.~45, no.~9, pp. 2807--2828, Sept. 2007.

\bibitem{MelesGPR2}
G.~A. Meles, J.~V. der Kruk, S.~A. Greenhalgh, J.~R. Ernst, H.~Maurer, and
  A.~G. Green, ``A new vector waveform inversion algorithm for simultaneous
  updating of conductivity and permittivity parameters from combination
  crosshole/borehole-to-surface \uppercase{GPR} data,'' \emph{IEEE Trans.
  Geosci. Remote Sensing}, vol.~48, no.~9, pp. 3391--3407, Sept. 2010.

\bibitem{InvDORT}
T.~Zhang, P.~C. Chaumet, E.~Mudry, A.~Sentenac, and K.~Belkebir,
  ``Electromagnetic wave imaging of targets buried in a cluttered medium using
  a hybrid inversion-\uppercase{DORT} method,'' \emph{Inv. Prob.}, vol.~28,
  no.~12, p. 125008, 2012.

\bibitem{DevaneyInvScatt}
M.~Dennison and A.~J. Devaney, ``Inverse scattering in inhomogeneous background
  media,'' \emph{Inv. Prob.}, vol.~19, no.~4, pp. 855--870, 2003.

\bibitem{BreastImagBook}
G.~A. Ybarra and Q.~H. Liu, \emph{Emerging Technologies in Breast Imaging and
  Mammography}.\hskip 1em plus 0.5em minus 0.4em\relax American Scientific
  Publishers, 2008, ch. 16: Breast Imaging Using Electrical Impedance
  Tomography.

\bibitem{BreastReview}
A.~Hassan and M.~El-Shenawee, ``Review of electromagnetic techniques for breast
  cancer detection,'' \emph{IEEE Rev. Biomed. Eng.}, vol.~4, no.~9, pp.
  103--118, 2011.

\bibitem{BayesInfBrain}
D.~Schmidt, J.~George, and C.~Wood, ``Bayesian inference applied to the
  electromagnetic inverse problem,'' \emph{Hum. Brain Map.}, vol.~7, pp.
  195--212, 1999.

\bibitem{HabashyBiomedInv}
C.~Gilmore, A.~Abubakar, W.~Hu, T.~M. Habashy, and P.~M. van~der Berg,
  ``Microwave biomedical data inversion using the finite-difference contrast
  source inversion method,'' \emph{IEEE Trans. Antennas Propag.}, vol.~57,
  no.~5, pp. 1528--1538, 2009.

\bibitem{HagnessBreast1}
D.~W. Winters, J.~D. Shea, E.~L. Madsen, G.~R. Frank, B.~D.~V. Veen, and S.~C.
  Hagness, ``Estimating the breast surface using \uppercase{UWB} microwave
  monostatic backscatter measurements,'' \emph{IEEE Trans. Biomed. Eng.},
  vol.~55, no.~1, pp. 247--256, 2008.

\bibitem{HagnessBreast2}
S.~K. Davis, B.~D.~V. Veen, S.~C. Hagness, and F.~Kelcz, ``Breast tumor
  characterization based on ultrawideband microwave backscatter,'' \emph{IEEE
  Trans. Biomed. Eng.}, vol.~55, no.~1, pp. 237--246, 2008.

\bibitem{thruwallSciencelong}
E.~J. Baranoski, ``Through-wall imaging: Historical perspective and future
  directions,'' \emph{J. Franklin Inst.}, vol. 345, pp. 556--569, September
  2008.

\bibitem{UWBthruwall}
A.~M. Attiya, A.~Bayram, A.~Safaai-Jazi, and S.~M. Riad, ``\uppercase{UWB}
  applications for through-wall detection,'' \emph{in Proc. IEEE Antennas
  Propag. Int. Symp.}, vol.~3, pp. 3079--3082, June 2004.

\bibitem{THRUWALLLI}
L.~Li, W.~Zhang, and F.~Li, ``A novel autofocusing approach for real-time
  through-wall imaging under unknown wall characteristics,'' \emph{IEEE Trans.
  Geosci. Remote Sensing}, vol.~48, pp. 423 -- 431, 2010.

\bibitem{EMNDT}
T.~Takagi, J.~R. Bowler, and Y.~Yoshida, \emph{Electromagnetic nondestructive
  evaluation}.\hskip 1em plus 0.5em minus 0.4em\relax Amsterdam, Netherlands:
  IOS Press Inc, 1997.

\bibitem{MwaveNDT1}
T.~Lasri and R.~Zoughi, ``Advances and applications in microwave and millimeter
  wave nondestructive evaluation,'' \emph{Subsurface Sensing Technol.
  Applicat.}, vol.~2, no.~4, Oct. 2001.

\bibitem{MwaveNDT2}
D.~Lesselier and J.~Bowler, ``Electromagnetic and ultrasonic nondestructive
  evaluation,'' \emph{Inv. Prob.}, vol.~18, no.~6, Dec. 2002.

\bibitem{TRGPRPlumb}
C.~J. Leuschen and R.~G. Plumb, ``A matched-filter-based reverse-time migration
  algorithm for ground-penetrating radar data,'' \emph{IEEE Trans. Geosci.
  Remote Sensing}, vol.~39, pp. 1257--1264, 2006.

\bibitem{TRGPRAsif}
F.~Foroozan and A.~Asif, ``Time-reversal ground-penetrating radar: Range
  estimation with cramér–rao lower bounds,'' \emph{IEEE Trans. Geosci.
  Remote Sensing}, vol.~48, pp. 3698 --3708, 2010.

\bibitem{RappaportStatisticalGPR}
X.~Xu, E.~L. Miller, C.~M. Rappaport, and G.~D. Sower, ``Statistical method to
  detect subsurface objects using array ground-penetrating radar data,''
  \emph{IEEE Trans. Geosci. Remote Sensing}, vol.~40, no.~4, pp. 963--976,
  2002.

\bibitem{PotinGPR}
D.~Potin, E.~Duflos, and P.~Vanheeghe, ``Landmines ground-penetrating radar
  signal enhancement by digital filtering,'' \emph{IEEE Trans. Geosci. Remote
  Sensing}, vol.~44, no.~9, pp. 2393--2406, 2006.

\bibitem{HabashyCrossWell}
M.~Li, A.~Abubakar, and T.~M. Habashy, ``Application of a two-and-a-half
  dimensional model-based algorithm to crosswell electromagnetic data
  inversion,'' \emph{Inv. Prob.}, vol.~26, no.~7, 2010.

\bibitem{ThreDEMBook}
B.~Spies and M.~Oristaglio, \emph{Three-Dimensional Electromagnetics}.\hskip
  1em plus 0.5em minus 0.4em\relax Tulsa, OK: Society Of Exploration
  Geophysicists, 1999.

\bibitem{TarantolaBook}
A.~Tarantola, \emph{Inverse Problem Theory and Methods for Model Parameter
  Estimation}.\hskip 1em plus 0.5em minus 0.4em\relax Philadelphia, PA: SIAM,
  2005.

\bibitem{BayesianInvProbBook}
J.~Idier, \emph{Bayesian Approach to Inverse Problems}.\hskip 1em plus 0.5em
  minus 0.4em\relax NJ: John Wiley and Sons, Inc., 2010.

\bibitem{LemmBook}
J.~C. Lemm, \emph{Bayesian Field Theory}.\hskip 1em plus 0.5em minus
  0.4em\relax Baltimore MD: Johns Hopkins University Press, 2003.

\bibitem{MassaContSourc}
G.~Oliveri, P.~Rocca, and A.~Massa, ``A
  \uppercase{B}ayesian-compressive-sampling-based inversion for imaging sparse
  scatterers,'' \emph{IEEE Trans. Geosci. Remote Sensing}, vol.~49, no.~10, pp.
  3993--4006, Oct. 2011.

\bibitem{MassaBorn}
L.~Poli, G.~Oliveri, and A.~Massa, ``Microwave imaging within the first-order
  \uppercase{B}orn approximation by means of the contrast-field
  \uppercase{B}ayesian compressive sensing,'' \emph{IEEE Trans. Antennas
  Propagat.}, vol.~60, no.~6, pp. 2865--2879, June 2012.

\bibitem{BayesExpDes}
K.~Chaloner and I.~Verdinelli, ``Bayesian experimental design: \uppercase{A}
  review,'' \emph{Statistical Science}, vol.~10, no.~3, pp. 237--304, 1995.

\bibitem{BCSCarin}
S.~Ji, Y.~Xue, and L.~Carin, ``Bayesian compressive sensing,'' \emph{IEEE
  Trans. Signal Processing}, vol.~56, no.~6, pp. 2346--2356, June 2008.

\bibitem{MTCarin}
S.~Ji, D.~Dunson, and L.~Carin, ``Multi-task compressive sensing,'' \emph{IEEE
  Trans. Signal Processing}, vol.~57, no.~1, pp. 92--106, Jan. 2009.

\bibitem{TippingSparse}
M.~E. Tipping, ``Sparse \uppercase{B}ayesian learning and the relevance vector
  machine,'' \emph{J. Machine Learning Res.,}, vol.~1, pp. 211--244, 2001.

\bibitem{TippingFastRVM}
M.~E. Tipping and A.~C. Faul, ``Fast marginal likelihood maximisation for
  sparse \uppercase{B}ayesian models,'' in \emph{Proc. of the 9th Int. Workshop
  Artificial Intelligence and Statistics}, Jan. 3-6 2003.

\bibitem{MarengoCS1}
E.~A. Marengo, R.~D. Hernandez, Y.~R. Citron, F.~K. Gruber, M.~Zambrano, and
  H.~Lev-Ari, ``Compressive sensing for inverse scattering,'' in \emph{Proc.
  XXIX URSI Gen. Assem.}, Jan. 7-16 2008.

\bibitem{MarengoCS2}
E.~A. Marengo, ``Compressive sensing and signal subspace methods for inverse
  scattering including multiple scattering,'' in \emph{Proc. of the IEEE
  Geoscience Remote Sensing Symp.}, Jul. 7-11 2008.

\bibitem{MarengoCS3}
------, ``Subspace and \uppercase{B}ayesian compressive sensing methods in
  imaging,'' in \emph{Proc. Prog. Electromagn. Res. Symp.}, Jul. 2-6 2008.

\bibitem{HaradaConjGrad}
H.~Harada, D.~J. Wall, T.~Takenaka, and T.~Tanaka, ``Conjugate gradient method
  applied to inverse scattering problems,'' \emph{IEEE Trans. Antennas
  Propag.}, vol.~43, no.~8, pp. 784--792, 1995.

\bibitem{Moghadam1}
M.~Moghadam and W.Chew, ``Nonlinear two-dimensional velocity profile inversion
  using time domain data,'' \emph{IEEE Trans. Geosci. Remote Sensing}, vol.~30,
  no.~1, pp. 147--156, 1992.

\bibitem{Weedon1}
W.~Weedon and W.~Chew, ``Time-domain inverse scattering using the local shape
  function (\uppercase{LSF}) method,'' \emph{Inv. Prob.}, vol.~9, pp. 551--564,
  1993.

\bibitem{MoraInv}
P.~Mora, ``Nonlinear two-dimensional elastic inversion of multi-offset seismic
  data,'' \emph{Geophysics}, vol.~52, no.~9, pp. 1211--1228, 1987.

\bibitem{WangDBIM}
Y.~Wang and W.~Chew, ``Reconstruction of two-dimensionsal permittivity
  distribution using the \uppercase{D}istorted \uppercase{B}orn
  \uppercase{I}terative \uppercase{M}ethod,'' \emph{IEEE Trans. Med. Imag.},
  vol.~9, no.~2, pp. 218--225, 1990.

\bibitem{PDFSampling}
A.~E. Gelfand and A.~F.~M. Smith, ``Sampling-based approaches to calculating
  marginal densities,'' \emph{Journal of the American Statistical Association},
  vol.~85, no. 410, pp. 398--409, 1990.

\bibitem{Gibbssampling}
B.~Walsh, ``Markov \uppercase{C}hain \uppercase{M}onte \uppercase{C}arlo and
  \uppercase{G}ibbs sampling,'' 2004, retrieved March 10, 2013, from
  http://web.mit.edu/~wingated/www/introductions/mcmc-gibbs-intro.pdf.

\bibitem{Fouda2}
A.~E. Fouda and F.~L. Teixeira, ``Statistical stability of ultrawideband
  time-reversal imaging in random media,'' \emph{IEEE Trans. Geosci. Remote
  Sensing}, vol.~52, no.~2, pp. 870--879, 2014.

\bibitem{Fouda4}
A.~E. Fouda, V.~Lopez-Castellanos, and F.~L. Teixeira, ``Experimental
  demonstration of statistical stability in ultrawideband time-reversal
  imaging,'' \emph{IEEE Geosci. Remote Sens. Lett.}, vol.~11, no.~1, pp. 29--33, 2014.

\bibitem{CandesCS}
E.~J. Candes and M.~B. Wakin, ``An introduction to compressive sampling,''
  \emph{IEEE Signal Process. Mag.}, vol.~25, no.~2, pp. 21--30, 2008.

\bibitem{PotterCS}
L.~C. Potter, E.~Ertin, J.~T. Parker, and M.~Cetin, ``Sparsity and compressed
  sensing in radar imaging,'' \emph{Proc. IEEE}, vol.~98, no.~6, pp.
  1006--1020, 2010.

\bibitem{TafloveBook}
A.~Taflove and S.~Hagness, \emph{Computational Electrodynamics: The
  Finite-Difference Time-Domain Method}, 3rd~ed.\hskip 1em plus 0.5em minus
  0.4em\relax Norwood, MA: Artech House, 2005.

\bibitem{CoverITBook}
T.~M. Cover and J.~A. Thomas, \emph{Elements of Information Theory}.\hskip 1em
  plus 0.5em minus 0.4em\relax NY: Wiley, 1991.

\bibitem{FinkTRmirrors}
M.~Fink, ``Time reversal mirrors,'' \emph{J. Phys. D.: Appl. Phys.}, vol.~26,
  no.~9, pp. 1333--1350, 1993.

\bibitem{YavuzSenstivPerturb}
M.~E. Yavuz and F.~L. Teixeira, ``On the sensitivity of time-reversal imaging
  techniques to model perturbations,'' \emph{IEEE Trans. Antennas Propagat.},
  vol.~56, pp. 834--843, 2008.

\bibitem{Fouda1}
A.~E. Fouda and F.~L. Teixeira, ``Imaging and tracking of targets in clutter
  using differential time-reversal techniques,'' \emph{Waves in Random and
  Complex Media}, vol.~22, no.~1, pp. 66--108, 2012.

\bibitem{TRinchangingmedia}
D.~Liu, S.~Vasudevan, J.~Krolik, G.~Bal, and L.~Carin, ``Electromagnetic
  time-reversal source localization in changing media: Experiment and
  analysis,'' \emph{IEEE Trans. Antennas Propagat.}, vol.~55, pp. 344--354,
  2007.

\bibitem{MoghadamDBIM}
M.~Moghadam, W.~Chew, and M.~Oristaglio, ``Comparison of the \uppercase{B}orn
  iterative method and \uppercase{T}arantola's method for an electromagnetic
  time-domain inverse problem,'' \emph{International Journal of Imaging Systems
  and Technology}, vol.~3, no.~4, pp. 318--333, 1991.

\bibitem{WangDBIM2}
Y.~Wang and W.~Chew, ``An iterative solution of the two-dimensional
  electromagnetic inverse scattering problem,'' \emph{International Journal of
  Imaging Systems and Technology}, vol.~1, no.~1, pp. 100--108, 1989.

\bibitem{teixeirLWD1}
G.-S. Liu, F.~L. Teixeira, and G.-J. Zhang, ``Analysis of directional logging
  tools in anisotropic and multieccentric cylindrically-layered
  \uppercase{E}arth formations,'' \emph{IEEE Trans. Antennas Propag.}, vol.~60,
  no.~1, pp. 318--327, 2012.

\bibitem{teixeirLWD2}
Y.-K. Hue and F.~L. Teixeira, ``Numerical mode-matching method for tilted coil
  antennas in cylindrically layered anisotropic media with multiple horizontal
  beds,'' \emph{IEEE Trans. Geosci. Remote Sens.}, vol.~45, no.~8, pp.
  2451--2462, 2007.

\bibitem{teixeirLWD3}
H.~O. Lee and F.~L. Teixeira, ``Cylindrical \uppercase{FDTD} analysis of
  \uppercase{LWD} tools through anisotropic dipping-layered \uppercase{E}arth
  media,'' \emph{IEEE Trans. Geosci. Remote Sens.}, vol.~45, no.~2, pp.
  383--388, 2007.

\end{thebibliography}

\end{document}